\documentclass[%
 aps,pre,
 amsmath,amssymb,
 reprint,%
eqsecnum]{revtex4-2}
\usepackage{pdfpages}
\usepackage{pgffor}
\makeatletter
\AtBeginDocument{\let\LS@rot\@undefined}
\makeatother

\usepackage{graphicx}
\usepackage{dcolumn}
\usepackage{bm}

\usepackage[utf8]{inputenc}
\usepackage[T1]{fontenc}
\usepackage{etoolbox}
\usepackage{mathtools}
\usepackage{xcolor}

\def\@email#1#2{%
 \endgroup
 \patchcmd{\titleblock@produce}
  {\frontmatter@RRAPformat}
  {\frontmatter@RRAPformat{\produce@RRAP{*#1\href{mailto:#2}{#2}}}\frontmatter@RRAPformat}
  {}{}
}%
\usepackage{wrapfig}
\usepackage{float}
\usepackage{physics}
\usepackage{bbold}
\usepackage{tabularx}
\usepackage[unicode=true,bookmarks=true,bookmarksnumbered=false,bookmarksopen=false, breaklinks=false,pdfborder={0 0 0},pdfborderstyle={},backref=false,colorlinks=true]{hyperref}

\hypersetup{pdfborderstyle={},pdfborderstyle={},pdfborderstyle={},pdfborderstyle={},pdfborderstyle={},pdfborderstyle={},linkcolor=blue,citecolor=blue}

\newcommand{\dif}{\mathrm{d}}
\newcommand{\cc}{\mathbf{c}}
\newcommand{\cca}{\mathbf{c}_1}
\newcommand{\ccb}{\mathbf{c}_2}
\newcommand{\vv}{\mathbf{v}}
\newcommand{\vva}{\mathbf{v}_1}
\newcommand{\vvb}{\mathbf{v}_2}
\newcommand{\vvab}{\mathbf{v}_{12}}
\newcommand{\ccab}{\mathbf{c}_{12}}

\newcommand{\Wwab}{\mathbf{W}_{12}}
\newcommand{\Wab}{W_{12}}
\newcommand{\wwa}{\mathbf{w}_1}
\newcommand{\wwb}{\mathbf{w}_2}

\newcommand{\cab}{c_{12}}

\newcommand{\ca}{c_{1}}
\newcommand{\wa}{w_{1}}
\newcommand{\cb}{c_{2}}
\newcommand{\wb}{w_{2}}
\newcommand{\vth}{v_{\mathrm{th}}}
\newcommand{\wth}{\omega_{\mathrm{th}}}
\newcommand{\oo}{\boldsymbol{\omega}}
\newcommand{\ooa}{\boldsymbol{\omega}_1}
\newcommand{\oob}{\boldsymbol{\omega}_2}
\newcommand{\ww}{\mathbf{w}}
\newcommand{\s}{\widehat{\boldsymbol{\sigma}}}
\newcommand{\dt}{{d_t}}
\newcommand{\dr}{{d_r}}
\newcommand{\Tt}{T_t}
\newcommand{\Trot}{T_r}
\newcommand{\llangle}{\langle\!\langle}
\newcommand{\rrangle}{\rangle\!\rangle}

\newcommand{\en}{\overline{\alpha}}
\newcommand{\et}{\overline{\beta}}
\newcommand{\een}{\alpha}
\newcommand{\eet}{\beta}
\newcommand{\ddab}{\boldsymbol{\Delta}_{12}}

\newcommand{\cuma}[3]{{a_{#1 #2}^{(#3)}}}

\newcommand{\bbop}{\mathcal{B}_{12,\s}}

\newcommand{\HCS}{\mathrm{H}}
\newcommand{\vom}{{\mathbf{\Gamma}}}
\newcommand{\cw}{{\widetilde{\mathbf{\Gamma}}}}
\newcommand{\JJ}{J}

\begin{document}

\title[Non-Gaussianities in freely evolving granular gases]{Translational and rotational non-Gaussianities in homogeneous freely evolving granular gases}

\author{Alberto Meg\'ias}
 \email{albertom@unex.es}
 \affiliation{Departamento de F\'isica, Universidad de Extremadura, E-06006 Badajoz, Spain}

\author{Andr\'es Santos}%
 \email{andres@unex.es}
 \affiliation{Departamento de F\'isica, Universidad de Extremadura, E-06006 Badajoz, Spain}
\affiliation{%
Instituto de Computaci\'on Cient\'ifica Avanzada (ICCAEx), Universidad de Extremadura, E-06006 Badajoz, Spain
}%

\date{\today}

\begin{abstract}
     The importance of roughness in the modeling of granular gases has been increasingly considered in  recent years. In this paper, a freely evolving homogeneous granular gas of inelastic and rough hard disks or spheres is studied under the assumptions of the Boltzmann kinetic equation. The homogeneous cooling state is studied from a theoretical point of view using a Sonine approximation, in contrast to a previous Maxwellian approach. A general theoretical description is done in terms of $\dt$  translational and $\dr$ rotational degrees of freedom, which accounts for the cases of spheres ($\dt=\dr=3$) and disks  ($\dt=2$, $\dr=1$) within a unified framework. The non-Gaussianities of the velocity distribution function of this state are determined by means of the first nontrivial cumulants and by the derivation of non-Maxwellian high-velocity tails. The results are validated by computer simulations using direct simulation Monte Carlo and event-driven molecular dynamics algorithms.
\end{abstract}
\maketitle

\section{Introduction}

Granular systems are themselves worth studying from mechanical, physical, and mathematical points of view. They are very commonly observed in nature, where different geometries can take place. Grains, from a dynamical point of view, move in a three-dimensional space, but constraints make two-dimensional problems become real and of special interest~\cite{CR91,FM02,PDB03,YEHLPG11,BNR15,GBM15,SP17,GBMV17,LGRAYV21,LCD21}.

We will focus on the description of granular systems at low-density fluidized states, where the assumptions underlying the Boltzmann equation apply~\cite{M93b,GS95,GZB97,SG98,D00,D01,GD02,BP03,GNB05b,SGNT06,VU09,VSG10,DB11,GS11,GSVP11a,GSVP11b,GMT13,G19}. The simplest collisional model for interactions in granular gaseous flows is the inelastic hard-sphere model, where the granular gas is assumed to be composed by inelastic and smooth identical hard disks, spheres, or hyperspheres in $\dt$ translational space dimensions~\cite{C90,PL01,G03,BP04,G19}. However, this description might be limiting and can be improved by considering rotational degrees of freedom, which may play an important role in the dynamics of granular gases by means of surface roughness. Here, we will use the simplest collisional model that implements roughening, the inelastic and rough hard-sphere model. In the latter model, the inelasticity is parameterized by a constant coefficient of normal restitution, $\een$ (in common with the inelastic hard-sphere model), and roughness is introduced by means of a coefficient of tangential restitution, $\eet$. Although, in general, the effective coefficient of tangential restitution depends on the impact angle because of friction~\cite{MBF76,LTL97}, here we adopt the simplest model with constant $\beta$, as frequently done in the literature~\cite{GS95,AHZ01,BP04,BPKZ07,SKG10,SKS11,VSK14,VSK14b,VS15,S18,G19,MS19,MS19b}.

Whereas in the inelastic hard-sphere model, the $\dt$-dimensional description of an inelastic gas of smooth and spinless hard (hyper)spheres is straightforward, in the case of rough spheres, where angular velocities come into play, rotational degrees of freedom, $\dr$, need to be introduced. A description in terms of $\dt$ and $\dr$ becomes highly dependent on the geometry and constraints of the system as antecedently reported~\cite{MS19,MS19b,MS21a,MS21b}. As in Ref.~\cite{MS21a}, we will derive the general description to be valid just for the two relevant cases of hard disks and hard spheres. For hard disks, angular velocities are constrained to the direction orthogonal to the plane of motion, so $\dt=2$ and $\dr=1$. For hard spheres, on the other hand, angular and translational velocities are vectors of a Euclidean three-dimensional vector space, i.e., $\dt=\dr=3$.

It is widely known and studied that the homogeneous Boltzmann equation---for both smooth and rough models---admits a scaling solution, in which the system cools down continuously and the whole evolution is driven by the granular temperature. This state is known as the \emph{homogeneous cooling state} (HCS), and has been of interest for the granular gas community in the last three decades~\cite{BP04,G19,BRC96,vNE98,BRC99,BP00,MS00,BP06ab,AP06,AP07,SM09,S18,MS19,KGS14,VSK14}. 
It is worth mentioning that, very recently, the HCS has been experimentally observed in microgravity experiments by Yu \emph{et al.}~\cite{YSS20}. In that paper, both Haff's cooling law and the exponential high-velocity tail of the velocity distribution function (VDF) as predicted by kinetic theory together with the inelastic hard-sphere model are confirmed. In Ref.~\cite{YSS20}, while the results were compared with the constant and velocity-dependent models for the coefficient of normal restitution, it was concluded that the latter model had a negligible influence on the results, supporting the approximation of constant coefficients of restitution. In fact, the system in Ref.~\cite{YSS20} is compatible with a constant $\een= 0.66$, highlighting that the latter approximation is not subjected only to quasielastic systems. Moreover, the authors proposed a possible influence of surface roughness in the collisional rules due to an overestimate of the relaxation time from the inelastic hard-sphere model as compared with the experimental outcomes. Thus, they claimed that the rotational degrees of freedom could be an answer to these deviations.

Theoretically, some of the early attempts to study the Gaussian deviations of the HCS VDF of a granular gas of rough particles were done in Refs.~\cite{GS95,AHZ01} using a Sonine expansion, that is, an isotropic expansion around a two-temperature---translational and rotational---Maxwellian VDF. However, although velocity correlations were not originally assumed for hard spheres, they were proved to be present~\cite{BPKZ07}. More recently, the first nontrivial velocity cumulants were studied for freely evolving hard spheres~\cite{VSK14}. Throughout the present paper, we will expose the results of those cumulants in a common frame for both disks and spheres from the collisional-moment point of view, and we will analyze results for hard disks.

In the case of freely cooling inelastic granular gases, deviations of the HCS VDF from a Maxwellian derived from the inelastic hard-sphere model are not only accounted for by the first nontrivial cumulants, but also an exponential high-velocity tail for this distribution was predicted by kinetic theory and computer simulations~\cite{EP97,vNE98,BRC99,MS00,PBF06} and satisfactorily observed experimentally not only for freely evolving granular gases~\cite{YSS20} as commented above, but also for uniformly heated systems~\cite{CZ22}. Results in Ref.~\cite{VSK14} put into manifest a highly populated tail for the marginal VDF of angular velocities in the inelastic and rough hard-sphere model, accompanied by high values of the fourth angular velocity cumulant for some values of the pair $(\alpha,\beta)$. However, this marginal distribution was interpreted as being consistent with an exponential form~\cite{VSK14},  similarly to what occurs in the inelastic hard-sphere model with the total (translational) VDF~\cite{EP97,vNE98,BRC99}. In this paper, we study the high-velocity tail for the marginal VDF of the translational and angular velocities, and for their product as well, where theory  indicates algebraic tails for the two latter marginal distributions.

The study of the non-Gaussianities of the HCS VDF is also motivated by recent research in nonhomogeneous states~\cite{KSG14,MS21a,MS21b} from Chapman--Enskog expansions around the HCS solution. Linear stability analyses of the homogeneous state hydrodynamics show that the Maxwellian approximation of the first-order VDF might not work for the hard-disk case \cite{MS21b}, and for some very small region of the parameter space for hard spheres, yielding a wrong prediction of a completely unstable region in the parameter space. Therefore, it was conjectured that non-Gaussianities might be crucial in the cited region of parameters, this effect being more important for disks than for spheres~\cite{MS21b}. This hypothesis was supported by high values of the first relevant cumulants in hard-sphere systems~\cite{VSK14} and by results from the smooth case, where the homogeneous VDF is generally more disparate from a Maxwellian for disks than for spheres~\cite{MS20}.

The present paper is structured as follows. In Sec.~\ref{sec:2}, the inelastic and rough hard-sphere model and the binary collisional rules are introduced. Afterwards, the framework of the homogeneous Boltzmann equation is described in Sec.~\ref{sec:3} and the hierarchy of evolution equations, the Sonine expansion of the VDF, and the definitions of cumulants are formally presented. Section~\ref{sec:3bis} is devoted to the Sonine approximation, where the infinite expansion is truncated beyond the first few nontrivial coefficients, the associated collisional moments are explicitly written for hard disks, and the HCS cumulants are obtained. Next, we study the forms of the marginal VDF from the Maxwellian and Sonine approximations, as well as their high-velocity tails in Sec.~\ref{sec:MVDFs} in the context of the Boltzmann equation. In Sec.~\ref{sec:Sim_Res}, we compare the theoretical predictions with direct simulation Monte Carlo (DSMC) and event-driven molecular dynamics (EDMD) computer simulations outcomes. Finally, concluding remarks and main results are summarized in Sec.~\ref{sec:conclusions}.

\section{\label{sec:2} Inelastic and Rough Hard Particles}

\subsection{System}

Let us consider a monodisperse dilute granular gas of hard disks or spheres, which are assumed to be inelastic and rough, their dynamics being described by their translational and angular velocities, $\vv$ and $\oo$, respectively. Whereas for spheres ($\dt=\dr=3$), both $\vv$ and $\oo$ are vectors in an Euclidean three-dimensional space, this is not the case for disks ($\dt=2$, $\dr=1$), where $\oo$ is a one-dimensional vector orthogonal to the two-dimensional vector space spanned by $\vv$. In general, however, all vector relations will be written in a three-dimensional Euclidean embedding space.

\begin{figure}[ht]
	\centering
	\includegraphics[width=0.24\textwidth]{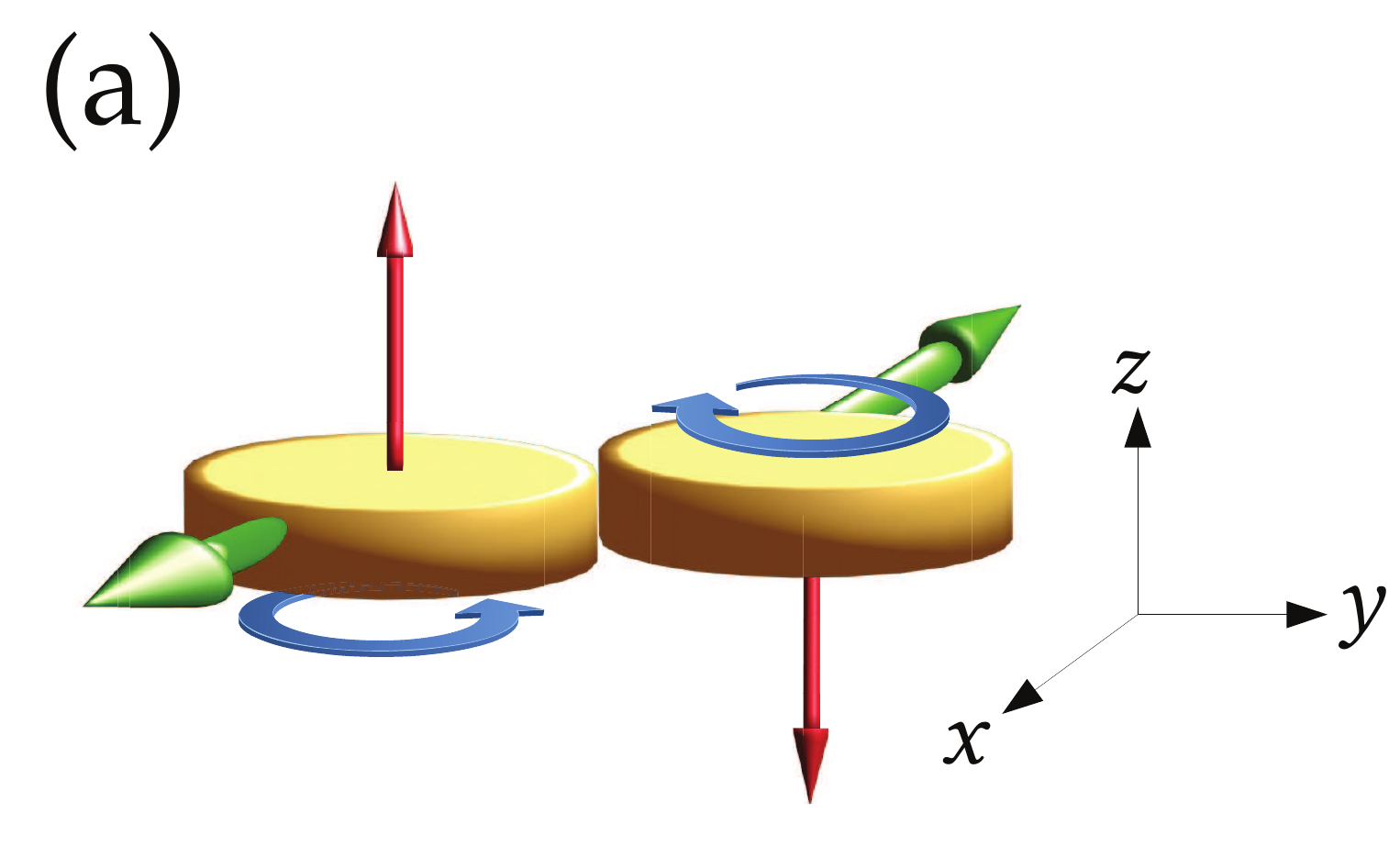}
	\includegraphics[width=0.23\textwidth]{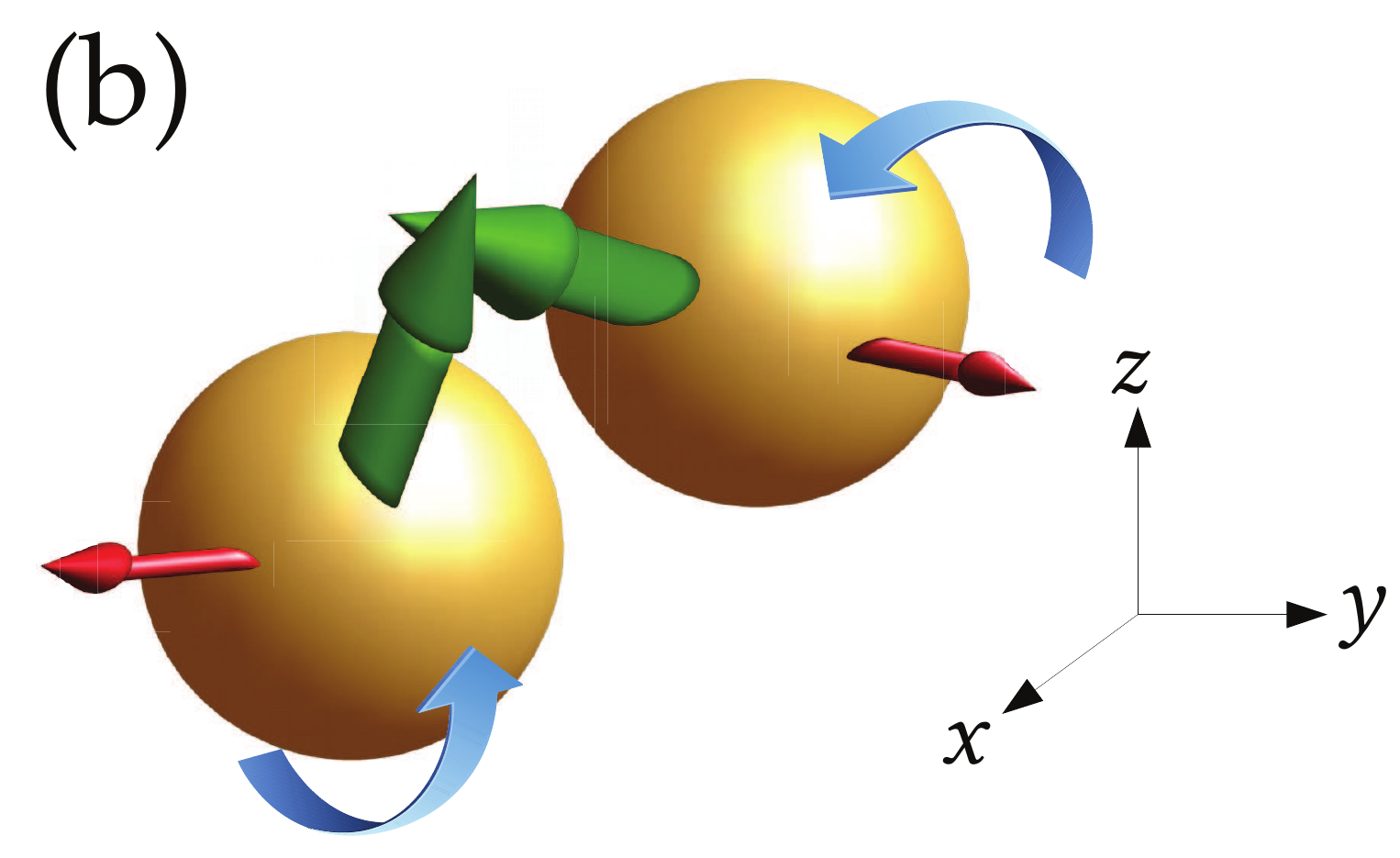}
	\caption{Illustration of a binary collision of (a) two hard disks and (b) two hard spheres. The (green) thick arrows represent the translational velocity vectors, while the (red) thin arrows depict the angular velocities complemented by the sense of rotation portrayed by the curved (blue) arrows. Notice that in (a) the translational velocities lie on the plane $xy$, while the angular velocities are constrained to the $z$ direction. }
	\label{fig:sketches}
\end{figure}

The gas is considered to be formed by a fixed large number of identical hard $\dt$ spheres with mass $m$, diameter $\sigma$, reduced moment of inertia $\kappa=4I/m\sigma^2$ ($I$ being the moment of inertia), and whose inelasticity and roughness are characterized by a coefficient of normal restitution, $\alpha$, and a coefficient of tangential restitution, $\beta$, respectively, both assumed to be constant, and defined by
\begin{equation}
    (\s\cdot\mathbf{g}^\prime_{12})=-\alpha (\s\cdot\mathbf{g}_{12}), \quad (\s\times\mathbf{g}^\prime_{12})=-\beta (\s\times\mathbf{g}_{12}),
\end{equation}
where $\s$ is the intercenter unit vector at contact,  $\mathbf{g}_{12}=\vv_{12}-\frac{\sigma}{2}\s\times (\oo_1+\oo_2)$ is the relative velocity of the contact points of particles 1 and 2 (with $\vv_{12}=\vv_1-\vv_2$), and primed quantities refer to postcollisional values. Note that, while $\alpha$ is nonnegative, $\beta$ can be either positive or negative. A negative value means that the postcollisional tangential component of the relative velocity maintains the same sign as the precollisional one, implying that the effect of surface friction is not dramatic. On the other hand, if the particles are sufficiently rough, the sign of the tangential component is inverted upon collision. Figure \ref{fig:sketches} presents a sketch illustrating a collision between (a) two hard disks and (b) two hard spheres.


\subsection{Direct collisional rules}

The direct binary collisional rules are obtained from the assumption of conservation of linear and angular momenta at the point of contact in each collision. They can be expressed by~\cite{BP04,VSK14,GSK18,S18,G19,MS19,MS19b,MS21a,MS21b}
\begin{subequations}
\begin{align}
    \cc_{1,2}^\prime\coloneqq\bbop\cc_{1,2}=&\cc_{1,2}\mp\ddab, \\
    \ww_{1,2}^\prime\coloneqq\bbop\ww_{1,2}=&\ww_{1,2}-\frac{1}{\sqrt{\kappa\theta}}\s\times\ddab, \label{eq:col_rules}
\end{align}
\end{subequations}
where $\bbop$ is the postcollisional operator acting on a dynamic quantity and giving the result after a collision, $\theta$ and $\ddab$ are defined below, and $\{\cc,\ww\}$ are the velocities reduced by their thermal value, that is,
\begin{align}
    \cc = \frac{\vv}{\vth(t)}, \qquad \ww = \frac{\oo}{\wth(t)}.
\end{align}
Here, $\vth(t) = \sqrt{{2\Tt(t)}/{m}}$, $\wth(t)=\sqrt{{2\Trot(t)}{/I}}$ are the thermal translational and angular velocities, $\Tt$ and $\Trot$ being the translational and rotational granular temperatures, respectively, which are defined by~\cite{BP04,VSK14,GSK18,S18,G19,MS19,MS19b,MS21a,MS21b}
\begin{equation}\label{eq:gran_Temperatures}
    \frac{\dt}{2}\Tt(t)=\frac{m}{2}\langle v^2\rangle, \quad
    \frac{\dr}{2}\Trot(t)=\frac{I}{2}\langle \omega^2\rangle,
\end{equation}
where $\langle \cdots\rangle=n^{-1}\int \dif\vv\int\dif\oo \medspace (\cdots) f(\vv,\oo;t)$ represents a one-body average value with respect to the VDF $f(\vv,\oo;t)$
normalized as
\begin{equation}
    n=\int \dif\vv\int\dif\oo \, f(\vv,\oo;t),
\end{equation}
$n$ being the particle number density. In Eq.~\eqref{eq:col_rules}, $\theta\equiv\Trot/\Tt$ is the rotational-to-translational granular temperature ratio. Moreover, the mean granular temperature is
\begin{equation}
	T(t) = \frac{\dt \Tt(t)+\dr \Trot(t)}{\dt+ \dr}.
\end{equation}
Finally,
the quantity
\begin{equation}\label{eq:ddab}
    \ddab = \en(\ccab\cdot\s)\s+\et\left[\ccab-(\ccab\cdot\s)\s-2\sqrt{\frac{\theta}{\kappa}}\s\times\Wwab\right]
\end{equation}
is the reduced impulse. In Eq.~\eqref{eq:ddab}, $\Wwab\equiv\frac{1}{2}(\ww_1+\ww_2)$ and
\begin{equation}
    \en = \frac{1+\alpha}{2}, \qquad \et = \frac{\kappa}{1+\kappa}\frac{1+\beta}{2}
\end{equation}
are effective coefficients of restitution.

Notice that the sets of vectors $\{\cc,\ww\}$ and $\{\vv,\oo\}$ span the same vector spaces, respectively. Therefore, the use of reduced quantities will be algebraically equivalent to the original velocity description.

\subsection{Inverse collisional rules}

The inverse collisional rules relating precollisional velocities $\{\cc_1^{\prime\prime},\ww_1^{\prime\prime},\cc_2^{\prime\prime},\ww_2^{\prime\prime}\}$  to postcollisional velocities $\{\cc_1,\ww_1\,\cc_2,\ww_2\}$ are~\cite{BP04,VSK14,GSK18,S18,G19,MS19,MS19b,MS21a,MS21b}
\begin{subequations}
\begin{align}\label{eq:col_rules_inv}
    \cc_{1,2}^{\prime\prime}\coloneqq\bbop^{-1}\cc_{1,2}=&\cc_{1,2}\mp\ddab^{-}, \\
    \ww_{1,2}^{\prime\prime}\coloneqq\bbop^{-1}\ww_{1,2}=&\ww_{1,2}-\frac{1}{\sqrt{\kappa\theta}}\s\times\ddab^{-},
\end{align}
\end{subequations}
with
\begin{equation}
    \ddab^{-} = \en\left(\frac{1}{\een}-\frac{1}{\eet}\right)(\ccab\cdot\s)\s+\frac{\ddab}{\eet}.
\end{equation}

From now on, throughout this paper, we will adopt the notation $\vom = \{\vv,\oo\}$ and $\cw = \{\cc,\ww\}$.

\section{\label{sec:3}Boltzmann equation}

\subsection{Basics}

We will carry out a description of the system under the assumption of molecular chaos or \emph{Stosszahlansatz}~\cite{GS03}, basing the analytical treatment on the homogeneous Boltzmann equation. As said before, we will generally derive the results keeping a dependence on the number of degrees of freedom, $\dt$ and $\dr$. The homogeneous Boltzmann equation reads
\begin{equation}\label{eq:BE}
    \frac{\partial f(\vom;t)}{\partial t}=\sigma^{\dt-1} \mathcal{I}_\vom [f,f],
\end{equation}
where   
\begin{equation}
\label{eq:I_coll}
    \mathcal{I}_{\vom_1} [f,f]=\int \dif\vom_2\int_{+} \dif \s\medspace (\vvab\cdot \s) \left(\frac{f_1^{''}f_2^{''}}{\alpha {\JJ}}-f_1 f_2 \right)
\end{equation}
is the collision operator.
Here, $f_{1,2}\equiv f(\vom_{1,2})$, $f_{1,2}''\equiv f(\vom_{1,2}'')$, the subscript $+$ designates the constraint $\vvab\cdot \s>0$, and $\JJ$ is the Jacobian due to the collisional change of velocities~\cite{MS19}, i.e.,
\begin{equation}
    \JJ=\left|\frac{\partial (\vva^\prime,\vvb^\prime,\ooa^\prime,\oob^\prime)}{\partial (\vva,\vvb,\ooa,\oob)}\right|=\left|\frac{\partial (\vva,\vvb,\ooa,\oob)}{\partial (\vva^{\prime\prime},\vvb^{\prime\prime},\ooa^{\prime\prime},\oob^{\prime\prime})}\right| = \een|\eet|^{2\dr/\dt}.
\end{equation}

Since the temporal change of the VDF is subjected only to collisions, it is convenient to  change from laboratory time, $t$, to \textit{collisional} time, $s$, as given by
\begin{equation}
    s(t) = \frac{1}{2}\int_0^t \dif t^\prime \medspace \nu(t^\prime),
\end{equation}
where $\nu(t)$ is the (nominal) collision frequency, defined by
\begin{align}
    \nu(t)=K n \sigma^{\dt-1}\vth(t),\quad K \equiv \frac{\sqrt{2}\pi^{\frac{\dt-1}{2}}}{\Gamma\left(\frac{\dt}{2}\right)}.
\end{align}
This variable $s(t)$ quantifies the accumulated average number of collisions per particle up to time $t$. Furthermore, the treatment based on the reduced velocities, $\cw$, allows us to define the reduced one-body VDF:
\begin{equation}
    \phi(\cw;s) = n^{-1}\vth^{\dt}(t)\wth^{\dr}(t)f(\vom;t).
\end{equation}
The homogeneous Boltzmann equation for the reduced VDF then reads
\begin{equation}\label{eq:red_boltz}
    \frac{K}{2}\partial_s\phi+\frac{\mu_{20}^{(0)}}{\dt}\frac{\partial}{\partial\cc}\cdot\left(\cc\phi\right)+\frac{\mu_{02}^{(0)}}{\dr}\frac{\partial}{\partial \ww}\cdot\left(\ww\phi\right)=\mathcal{I}_{\cw}[\phi,\phi],
\end{equation}
where
\begin{align}
\label{eq:col_mom_def2}
    \mu_{pq}^{(r)} =& -\int \dif \cw \medspace c^p w^q (\cc\cdot\ww)^r \mathcal{I}_{\cw}[\phi,\phi]\nonumber\\
     =& -\frac{1}{2}\int\dif\cw_1\int\dif\cw_2\int_{+}\dif\s\medspace (\ccab\cdot\s) \phi(\cw_1)\phi(\cw_2)\nonumber \\
     &\times (\bbop-1)\left[\ca^p\wa^q(\cca\cdot\wwa)^r+\cb^p\wb^q(\ccb\cdot\wwb)^r\right]
     \end{align}
are (reduced) collisional moments.
Note that, in the particular case of disks on a plane, the index $r$ is meaningless due to the orthogonality between the vector spaces spanned by translational and angular velocities [see Fig.~\ref{fig:sketches}(a)]. However, from a general point of view, the three-dimensional vector forms will be maintained.

Upon derivation of Eq.~\eqref{eq:red_boltz}, use has been made of the evolution equations for the translational and rotational temperatures,
\begin{equation}
\label{eq:ev_Temp}
    \frac{K}{2}\partial_s \Tt = -\frac{2}{\dt}\mu_{20}^{(0)}\Tt, \quad
    \frac{K}{2}\partial_s \Trot = -\frac{2}{\dr}\mu_{02}^{(0)}\Trot,
\end{equation}
which imply
\begin{subequations}
\label{eq:ev_theta,Temp_tot_ev}
\begin{equation}
\frac{K}{2}\partial_s\ln \theta=2\left[\frac{\mu_{20}^{(0)}}{\dt}-\frac{\mu_{02}^{(0)}}{\dr}\right], \label{eq:ev_theta}
\end{equation}
\begin{equation}\label{eq:Temp_tot_ev}
    \frac{K}{2}\partial_s T =  -\zeta^* T,
\end{equation}
\end{subequations}
where $\zeta^*\equiv 2(\mu_{20}^{(0)}+\mu_{02}^{(0)}\theta)/(\dt+\dr\theta)$ is the (reduced) cooling rate, and thus Eq.~\eqref{eq:Temp_tot_ev} represents Haff's cooling law~\cite{H83} for the inelastic and rough hard-sphere model.

From Eq.~\eqref{eq:red_boltz}, one can directly derive the hierarchy equations for the evolution of the velocity moments $M_{pq}^{(r)}\equiv\langle c^p w^q (\cc\cdot\ww)^r\rangle$:
\begin{equation}\label{eq:ev_moments}
    \frac{K}{2}\frac{\partial \ln M_{pq}^{(r)}}{\partial s}-\frac{(p+r)\mu_{20}^{(0)}}{\dt}-\frac{(q+r)\mu_{02}^{(0)}}{\dr}=-\frac{\mu_{pq}^{(r)}}{M_{pq}^{(r)}}.
\end{equation}

\subsection{Collisional moments}
The collisional change of a certain velocity function can be obtained by application of the operator $\delta\bbop\equiv\bbop-1$ on the function. For instance,
\begin{subequations}
\begin{align}\label{eq:coll_change_c2_w2}
	\delta\bbop\left(c_1^2+c_2^2\right)=&2\en(\en-1)(\ccab\cdot\s)^2+2\et(\et-1)\nonumber \\
	&\times\left(\s\times\ccab\right)^2+8\et^2\frac{\theta}{\kappa}(\s\times\Wwab)^2\nonumber\\
	&-4\et(2\et-1)\sqrt{\frac{\theta}{\kappa}}\ccab\cdot(\s\times \Wwab),\\
	\delta\bbop \left(w_1^2+w_2^2\right)=& \frac{2\et^2}{\kappa\theta}\left(\s\times\ccab\right)^2+8\frac{\et}{\kappa}\left(\frac{\et}{\kappa}-1\right)\nonumber \\
	&\times \left(\s\times\Wwab\right)^2+4\frac{\et}{\sqrt{\kappa\theta}}\left(2\frac{\et}{\kappa}-1\right)\nonumber \\
	&\times \Wwab\cdot(\s\times\ccab) .
\end{align}
\end{subequations}
The results for $\delta\bbop\left(c_1^4+c_2^4\right)$, $\delta\bbop\left(w_1^4+w_2^4\right)$, $\delta\bbop\left(c_1^2w_1^2+c_2^2w_2^2\right)$, and $\delta\bbop\left[(\cca\cdot\wwa)^2+(\ccb\cdot\wwb)^2\right]$ can be found in the Supplemental Material~\cite{note_23_01}.

Inserting the collisional changes into Eq.~\eqref{eq:col_mom_def2}, the collisional moments $\mu_{pq}^{(r)}$ can be formally expressed  in terms of \textit{two-body} averages of the form
\begin{equation}
	\llangle \psi\rrangle = \int \dif\cw_1\int\dif\cw_2\, \psi(\cw_1,\cw_2)\phi(\cw_1)\phi(\cw_2).
\end{equation}
In particular,
\begin{subequations}
\label{eq:cum_2body}
\begin{align}
	\mu_{20}^{(0)}=&\frac{B_3}{2}\biggl\{\left[\en(1-\en)+\frac{\dt-1}{2}\et(1-\et)\right]\llangle \cab^3\rrangle\nonumber \\
	& -2\et^2\frac{\theta}{\kappa}\left[3\llangle \cab\Wab^2\rrangle-\llangle \cab^{-1}(\ccab\cdot\Wwab)^2\rrangle \right]\biggr\},\\
	\mu_{02}^{(0)}=& \frac{B_3}{2}\frac{\et}{\kappa}\Bigg\{-\frac{\et}{\theta}\frac{\dt-1}{2}\llangle \cab^3\rrangle+2\left(1-\frac{\et}{\kappa}\right)\nonumber \\
	&\times \left[3\llangle \cab\Wab^2\rrangle-\llangle \cab^{-1}(\ccab\cdot\Wwab)^2\rrangle \right] \Bigg\},
\end{align}
\end{subequations}
where the factor $B_3= \pi^{\frac{\dt-1}{2}}/\Gamma\left( \frac{\dt+3}{2}\right)$ comes from an angular integral. The formally exact expressions of the collisional moments $\mu_{40}^{(0)}$, $\mu_{04}^{(0)}$, and $\mu_{22}^{(0)}$   in terms of two-body averages are given in the Supplemental Material~\cite{note_23_01},
where also some related tests for the simulation data are included.

\subsection{Sonine expansion}\label{sec:Sonine}
Assuming isotropy, $\phi(\cw;s)$
must depend on velocity only through  three scalars: $c^2$, $w^2$, and $(\cc\cdot\ww)^2$. This can be made explicit by the polynomial expansion
\begin{equation}\label{eq:phi_Sonine_expansion}
    \phi(\cw) = \phi_{\text{M}}(\cw)\sum_{j=0}^\infty\sum_{k=0}^\infty
    \sum_{\ell=0}^\infty a_{jk}^{(\ell)}\Psi_{jk}^{(\ell)}(\cw),
\end{equation}
where
\begin{equation}\label{eq:MA}
\phi_{\text{M}}(\cw)=\pi^{-(\dt+\dr)/2}e^{-c^2-w^2}
\end{equation}
is the (two-temperature) Maxwellian distribution, $a_{jk}^{(\ell)}$ are Sonine coefficients, and the functions
\begin{equation}
    \Psi_{jk}^{(\ell)} =L_j^{(2\ell+\frac{\dt}{2}-1)}(c^2)L_k^{(2\ell+\frac{\dr}{2}-1)}(w^2)(c^2w^2)^\ell P_{2\ell}(u)
\end{equation}
form a complete set of orthogonal polynomials~\cite{VSK14}.
Here, $L_j^{(\ell)}(x)$ are associated Laguerre polynomials, $u\equiv (\cc\cdot\ww)/cw$ is the cosine of the angle formed by the vectors $\cc$ and $\ww$, and $P_\ell(u)$ are Legendre polynomials~\cite{AS72}. The orthogonality condition is
\begin{subequations}
\begin{equation}
\label{eq:ortho}
    \braket{\Psi_{jk}^{(\ell)}}{\Psi_{j^\prime k^\prime}^{(\ell^\prime)}} = N_{jk}^{(\ell)}\delta_{j j^\prime}\delta_{k k^\prime}\delta_{\ell \ell^\prime},
\end{equation}
\begin{equation}
    N_{jk}^{(\ell)} =
        \frac{\Gamma\left(2\ell+\frac{\dt}{2}+j\right)\Gamma\left(2\ell+\frac{\dr}{2}+k\right)}{\Gamma\left(\frac{\dt}{2}\right)\Gamma\left(\frac{\dr}{2}\right)(4\ell+1)j!k!},
\end{equation}
\end{subequations}
where the inner product of two arbitrary  real functions $\Phi_1(\cw)$ and $\Phi_2(\cw)$ is defined as
\begin{equation}
    \braket{\Phi_1}{\Phi_2} = \int \dif\cw\medspace \phi_{\text{M}}(\cw)\Phi_1(\cw)\Phi_2(\cw).
\end{equation}
Note that $\langle \Phi\rangle=\braket{\Phi}{\phi/\phi_{\text{M}}}$. Using Eq.~\eqref{eq:ortho} in Eq.~\eqref{eq:phi_Sonine_expansion}, one can express the Sonine coefficients as
\begin{equation}\label{eq:cuma}
    a_{jk}^{(\ell)} = \frac{\langle \Psi_{jk}^{(\ell)}\rangle }{N_{jk}^{(\ell)}}.
\end{equation}
In particular, $a_{00}^{(0)}=1$, $a_{10}^{(0)}=a_{01}^{(0)}=0$, while
\begin{subequations}
\label{eq:cum_dim}
\begin{align}
    \cuma{2}{0}{0} &= \frac{4\langle c^4\rangle}{\dt(\dt+2)} -1, \quad \cuma{0}{2}{0} = \frac{4\langle w^4\rangle}{\dr(\dr+2)} -1,  \\
    \cuma{1}{1}{0} &= \frac{4\langle c^2 w^2\rangle}{\dt\dr} -1, \quad \cuma{0}{0}{1} = \frac{8}{15}\left[\langle (\cc\cdot\ww)^2\rangle -\frac{1}{3}\langle c^2 w^2\rangle \right]
\end{align}
\end{subequations}
are fourth-order cumulants.
Notice that $\cuma{0}{0}{1}$ is only meaningful in the hard-sphere case and thus it is not expressed in terms of the number of degrees of freedom.

The evolution equations for the cumulants defined by Eqs.~\eqref{eq:cum_dim} can be easily obtained from the moment hierarchy, Eq.~\eqref{eq:ev_moments}, as
\begin{subequations}
\label{eq:ev_eqs}
\begin{equation}
\label{eq:ev_cum_theta2}
    \frac{K}{2}\partial_s\ln \left(1+a_{20} \right)=\frac{4}{\dt(\dt+2)}\left[(\dt+2)\mu_{20}-\frac{\mu_{40}}{1+a_{20}}\right],
    \end{equation}
\begin{equation}
\label{eq:ev_cum_theta3}
    \frac{K}{2}\partial_s\ln \left(1+a_{02} \right)=\frac{4}{\dr(\dr+2)}\left[(\dr+2)\mu_{02}-\frac{\mu_{04}}{1+a_{02}}\right],
\end{equation}
\begin{equation}
\label{eq:ev_cum_theta4}
    \frac{K}{2}\partial_s\ln \left(1+a_{11} \right)=\frac{4}{\dr\dt}\left[\frac{\dr}{2}\mu_{20}+\frac{\dt}{2}\mu_{02}-\frac{\mu_{22}}{1+a_{11}}\right],
\end{equation}
\begin{align}
\label{eq:ev_cum_theta5}
    \frac{K}{2}\partial_s\ln \left[1+a_{11}+\frac{5}{2}a_{00}^{(1)} \right]=&\frac{4}{3}\Bigg[\frac{1}{2}\mu_{20}+\frac{1}{2}\mu_{02}\nonumber \\ &-\frac{\mu_{00}^{(2)}}{1+a_{11}+\frac{5}{2}a_{00}^{(1)}}\Bigg],
\end{align}
\end{subequations}
where henceforth we simplify the notation as $\cuma{j}{k}{0}\to a_{jk}$ and $\mu_{jk}^{(0)}\to \mu_{jk}$.

\subsection{\label{sec:HCS}Homogeneous cooling state}
The scaling method in the description of the kinetic equation suggests that a stationary solution, $\phi=\phi^\HCS$, of Eq.~\eqref{eq:red_boltz} applies for long times (hydrodynamic limit). This is the HCS, in which the temperature ratio, $\theta^\HCS$, is constant and the whole time dependence of the unscaled VDF $f^\HCS(\vom;t)$ takes place through the mean temperature $T(t)$ only.
On the other hand, this stationary solution $\phi=\phi^\HCS$ is not exactly known.

From Eqs.~\eqref{eq:ev_theta} and \eqref{eq:ev_eqs}, it follows that, in the HCS,
\begin{subequations}
\label{eq:HCS}
\begin{align}
    \mu^\HCS_{20} &= \frac{\dt}{\dr} \mu^\HCS_{02}, \quad  (\dt+2)\mu^\HCS_{20} = \frac{\mu^\HCS_{40}}{1+a_{20}^\HCS}, \label{eq:HCS1}\\ (\dr+2)\mu^\HCS_{02} &= \frac{\mu^\HCS_{04}}{1+a_{02}^\HCS}, \quad
    \dr\mu^\HCS_{20}=\frac{\mu^\HCS_{22}}{1+a_{11}^\HCS},  \label{eq:HCS2}\\
    \mu^\HCS_{20} &= \frac{\mu_{00}^{(2)\HCS}}{1+a_{11}^\HCS+\frac{5}{2}a_{00}^{(1)\HCS}}.\label{eq:a001_HCS}
\end{align}
\end{subequations}
Notice that, as expected, Eq.~\eqref{eq:a001_HCS} is only meaningful for spheres ($\dt=\dr=3$).

\begin{table*}[ht]
\caption{Relevant collisional moments from the Sonine approximation in the hard-disk case.}
\label{table:2}
\begin{ruledtabular}
\begin{tabular}{rcl}
$(p,q)$     &&  \hspace{0.4cm}$\mu_{pq}/\sqrt{2\pi}$\\ \hline
    $(2,0)$ && $\displaystyle{\left[2\en(1-\en)+\et(1-\et)\right]\left(1+\frac{3}{16}a_{20}\right)-\theta\frac{\et^2}{\kappa}\left(1-\frac{a_{20}}{16}+\frac{a_{11}}{4}\right)}$ \\
    $(0,2)$ && $\displaystyle{\frac{\et}{\kappa}\left(1-\frac{\et}{\kappa} \right)\left(1-\frac{a_{20}}{16}+\frac{a_{11}}{4}\right)-\frac{\et^2}{\kappa\theta}\left(1+\frac{3}{16}a_{20}\right)}$ \\
    $(4,0)$ &&  $\displaystyle{8\en^3(2-\en)\left(1+\frac{15}{16}a_{20}\right)+3\et^3(2-\et)\left(1+\frac{15}{16}a_{20}\right)+\left(\en+\frac{\et}{2}\right)\left(9+\frac{223}{16}a_{20}\right)-\en^2\left(17+\frac{327}{16}a_{20}\right)}$\\
    &&$\displaystyle{-\et^2\left(15+\frac{281}{16}a_{20}\right)-4\en\et(\en\et-\en-\et)\left(1+\frac{15}{16}a_{20}\right)-4\en\et\left(1+\frac{23}{16}a_{20} \right)-\frac{\et^2\theta}{2\kappa}\Bigg\{9+\frac{35}{16}a_{20}+\frac{27}{4}a_{11}}$\\
    &&$\displaystyle{-4\left[2\en(1-\en)+3\et(1-\et) \right]\left(1+\frac{3}{16}a_{20}+\frac{3}{4}a_{11} \right)+6\frac{\et^2\theta}{\kappa}\left(1-\frac{1}{16}a_{20}+\frac{1}{2}a_{02}+\frac{1}{2}a_{11} \right) \Bigg\}}$\\
    $(0,4)$ && $\displaystyle{\frac{\et}{\kappa}\left[3\left(1-\frac{1}{16}a_{20}+a_{02}+\frac{1}{2}a_{11}\right)-3\frac{\et}{\kappa}\left(2-2\frac{\et}{\kappa}+\frac{\et^2}{\kappa^2}\right)\left(1-\frac{1}{16}a_{20}+\frac{1}{2}a_{02}+\frac{1}{2}a_{11} \right)-\frac{3}{2}\frac{\et}{\kappa}a_{02}\right.}$\\
    && $\displaystyle{\left.-3\frac{\et}{\theta}\left(1-2\frac{\et}{\kappa}+2\frac{\et^2}{\kappa^2} \right)\left(1+\frac{3}{16}a_{20}+\frac{3}{4}a_{11} \right)-3\frac{\et^3}{\kappa\theta^2}\left(1+\frac{15}{16}a_{20} \right) \right]}$\\
    $(2,2)$ && $\displaystyle{\left[\en(1-\en)+\frac{\et}{2}(1-\et)\right]\left(1+\frac{3}{16}a_{20}+\frac{3}{4}a_{11}\right)+\left(\en+\frac{\et}{2}\right)\frac{a_{11}}{2}-\frac{\et^2\theta}{\kappa} \left(1-\frac{1}{16}a_{20}+\frac{1}{2}a_{11}+\frac{3}{4}a_{02} \right)}$\\
    && $\displaystyle{+\frac{\et}{\kappa}\left\{\frac{5}{4}+\frac{23}{64}a_{20}+\frac{27}{16}a_{11}-2\left[\en(1-\en)+\et\left(1-\frac{3}{2}\et\right)\right]\left(1+\frac{3}{16}a_{20}+\frac{3}{4}a_{11}\right)-(\en+\et)a_{11}\right\}+3\frac{\et^3\theta}{\kappa^2}}$\\
    && $\displaystyle{\times\left(1-\frac{a_{20}}{16}+\frac{a_{02}}{2}+\frac{a_{11}}{2} \right)-\frac{7}{4}\frac{\et^2}{\kappa\theta}\left(1+\frac{129}{112}a_{20} \right)-\frac{5}{4}\frac{\et^2}{\kappa^2}\left(1+\frac{23}{80}a_{20}+\frac{3}{4}a_{11}\right)+\frac{\et^2}{\kappa\theta}\left[2\en(1-\en)+3\et(1-\et)\right] }$\\
    && $\displaystyle{\times\left(1+\frac{15}{16}a_{20}\right)+\frac{\et^2}{\kappa^2}\left[2\en(1-\en)+3\et(1-2\et)\right]\left(1+\frac{3}{16}a_{20}+\frac{3}{4}a_{11}\right)+3\frac{\et^4\theta}{\kappa^3}\left(1-\frac{a_{20}}{16}+\frac{a_{02}}{2}+\frac{a_{11}}{2}\right)}$\\
\end{tabular}
\end{ruledtabular}
\end{table*}

\section{Approximate schemes}
\label{sec:3bis}
All the equations presented in Sec.~\ref{sec:3} are formally exact within the framework of the Boltzmann equation. However, no explicit results can be obtained unless one makes use of approximations.

\subsection{Maxwellian approximation}

The simplest approximation is the Maxwellian one, i.e., $\phi(\cw)\to\phi_{\text{M}}(\cw)$. In that case~\cite{MS21a},
\begin{subequations}
\label{6abc}
  \begin{align}
  \label{eq:mu20M}
   \mu_{20}\to&\frac{K}{2}\biggl\{1-\alpha^2+\frac{2\dr\kappa(1+\beta)}{\dt(1+\kappa)^2 }\biggl[1-\theta+\frac{\kappa(1-\beta)}{2}
   \nonumber\\
&\times\left(1
  +\frac{\theta}{\kappa}\right)\biggr]\biggr\},
\end{align}
\begin{equation}
\mu_{02}\to K\frac{\dr \kappa(1+\beta)}{\dt(1+\kappa)^2}\left[1-\frac{1}{\theta}+
\frac{1-\beta}{2}\left(\frac{1}{\theta}+\frac{1}{\kappa}\right)\right],
\end{equation}
\begin{equation}
\label{eq:38}
    \zeta^*\to \frac{K}{\dt+\dr\theta}\left[1-\een^2+\frac{\dr}{\dt}\frac{1-\eet^2}{1+\kappa}\left({\kappa}+{\theta}\right)\right].
\end{equation}
\end{subequations}
In this Maxwellian approximation, Eqs.~\eqref{eq:ev_Temp} and \eqref{eq:ev_theta,Temp_tot_ev} can be solved to get the evolution of the partial and mean temperatures, as well as the HCS value of the temperature ratio $\theta^\HCS$. However, by construction, the Maxwellian approximation is unable to account for the non-Gaussianities of the VDF, either in the transient evolution to the HCS or in the HCS itself.

\subsection{Sonine approximation}
The basic quantities measuring  non-Gaussianities are the cumulants defined in Eqs.~\eqref{eq:cum_dim}. Therefore, as the simplest scheme to capture those cumulants,
we introduce the Grad--Sonine methodology~\cite{CC70,VSK14,G19} and truncate the infinite Sonine expansion, Eq.~\eqref{eq:phi_Sonine_expansion},   after $j+k+2\ell\geq 3$, i.e.,
\begin{align}\label{eq:41}
    \phi\to\phi_{\text{S}}=&\phi_{\text{M}}\left[1+a_{20}\Psi_{20}^{(0)}+a_{02}\Psi_{02}^{(0)}+a_{11}\Psi_{11}^{(0)}\right.\nonumber\\
    &\left.+a_{00}^{(1)}\Psi_{00}^{(1)}\right],
\end{align}
where the term $a_{00}^{(1)}\Psi_{00}^{(1)}$ is not present in the hard-disk case.
With the replacement given by Eq.~\eqref{eq:41}, the two-body averages appearing in the collisional moments [see, for instance, Eqs.~\eqref{eq:cum_2body}] can be explicitly calculated as linear and quadratic functions of the cumulants. Next, our Sonine approximation is constructed by neglecting quadratic terms, so only linear terms are retained.

By particularizing to the hard-sphere case ($\dt=\dr=3$), previous results are recovered~\cite{VSK14}. Moreover, we obtain novel expressions for hard disks ($\dt=2$, $\dr=1$), which are displayed in Table~\ref{table:2}. Further details about some of the computations are available in the Supplemental Material~\cite{note_23_01}.

For consistency with the truncation and linearization steps carried out in the Sonine approximation, the evolution equations in the hard-disk case are obtained by inserting the expressions in Table~\ref{table:2} into Eqs.~\eqref{eq:ev_theta} and \eqref{eq:ev_cum_theta2}--\eqref{eq:ev_cum_theta4}, and linearizing the bracketed quantities. This gives a closed set of four differential equations, which are linear in the cumulants and nonlinear in the temperature ratio. Likewise, the HCS values are obtained by linearizing Eqs.~\eqref{eq:HCS1} and \eqref{eq:HCS2} with respect to the cumulants.
The linear stability of the HCS versus uniform and isotropic perturbations is proved in Appendix~\ref{sec:LSA}.

\begin{figure}[h!]
    \centering
    \includegraphics[width=0.23\textwidth]{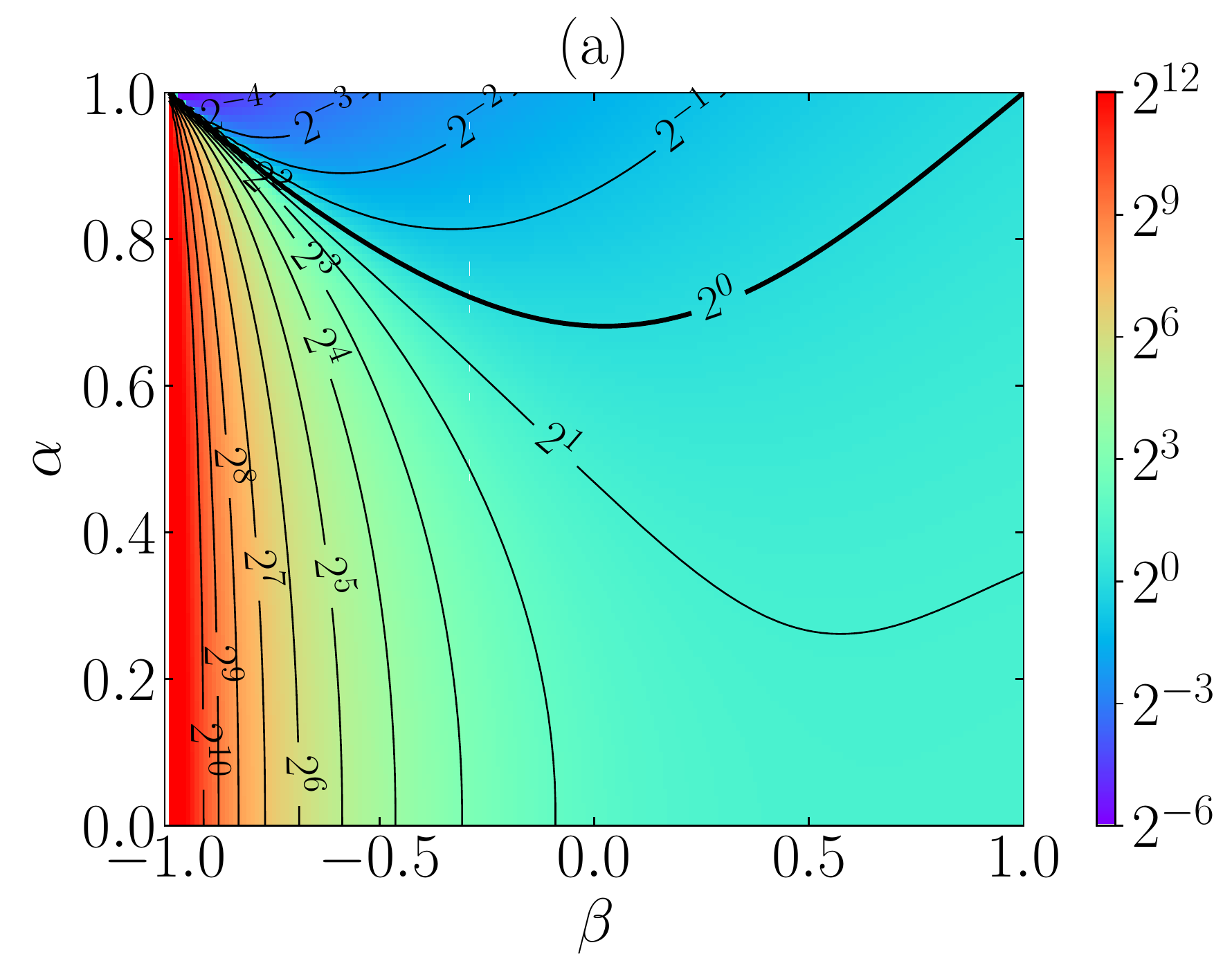}
    \includegraphics[width=0.23\textwidth]{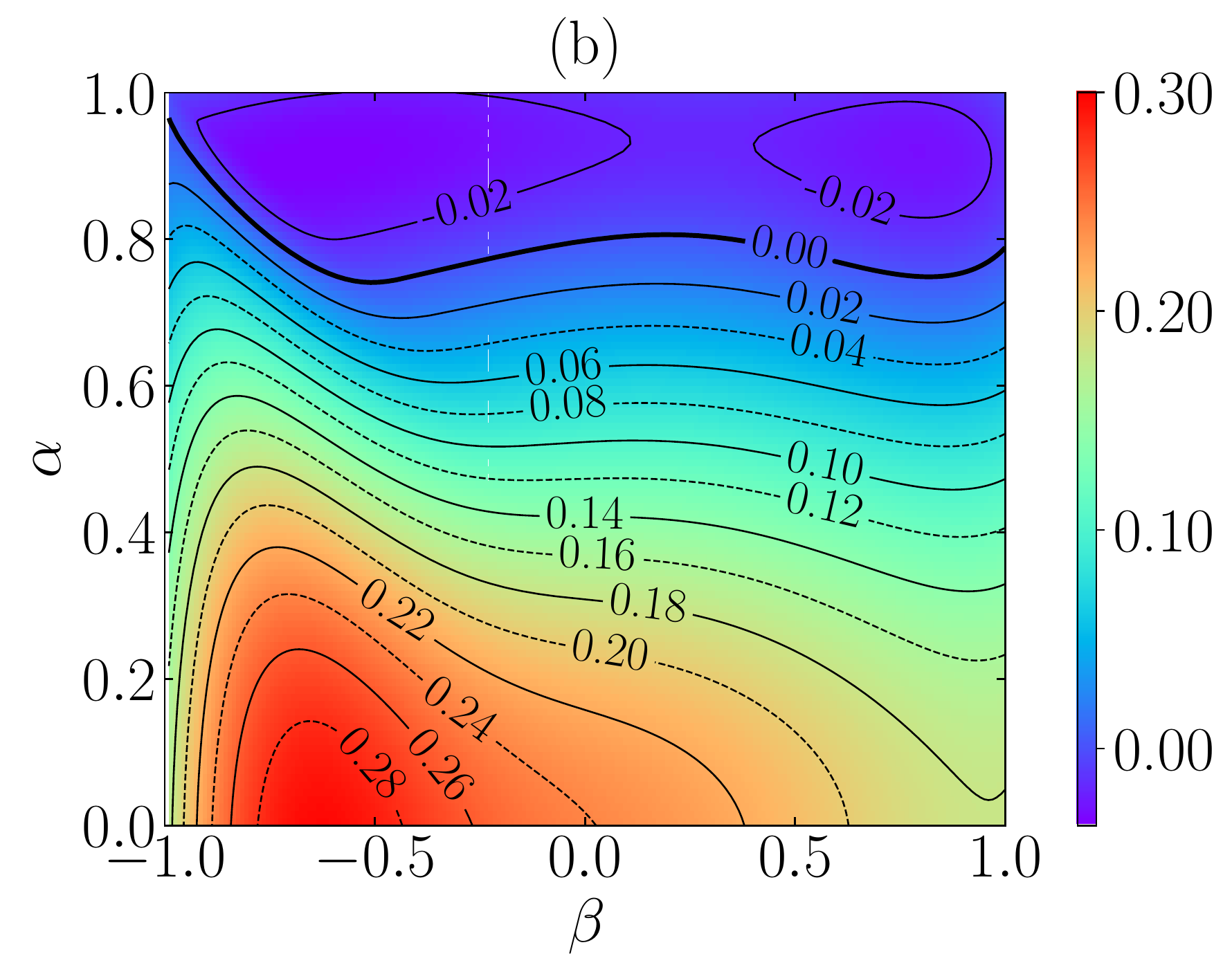}\\
    \includegraphics[width=0.23\textwidth]{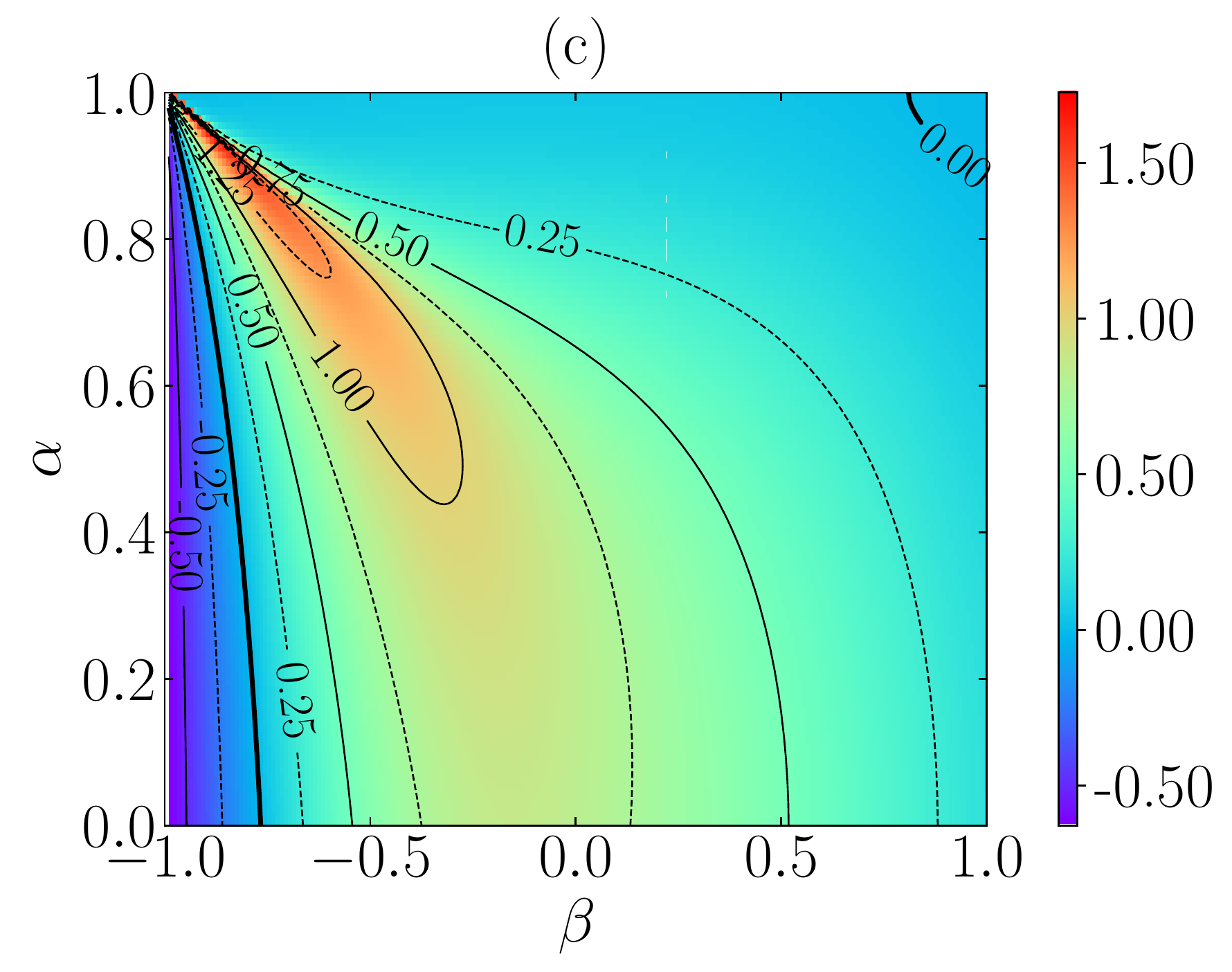}
    \includegraphics[width=0.23\textwidth]{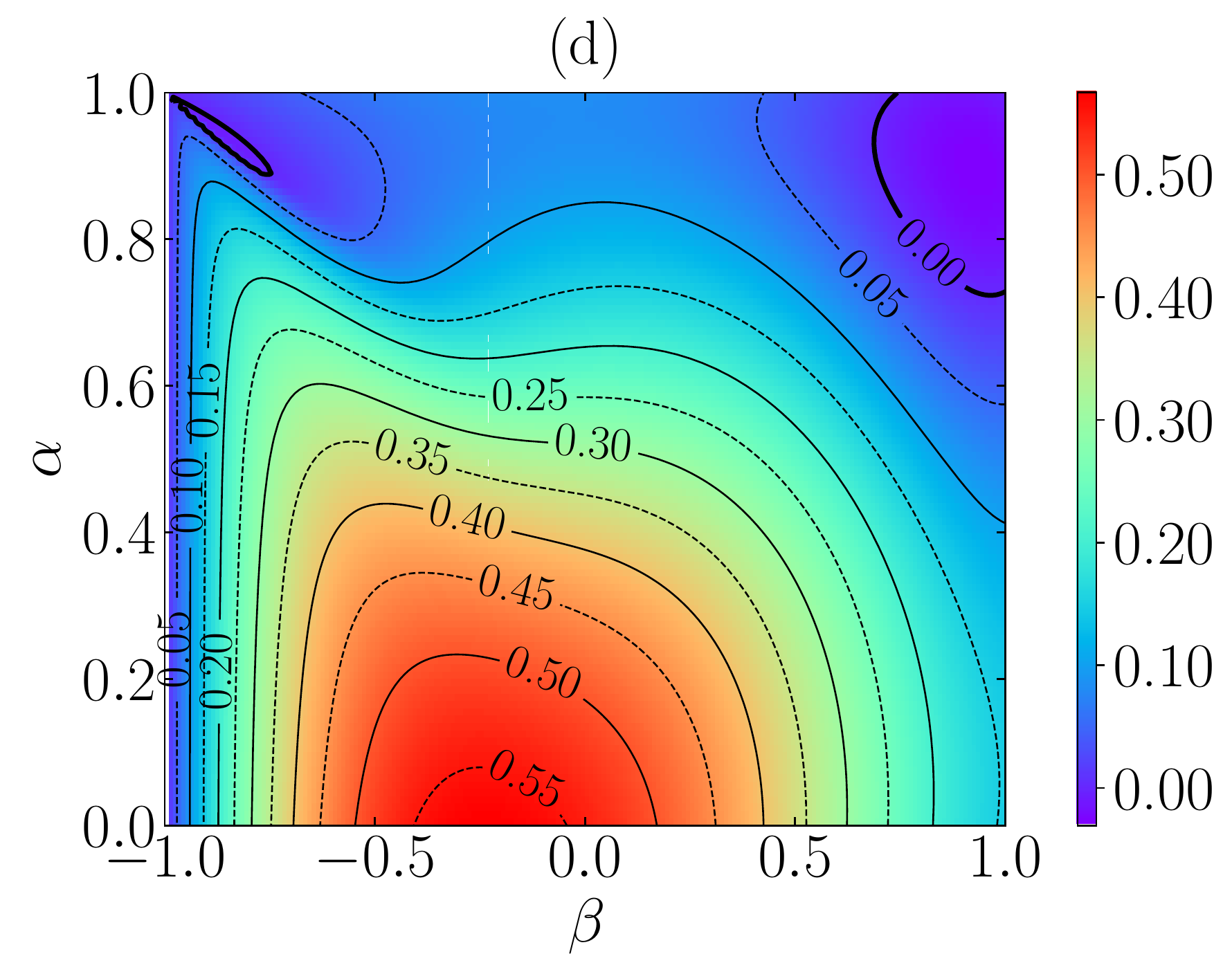}
    \caption{Theoretical values of (a) $\theta^\HCS$, (b) $a_{20}^\HCS$, (c) $a_{02}^\HCS$, and (d) $a_{11}^\HCS$ as functions of the coefficients of restitution, $\alpha$ and $\beta$, for uniform disks ($\kappa=\frac{1}{2}$) in the Sonine approximation.}
    \label{fig:HCS_HD}
\end{figure}

\begin{figure}[h!]
    \centering
    \includegraphics[width=0.23\textwidth]{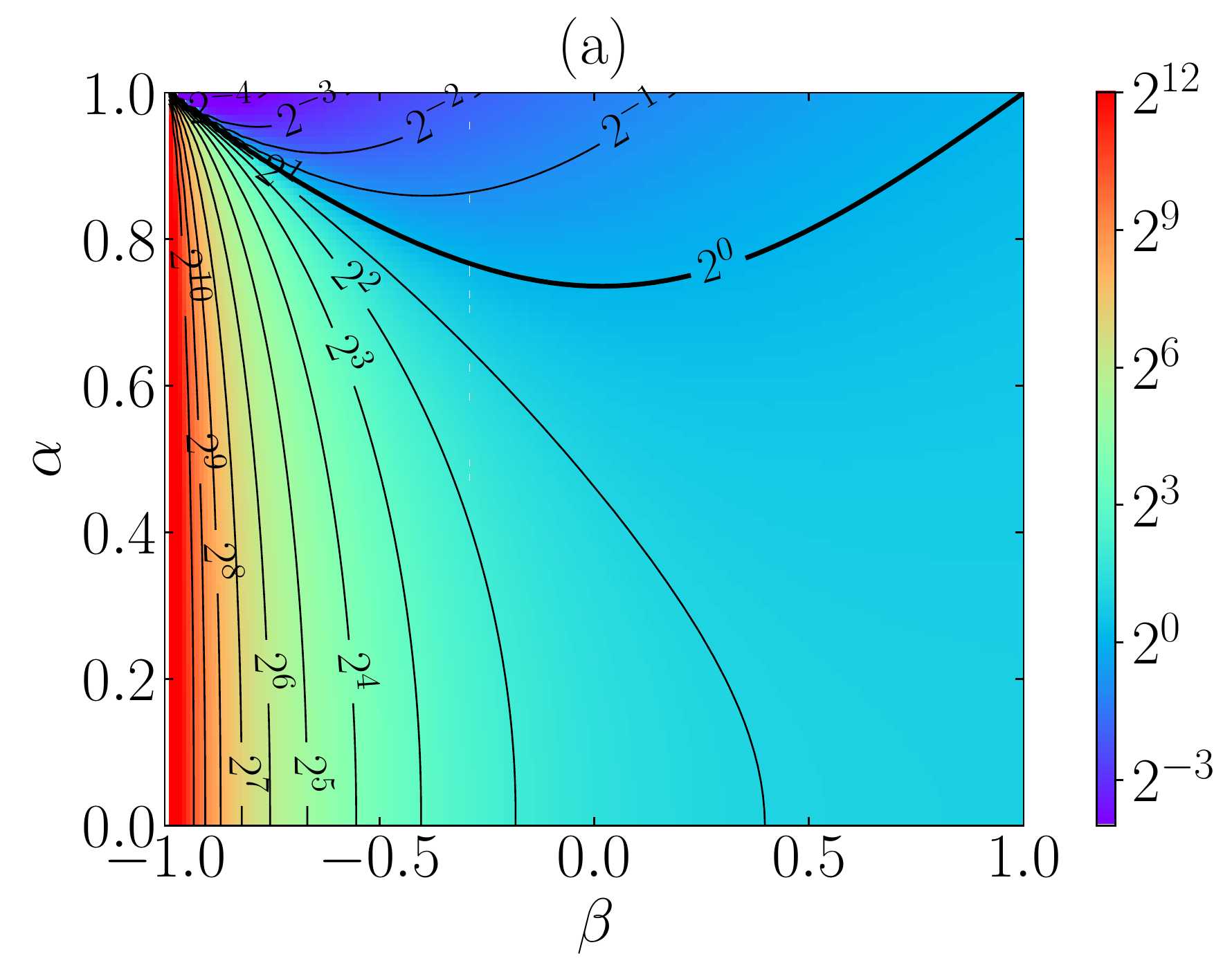}
    \includegraphics[width=0.23\textwidth]{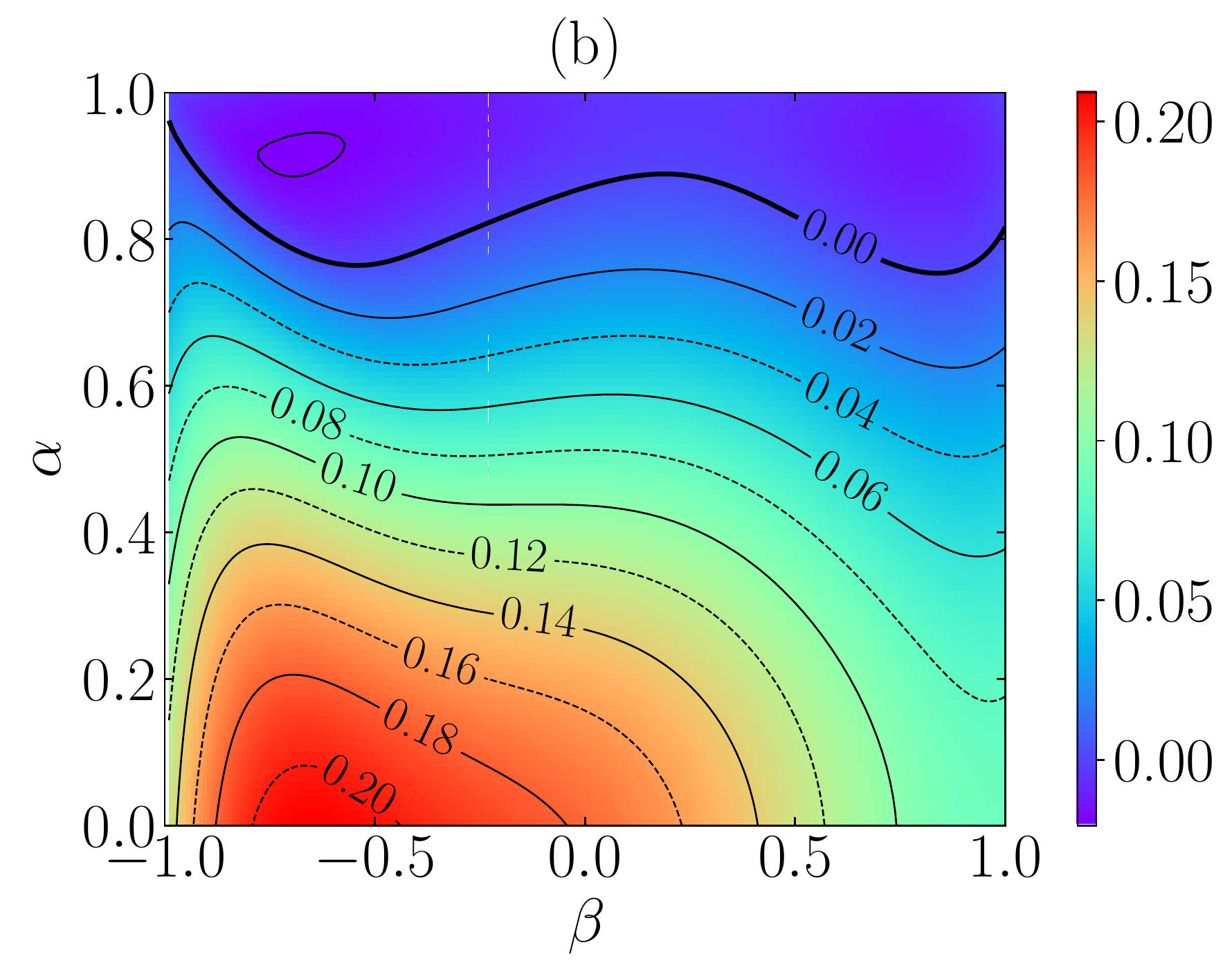}\\
    \includegraphics[width=0.23\textwidth]{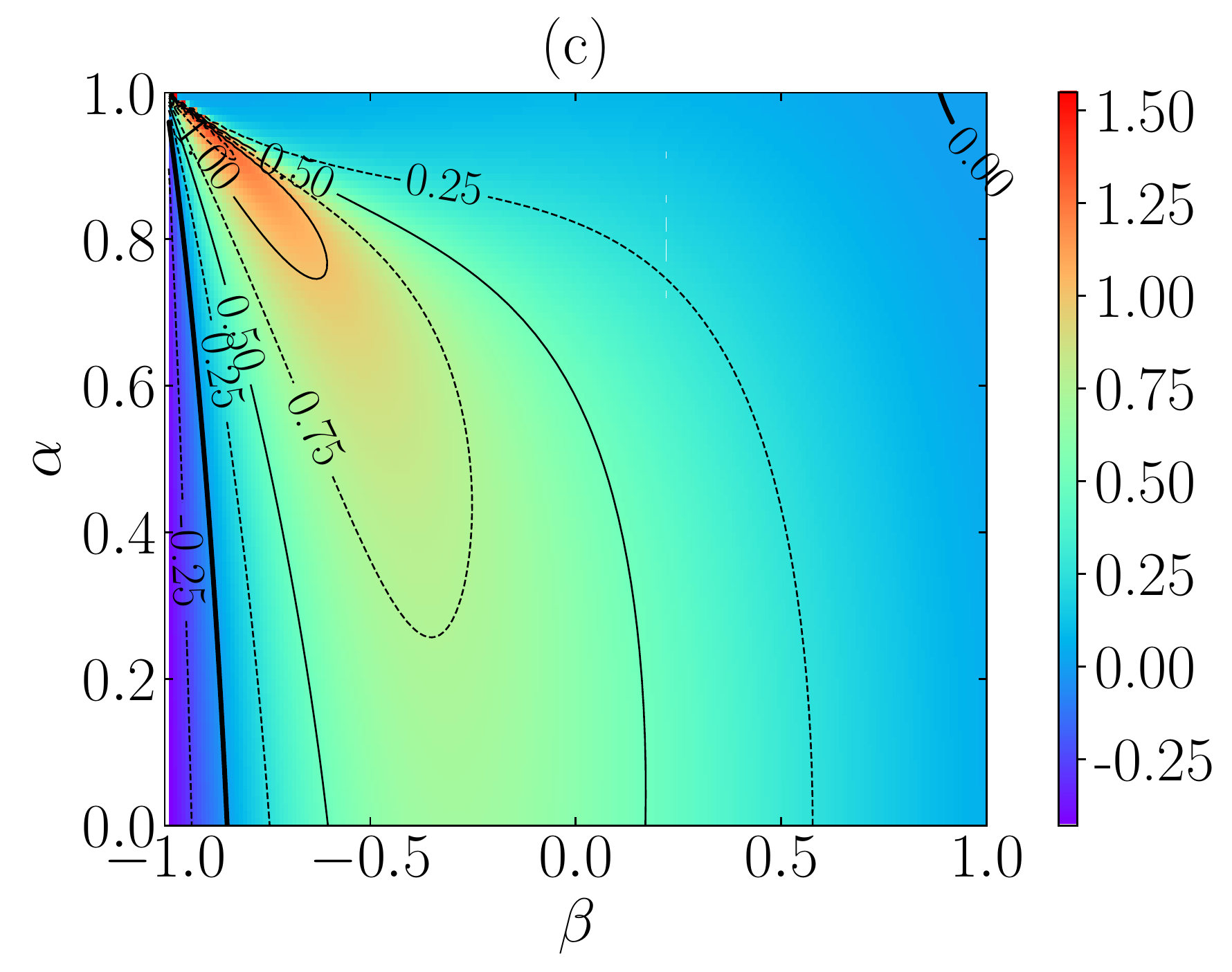}
    \includegraphics[width=0.23\textwidth]{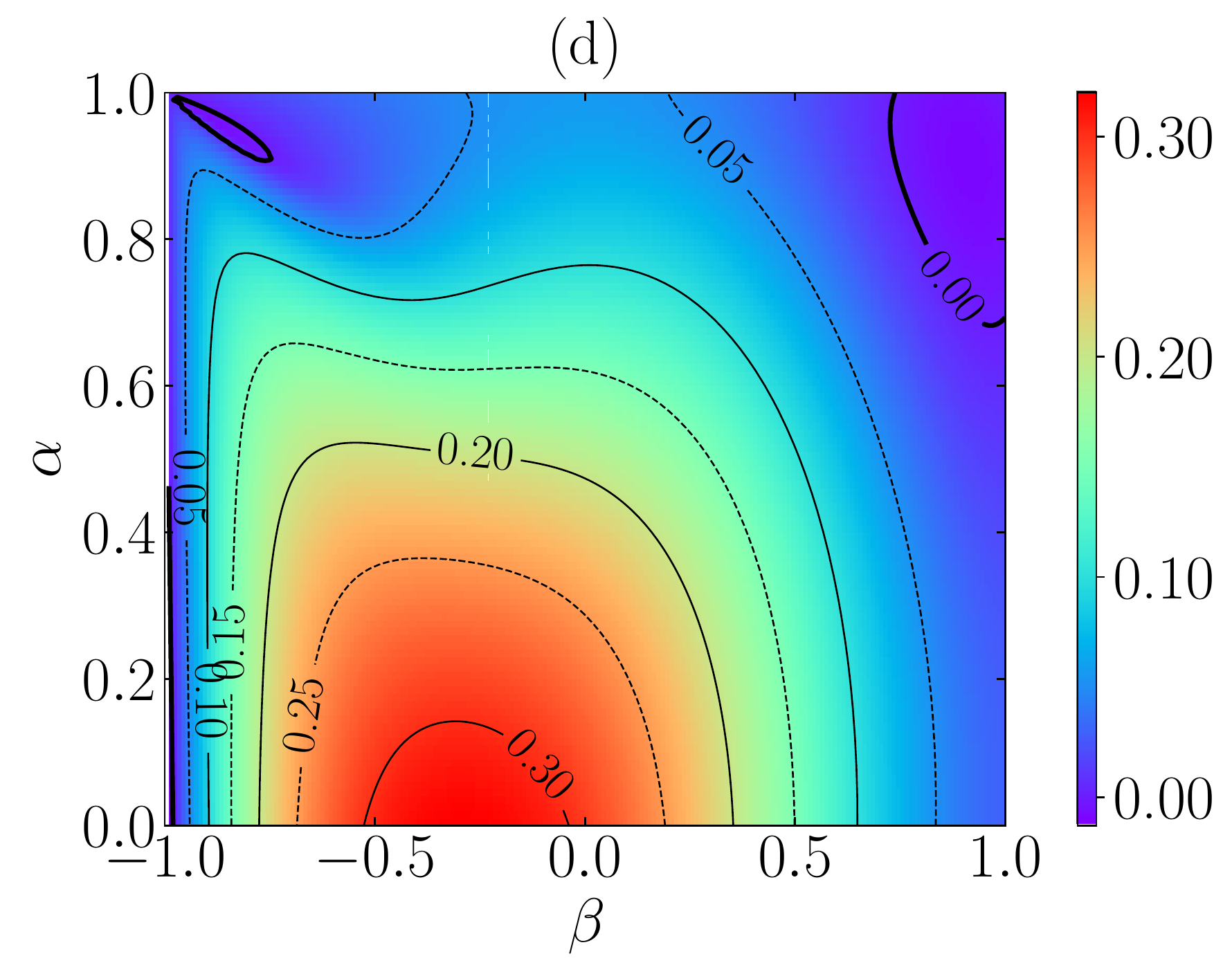}\\
    \includegraphics[width=0.23\textwidth]{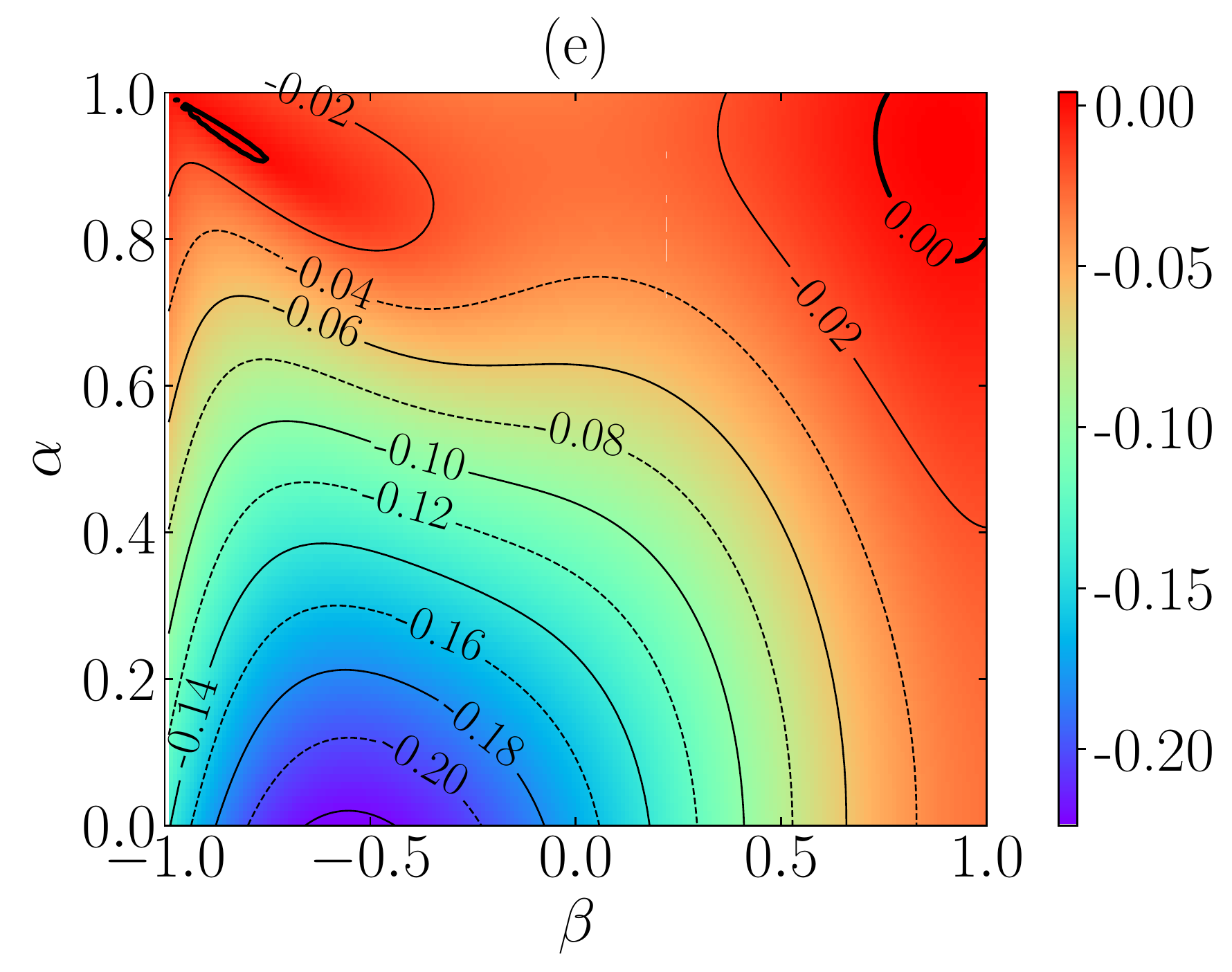}
    \caption{Theoretical values of (a) $\theta^\HCS$, (b) $a_{20}^\HCS$, (c) $a_{02}^\HCS$, (d) $a_{11}^\HCS$, and  (e) $a_{00}^{(1)\HCS}$ as functions of the coefficients of restitution, $\alpha$ and $\beta$, for uniform spheres ($\kappa=\frac{2}{5}$) in the Sonine approximation.}
    \label{fig:HCS_HS}
\end{figure}

Figures~\ref{fig:HCS_HD} and \ref{fig:HCS_HS} show the HCS quantities $\theta^\HCS$, $a_{20}^\HCS$, $a_{02}^\HCS$, and $a_{11}^\HCS$, obtained from the Sonine approximation for uniform disks ($\kappa=\frac{1}{2}$) and spheres ($\kappa=\frac{2}{5}$), respectively, as functions of the coefficients of restitution $\alpha$ and $\beta$. In the hard-sphere case, the cumulant $a_{00}^{(1)\HCS}$ is also included. It can be observed that, typically, hard-disk systems depart from the Maxwellian state more than hard-sphere systems. Interestingly, both hard-disk and hard-sphere systems present relatively large values of $a_{02}^\HCS$ and $a_{11}^\HCS$, thus signaling a possible quantitative breakdown of the Sonine approximation, which implicitly assumes small deviations from the Maxwellian VDF.

\section{\label{sec:MVDFs}Marginal distribution functions and High-velocity tails in the Homogeneous Cooling State}

\subsection{Marginal distribution functions}
As said before, the reduced VDF $\phi(\cw)$ in isotropic states depend on the three scalars $c^2$, $w^2$, and $c^2w^2$ [plus $(\cc\cdot\ww)^2$ only for spheres]. To disentangle those dependencies, it is convenient to define the following \emph{marginal} distributions~\cite{VSK14,VS15}:
\begin{subequations}\label{eq:marginal_defs}
    \begin{align}
        \phi_\cc(\cc) =& \int\dif\ww \,\phi(\cw), \label{eq:marginal_defs_c}\\
        \phi_\ww(\ww) =& \int\dif\cc \,\phi(\cw),\label{eq:marginal_defs_w}\\
        \phi_{cw}(x) =& \int \dif\cw\, \delta(c^2w^2-x)\phi(\cw), \label{eq:marginal_defs_cw}
    \end{align}
\end{subequations}
where $x$ represents the product $c^2w^2$. Note that, by isotropy, $\phi_\cc(\cc)$ and $\phi_\ww(\ww)$ depend only on the moduli $c$ and $w$, respectively. Moreover, the marginal distributions in Eqs.~\eqref{eq:marginal_defs} are directly related to the cumulants $a_{20}^{(0)}$, $a_{02}^{(0)}$, and $a_{11}^{(0)}$ defined by Eqs.~\eqref{eq:cum_dim}, namely
\begin{subequations}
\begin{align}
\int \dif\cc\,c^4  \phi_\cc(\cc)=&\frac{\dt(\dt+2)}{4}\left[1+a_{20}^{(0)}\right],\\
\int \dif\ww\,w^4  \phi_\ww(\ww)=&\frac{\dr(\dr+2)}{4}\left[1+a_{02}^{(0)}\right],\\
\int_0^\infty \dif x\, x \phi_{cw}(x)=&\frac{\dt\dr}{4}\left[1+a_{11}^{(0)}\right].
\end{align}
\end{subequations}

The Maxwellian expressions for these functions are
\begin{subequations}
\begin{align}\label{eq:marginal_MA}
    \phi_{\cc,\mathrm{M}}(\cc) =&  \pi^{-\dt/2} e^{-c^2},\\
    \phi_{\ww,\mathrm{M}}(\ww) =&  \pi^{-\dr/2} e^{-w^2},\\
    \phi_{cw,\mathrm{M}}(x) =& \frac{1}{2}\Omega_\dt\Omega_\dr \pi^{-\frac{\dt+\dr}{2}} x^{\frac{\dt+\dr}{4}-1}K_{\frac{\dt-\dr}{2}}(2\sqrt{x}),
\end{align}
\end{subequations}
where   $\Omega_{d}={2\pi^{d/2}}/{\Gamma\left(\frac{d}{2}\right)}$ is the $d$-dimensional solid angle and $K_a(x)$ is the modified Bessel function of the second kind.
In the Sonine approximation defined by Eq.~\eqref{eq:41}, one has
\begin{subequations}
\label{eq:marginal_SA}
\begin{align}
    \frac{\phi_{\cc,\mathrm{S}}(\cc)}{\phi_{\cc,\mathrm{M}}(\cc)} =& 1+a_{20} \frac{4c^4-4(\dt+2)c^2+\dt(\dt+2)}{8},\\
    \frac{\phi_{\ww,\mathrm{S}}(\ww)}{\phi_{\ww,\mathrm{M}}(\ww)} =& 1+a_{02} \frac{4w^4-4(\dr+2)w^2+\dr(\dr+2)}{8},\\
    \frac{\phi_{cw,\mathrm{S}}(x)}{\phi_{cw,\mathrm{M}}(x)} =& 1 +\frac{a_{20}+2a_{11}+a_{02}}{2}x+a_{20}\frac{\dt(\dt+2)}{8} \nonumber \\
    &+\frac{\dt\dr}{4}a_{11}+a_{02}\frac{\dr(\dr+2)}{8}\nonumber\\
    &-\sqrt{x}\frac{K_{1-\frac{\dt-\dr}{2}}(2\sqrt{x})}{K_{\frac{\dt-\dr}{2}}(2\sqrt{x})}\left[\frac{a_{20}+a_{02}}{2}\right.\nonumber\\
    &\left.+\frac{\dt+\dr}{4}(a_{20}+2a_{11}+a_{02}) \right].
\end{align}
\end{subequations}
While Eqs.~\eqref{eq:marginal_SA} may reproduce the correct behavior of the HCS in the \emph{thermal} domain, it is known  from the smooth case~\cite{EP97,vNE98,BRC99} and from hard-sphere results~\cite{VSK14} that they are unable to account for the high-velocity tail.

\subsection{\label{subsec:HET} High-velocity tails}

Let us now study the high-velocity tail for the marginal VDF in the HCS, in analogy to previous works for the smooth case~\cite{EP97,vNE98}.

To carry out this asymptotic analysis, we start from the homogeneous Boltzmann equation, Eq.~\eqref{eq:red_boltz}, and split the collisional operator into a loss and a gain term, that is~\cite{EP97,vNE98},
\begin{equation}
    \mathcal{I}_{\cw}[\phi,\phi] = \mathcal{I}_{\cw}^{\mathrm{G}}[\phi,\phi]-\mathcal{I}_{\cw}^{\mathrm{L}}[\phi,\phi],
\end{equation}
where the loss term can be written as
\begin{equation}
\label{eq:loss}
    \mathcal{I}_{\cw_1}^{\mathrm{L}}[\phi,\phi] = B_1\phi(\cw_1) \int\dif\cw_2\, c_{12} \phi(\cw_2),
    \end{equation}
with $B_1= \pi^{\frac{\dt-1}{2}}/\Gamma\left( \frac{\dt+1}{2}\right)$. The gain term accounts for all the particles that after a collision have velocities $\cw_1$. In contrast, the loss term takes into account the amount of particles with $\cw_1$ that, after a collision, are not contributing any more to these velocities. 

Intuitively, one would expect that escaping from the rapid regime is easier than entering the high-velocity limit, given the low likelihood of encountering rapid particles compared to thermal ones. Thus, the main assumption we will use is that, for high velocities of the HCS, the loss term prevails over the gain term. From Eq.~\eqref{eq:I_coll}, and following the case of smooth particles~\cite{vNE98}, the assumption above can be expressed as
\begin{equation}
\label{eq:HVT_test}
    \lim_{c_1\to\infty \text{ or } w_1\to\infty}\frac{\phi^\HCS(\cw_1^{\prime\prime})\phi^\HCS(\cw_2^{\prime\prime})}{\phi^\HCS(\cw_1)\phi^\HCS(\cw_2)}=0.
\end{equation}

\subsubsection{Tail of $\phi_\cc^\HCS(\cc)$}
Integrating over $\ww$ on both sides of the stationary version of Eq.~\eqref{eq:red_boltz}, neglecting the gain term, replacing $c_{12}\to c_1$ in Eq.~\eqref{eq:loss}, and taking the limit $c\gg 1$, we get the linear differential equation
\begin{equation}
    \frac{\mu_{20}^\HCS}{\dt}\frac{\partial}{\partial c}\phi_\cc^\HCS(\cc) \approx - B_1 \phi_\cc^\HCS(\cc),
\end{equation}
whose solution is
\begin{equation}
\label{eq:HVT_c}
    \phi_{\cc}^\HCS(\cc)\approx \mathcal{A}_c e^{-\gamma_c c},\quad  \gamma_c = \frac{\dt B_1}{\mu_{20}^\HCS},
\end{equation}
where $\mathcal{A}_c$ is an integration constant. This is equivalent to the result in the smooth case~\cite{EP97,vNE98}, except that now  $\mu_{20}^\HCS$ takes into account the influence of surface roughness.
The exponential decay of  $\phi_{\cc}^\HCS(\cc)$ implies that all the cumulants of the form $a_{j0}^\HCS$ are finite.

\subsubsection{Tail of $\phi_\ww^\HCS(\ww)$}
Now we integrate  over $\cc$ on both sides of  Eq.~\eqref{eq:red_boltz} and neglect again the gain term. This yields
\begin{equation}\label{eq:w_tail_eq}
    \mu_{02}^\HCS\phi_\ww^\HCS(\ww)+\frac{\mu_{02}^\HCS}{\dr}w\frac{\partial}{\partial w}\phi_\ww^\HCS(\ww) \approx - B_1 \overline{c_{12}}^\HCS(\ww)\phi_\ww^\HCS(\ww),
\end{equation}
where
\begin{equation}\label{eq:c12over}
    \overline{c_{12}}(\ww_1) = \int \dif\cca\int\dif\cw_2\, c_{12} \phi_{\cc|\ww}(\cca|\wwa)\phi(\cw_2).
\end{equation}
Here,  $\phi_{\cc|\ww}(\cc|\ww)$ is a conditional probability distribution function defined as
\begin{equation}
    \phi_{\cc|\ww}(\cc|\ww)\phi_{\ww}(\ww)=\phi(\cc,\ww).
\end{equation}
The quantity $\overline{c_{12}}(\ww)$ represents the average relative translational velocity of those particles with an angular velocity $\ww$. It is a functional of the whole VDF $\phi(\cw)$, so Eq.~\eqref{eq:w_tail_eq} is not a closed equation for the marginal distribution $\phi_\ww^\HCS(\ww)$.

The positive values observed in Figs.~\ref{fig:HCS_HD} and \ref{fig:HCS_HS} for the cumulant $a_{11}^\HCS$ imply that  high angular velocities are positively correlated to high translational velocities, so $\overline{c_{12}}^\HCS(\ww)$ is
expected to increase with $w$. However, to estimate the tail of $\phi_\ww^\HCS(\ww)$, we further assume that the dependence of $\overline{c_{12}}^\HCS(\ww)$ on $w$ is weak enough as to take  $\overline{c_{12}}^\HCS(\ww)\approx \llangle c_{12}\rrangle^\HCS$. With this adiabaticlike approximation, Eq.~\eqref{eq:w_tail_eq} becomes a closed linear equation whose solution is
\begin{equation}
\label{eq:phi_w(w)}
    \phi_\ww^\HCS(\ww)\approx \mathcal{A}_w w^{-\gamma_w},\quad \gamma_w=\dr+\gamma_c \llangle c_{12}\rrangle^\HCS,
\end{equation}
where $\mathcal{A}_w$ is the associated integration constant. In the expression of $\gamma_w$, we have made use of the HCS  condition $\mu_{02}^\HCS/\dr=\mu_{20}^\HCS/\dt$ [see Eqs.~\eqref{eq:HCS1}].

While, in principle, Eqs.~\eqref{eq:phi_w(w)} are approximate because of the ansatz $\overline{c_{12}}^\HCS(\ww)\approx \llangle c_{12}\rrangle^\HCS$, it accounts for an algebraic decay of $\phi_\ww^\HCS(\ww)$  explaining the relatively high values attained by $a_{02}^\HCS$. In fact, Eqs.~\eqref{eq:phi_w(w)} imply that the coefficients of the form $a_{0k}^\HCS$ diverge if $2k\geq \gamma_w-1$.

\subsubsection{\label{subsubsec:cw} Tail of $\phi_{cw}(x)$}
Whereas  the derivation of the high-velocity tail for $\phi_{\cc}$ is clean, and the one for $\phi_{\ww}$, although approximate, is reasonable, in the case of the distribution $\phi_{cw}$ the reasoning is somewhat more speculative.
Let us start by introducing the marginal probability distribution function of the variable $w^2$, $\phi_{w^2}(w^2)=(\Omega_{\dr}/2)w^{\dr-2}\phi_\ww(\ww)$. According to Eqs.~\eqref{eq:phi_w(w)}, the high-velocity tail of $\phi_{w^2}^\HCS(w^2)$ is
\begin{equation}
\label{eq:phi_w2(w2)}
    \phi_{w^2}^\HCS(w^2)\approx \mathcal{A}_w\frac{\Omega_{\dr}}{2} (w^2)^{\frac{\dr-\gamma_w}{2}-1}.
\end{equation}
As can be inferred from Eqs.~\eqref{eq:HVT_c} and \eqref{eq:phi_w(w)}, the tail of angular velocities is much more populated than that of translational velocities. Therefore, it is reasonable to expect that the main contribution to $\phi_{cw}(c^2w^2)$ comes essentially from particles with thermal translational velocities ($c\sim 1$) and high angular velocities ($w\gg 1$). Thus, in view of Eq.~\eqref{eq:phi_w2(w2)}, we  conjecture that
\begin{equation}
\label{eq:sol_cw_1}
    \phi^\HCS_{cw}(x)\approx \mathcal{A}_{cw} x^{-\gamma_{cw}},\quad \gamma_{cw} =1+ \frac{\gamma_w-\dr}{2}.
\end{equation}
This algebraic decay   would be responsible for the relatively large values of $a_{11}^\HCS$ and implies the divergence of the coefficients of the form $a_{jj}^\HCS$  if $j\geq \gamma_{cw}-1$.
An alternative justification of Eqs.~\eqref{eq:sol_cw_1} is provided in the Supplemental Material~\cite{note_23_01}.

While, according to Eq.~\eqref{eq:HVT_c}, the asymptotic decay of $\phi_\cc(\cc)$ is governed by a velocity scale $c\sim \gamma_c^{-1}$, Eqs.~\eqref{eq:phi_w(w)} and \eqref{eq:sol_cw_1} show that the decays of $\phi_\ww(\ww)$ and $\phi_{cw}(x)$ are scale-free. It can be checked that the exponents $\gamma_c$, $\gamma_w$, and $\gamma_{cw}$ are generally smaller for disks than for spheres, meaning that the high-velocity tails are fatter in the former case than in the latter. Apart from that, they exhibit a similar qualitative dependence on the coefficients of restitution.

The consistency of Eq.~\eqref{eq:HVT_test}  with the tails obtained here is discussed in Appendix~\ref{ap:ansatz_het}.

\begin{figure*}[ht]
    \centering
    \includegraphics[width=0.3\textwidth]{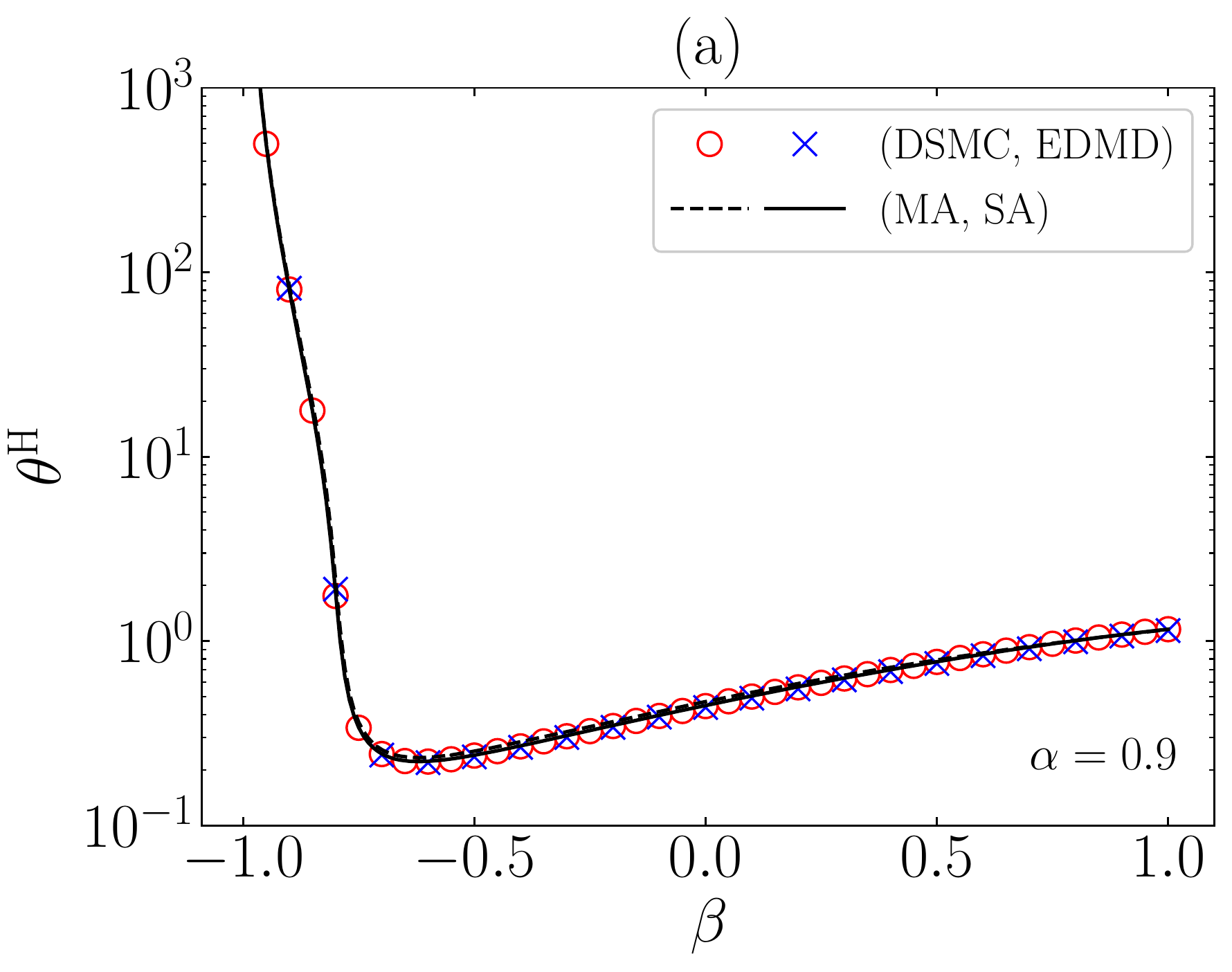}
    \includegraphics[width=0.3\textwidth]{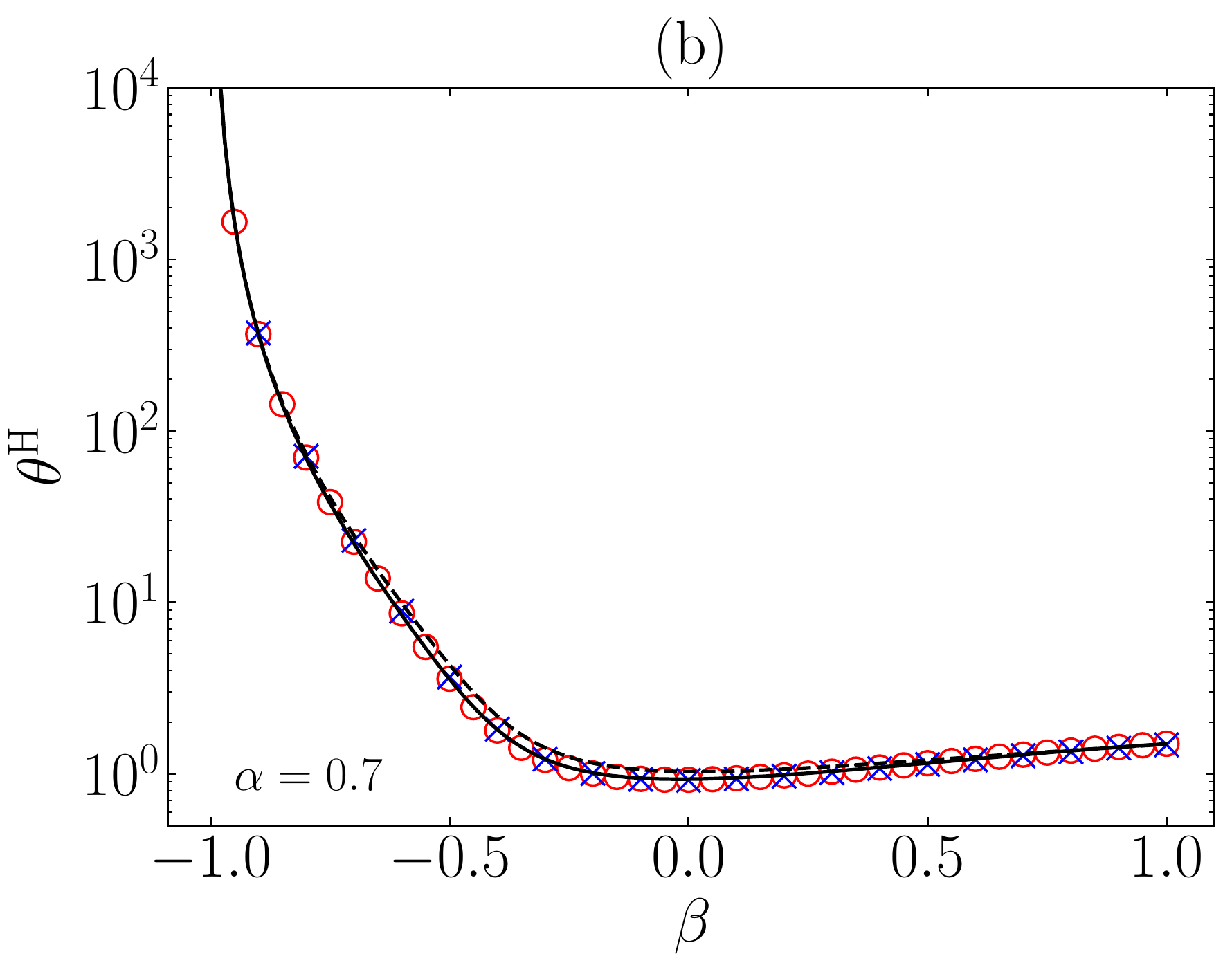}
    \includegraphics[width=0.3\textwidth]{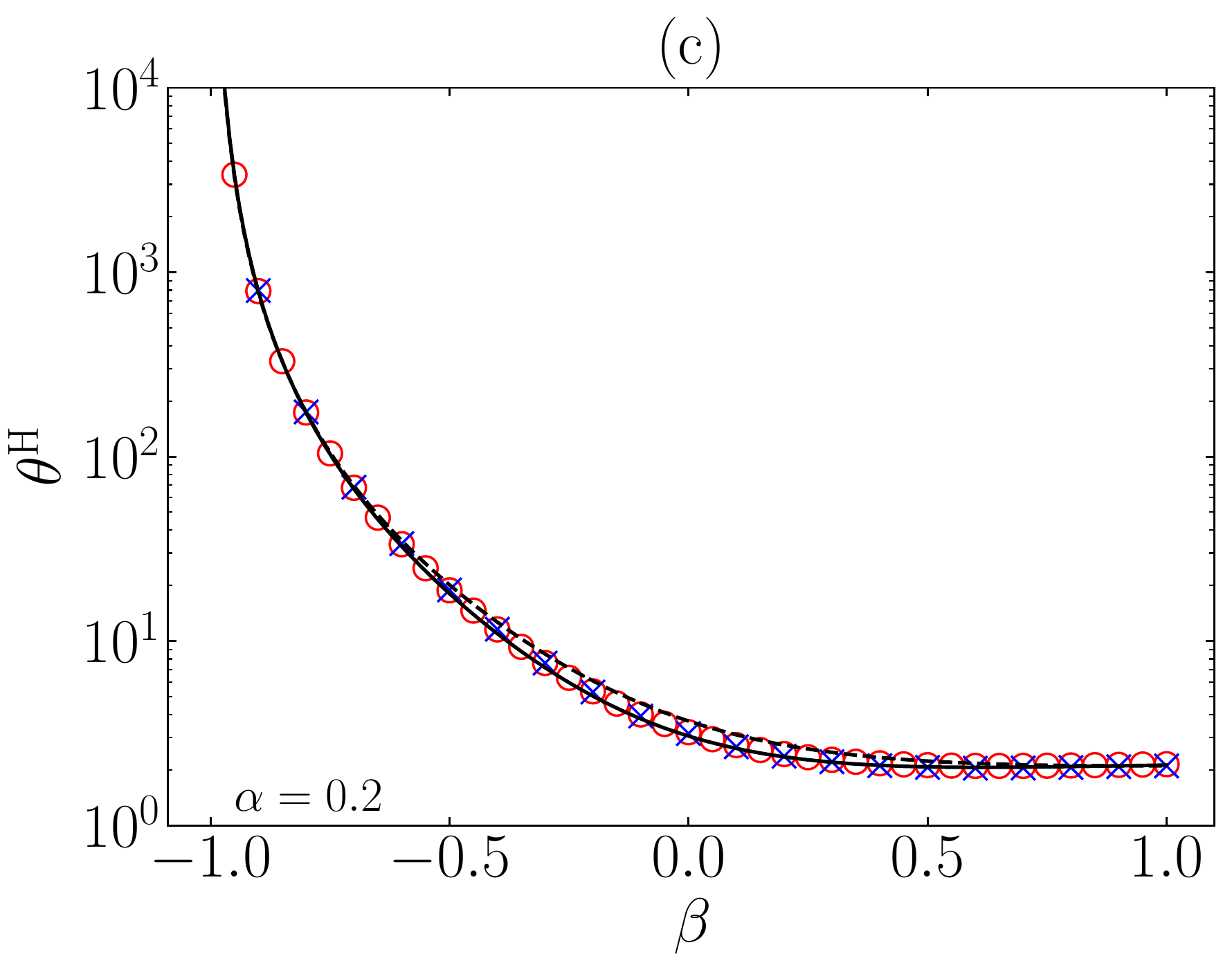}\\
    \includegraphics[width=0.3\textwidth]{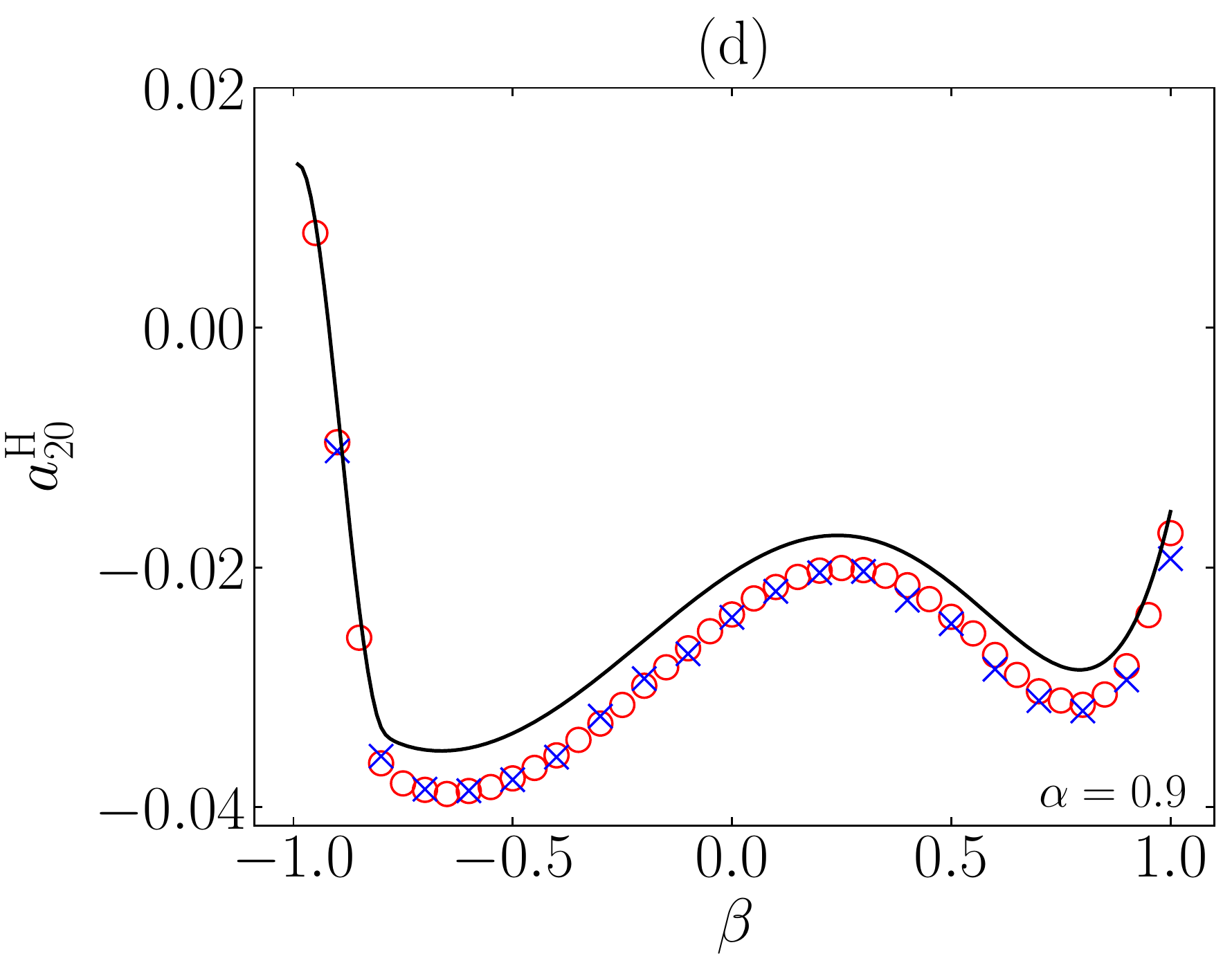}
    \includegraphics[width=0.3\textwidth]{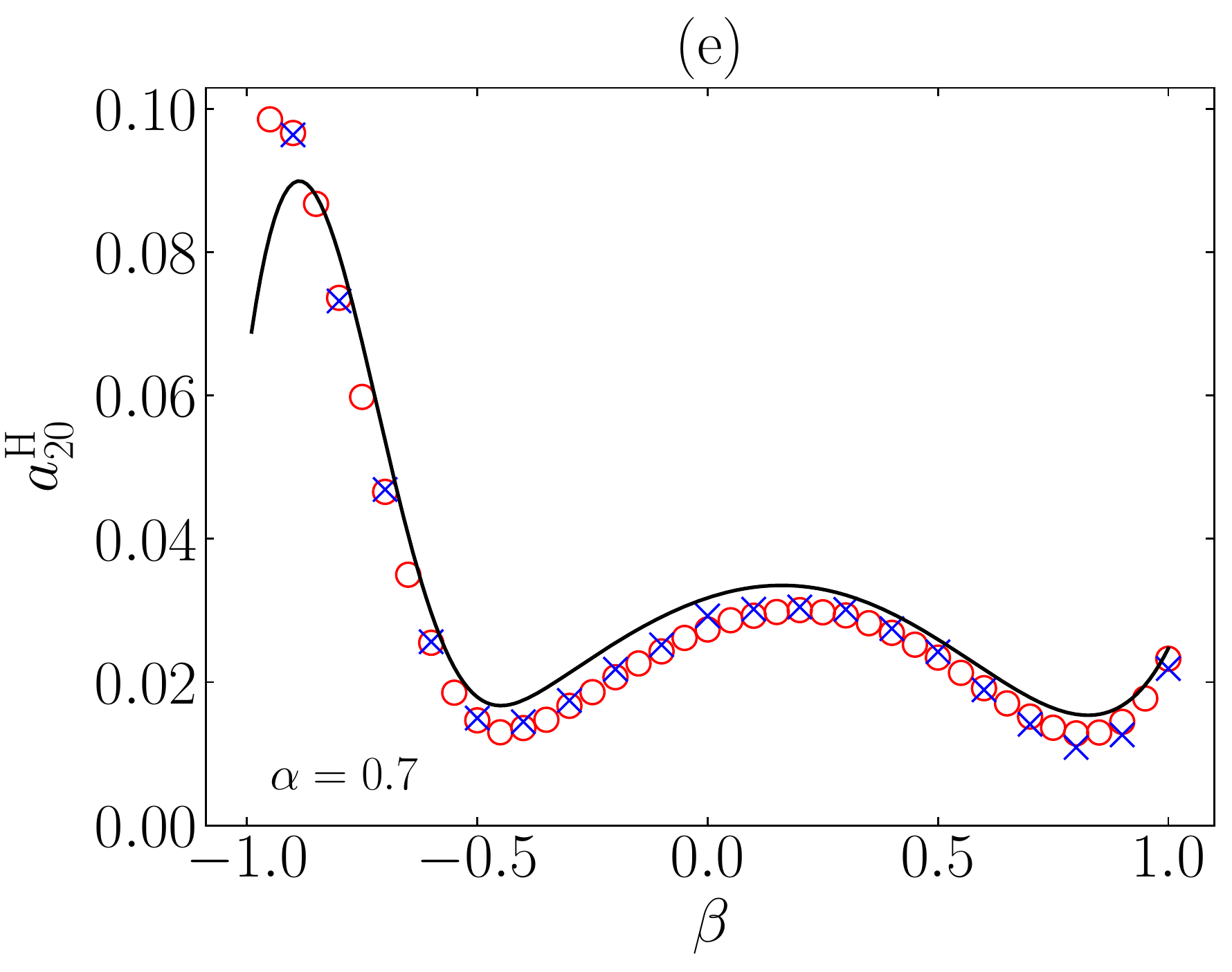}
    \includegraphics[width=0.3\textwidth]{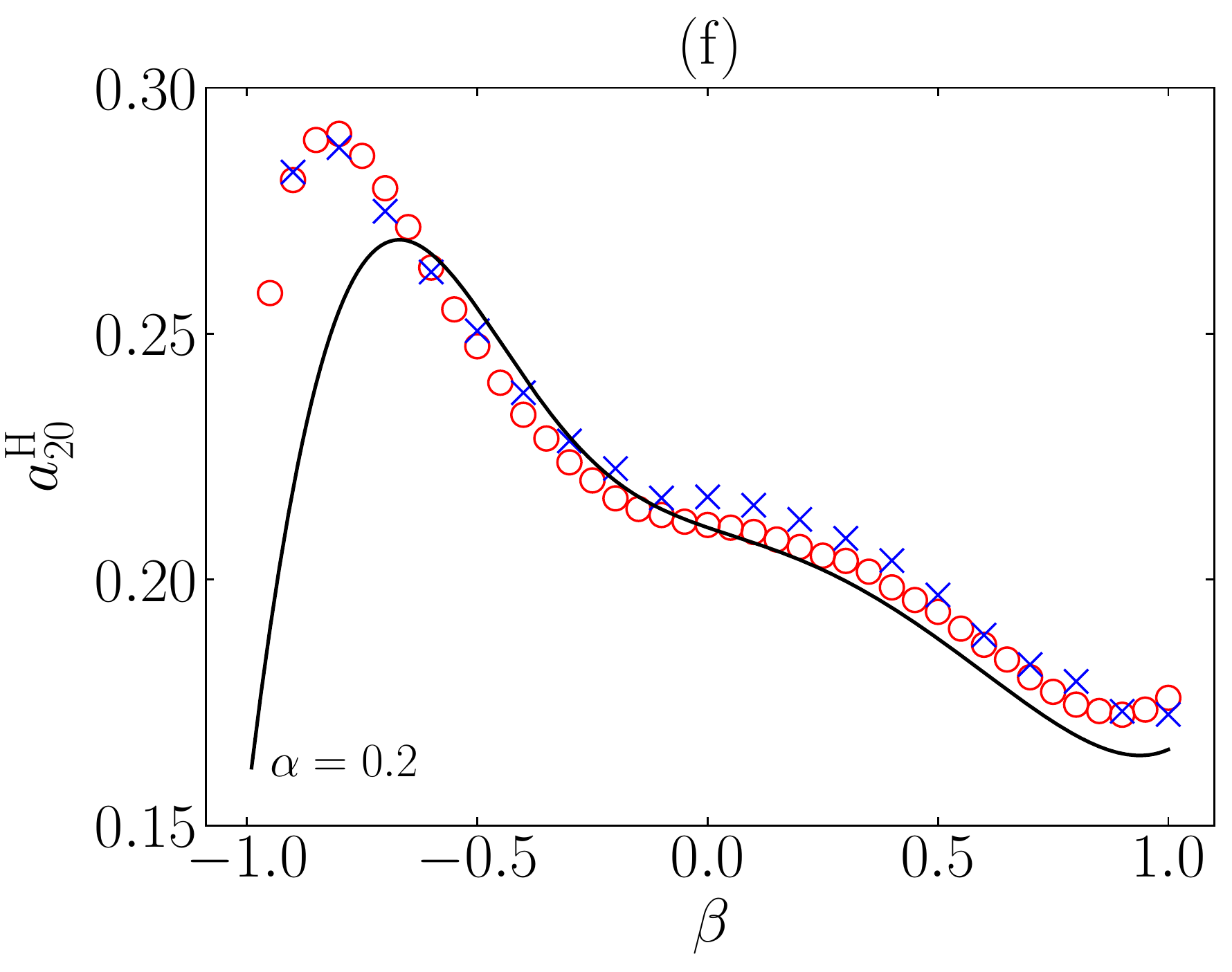}\\
    \includegraphics[width=0.3\textwidth]{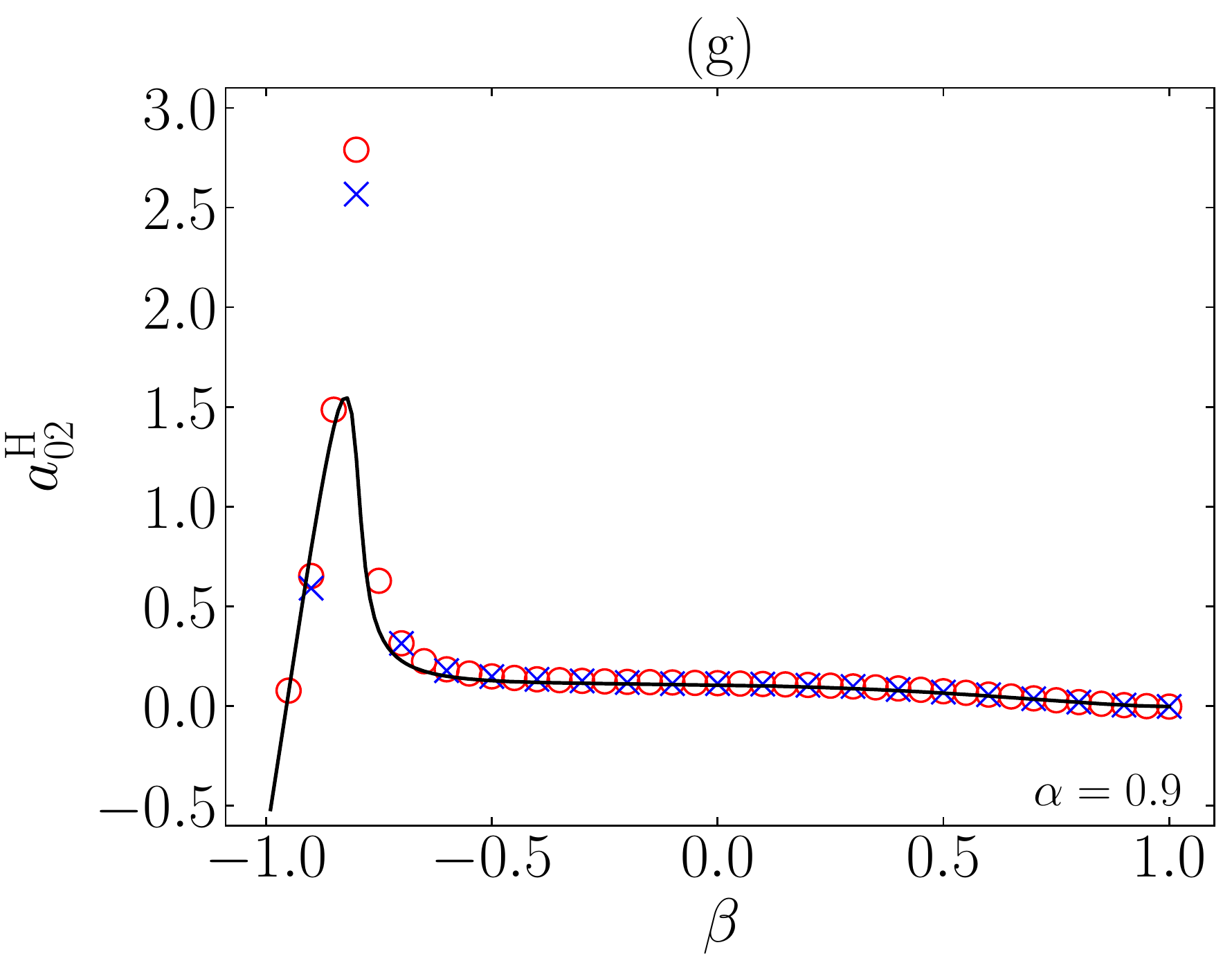}
    \includegraphics[width=0.3\textwidth]{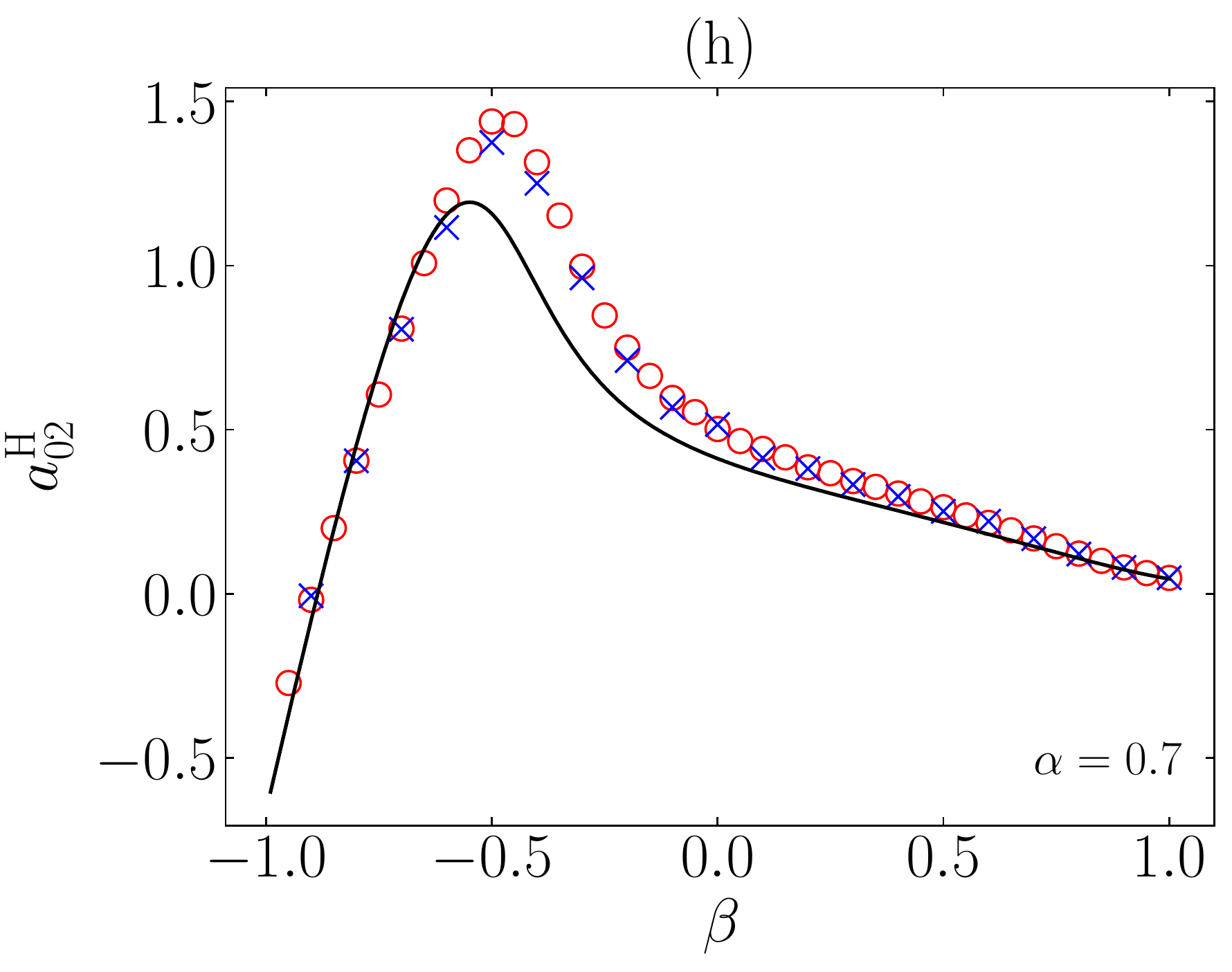}
    \includegraphics[width=0.3\textwidth]{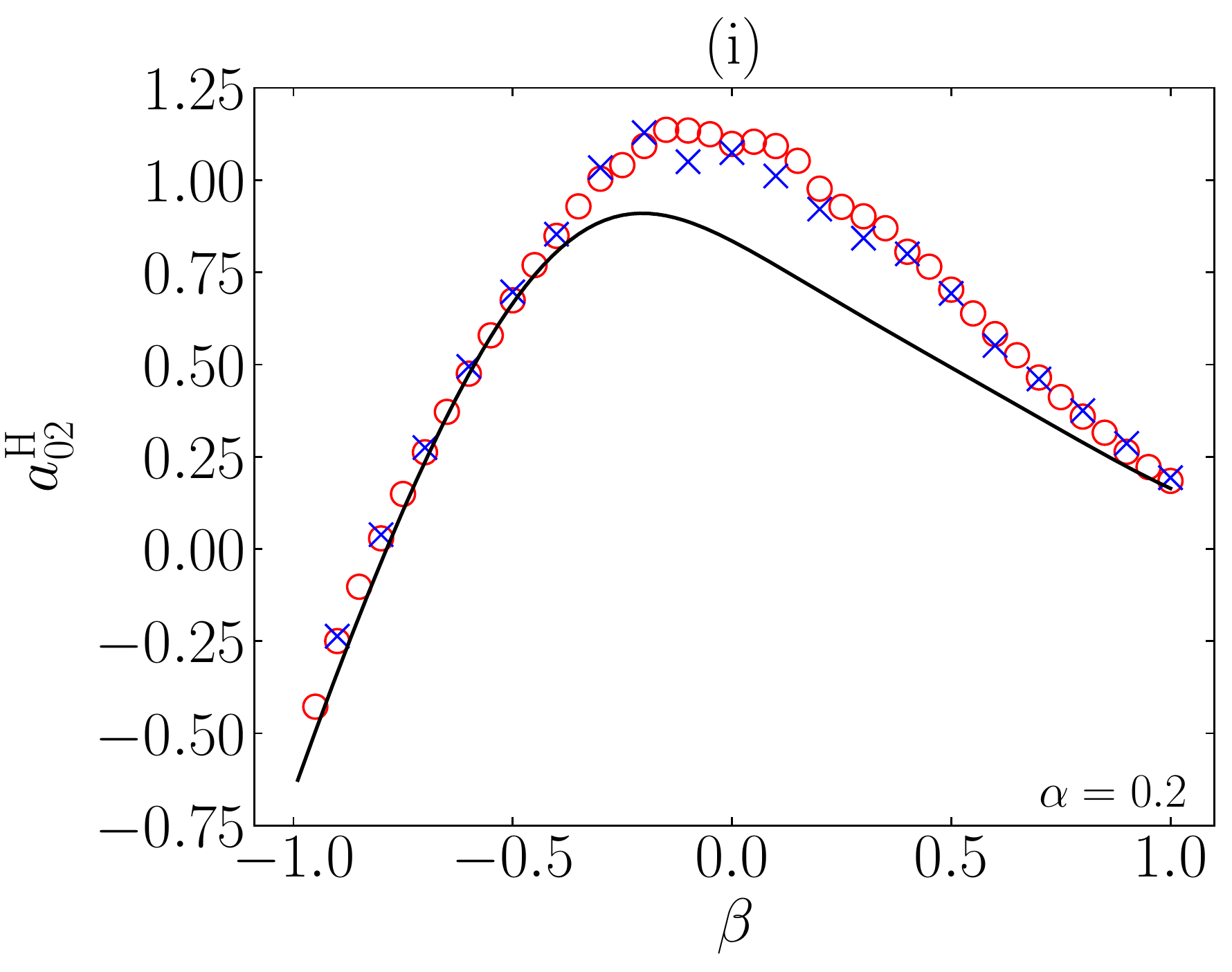}\\
    \includegraphics[width=0.3\textwidth]{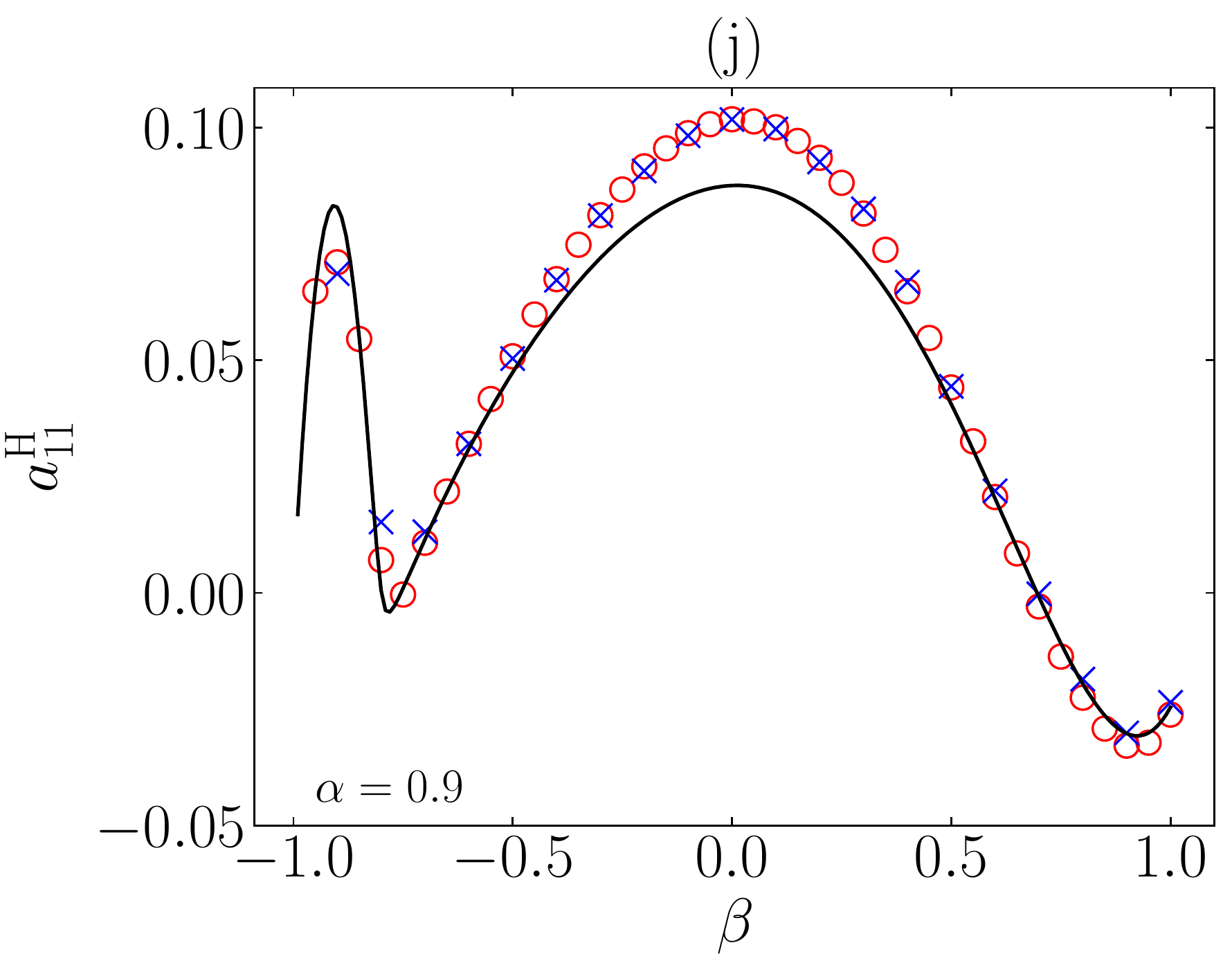}
    \includegraphics[width=0.3\textwidth]{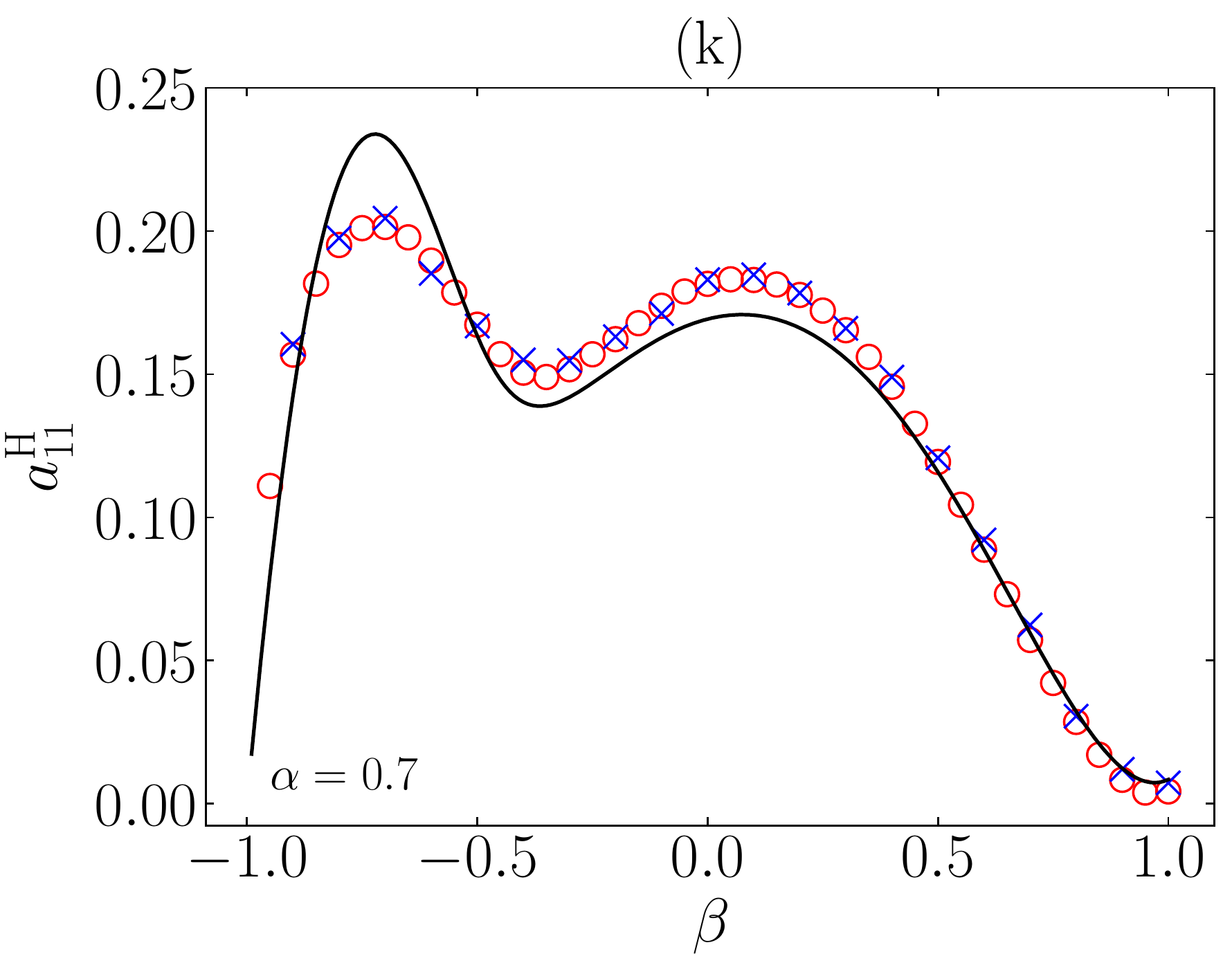}
    \includegraphics[width=0.3\textwidth]{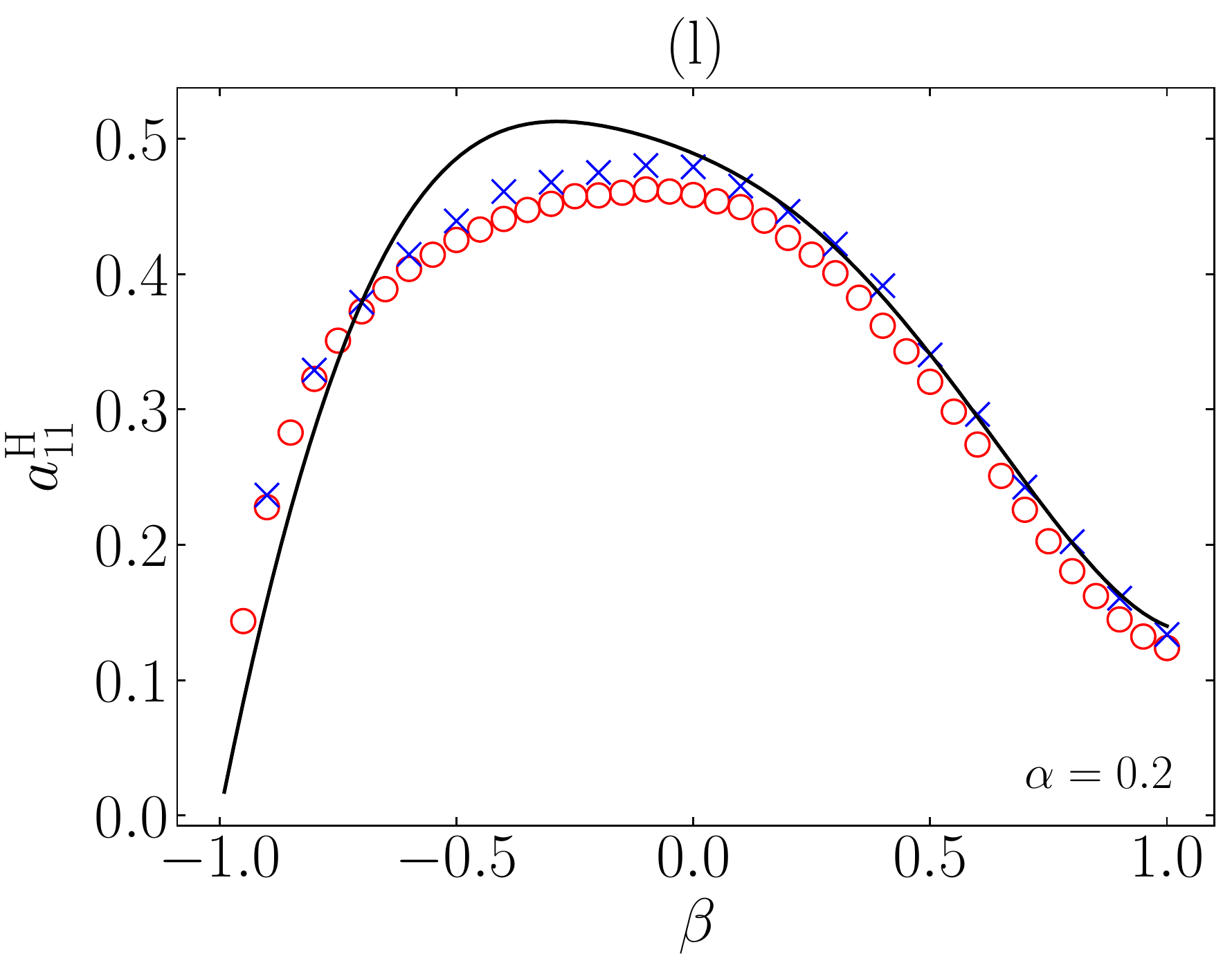}
    \caption{Plots of (a)--(c) the temperature ratio $\theta^\HCS$, (d)--(f) the cumulant $a_{20}^\HCS$, (g)--(i) the cumulant $a_{02}^\HCS$, and (j)--(l) the cumulant $a_{11}^\HCS$, for uniform disks ($\kappa=\frac{1}{2}$), as functions of the coefficient of tangential restitution $\eet$. The left [(a), (d), (g), (j)], middle [(b), (e), (h), (k)], and right [(c), (f), (i), (l)] panels correspond to $\een=0.9$, $0.7$, and $0.2$, respectively. Symbols represent DSMC ($\circ$) and EDMD ($\times$) results, while the solid lines are theoretical predictions from the Sonine approximation (SA). Additionally, the dashed lines in (a)--(c) represent the Maxwellian approximation (MA) for the temperature ratio. Note that a vertical logarithmic scale is used in (a)--(c).}
    \label{fig:sim_res_1}
\end{figure*}
\begin{figure*}[ht]
    \centering
    \includegraphics[width=0.3\textwidth]{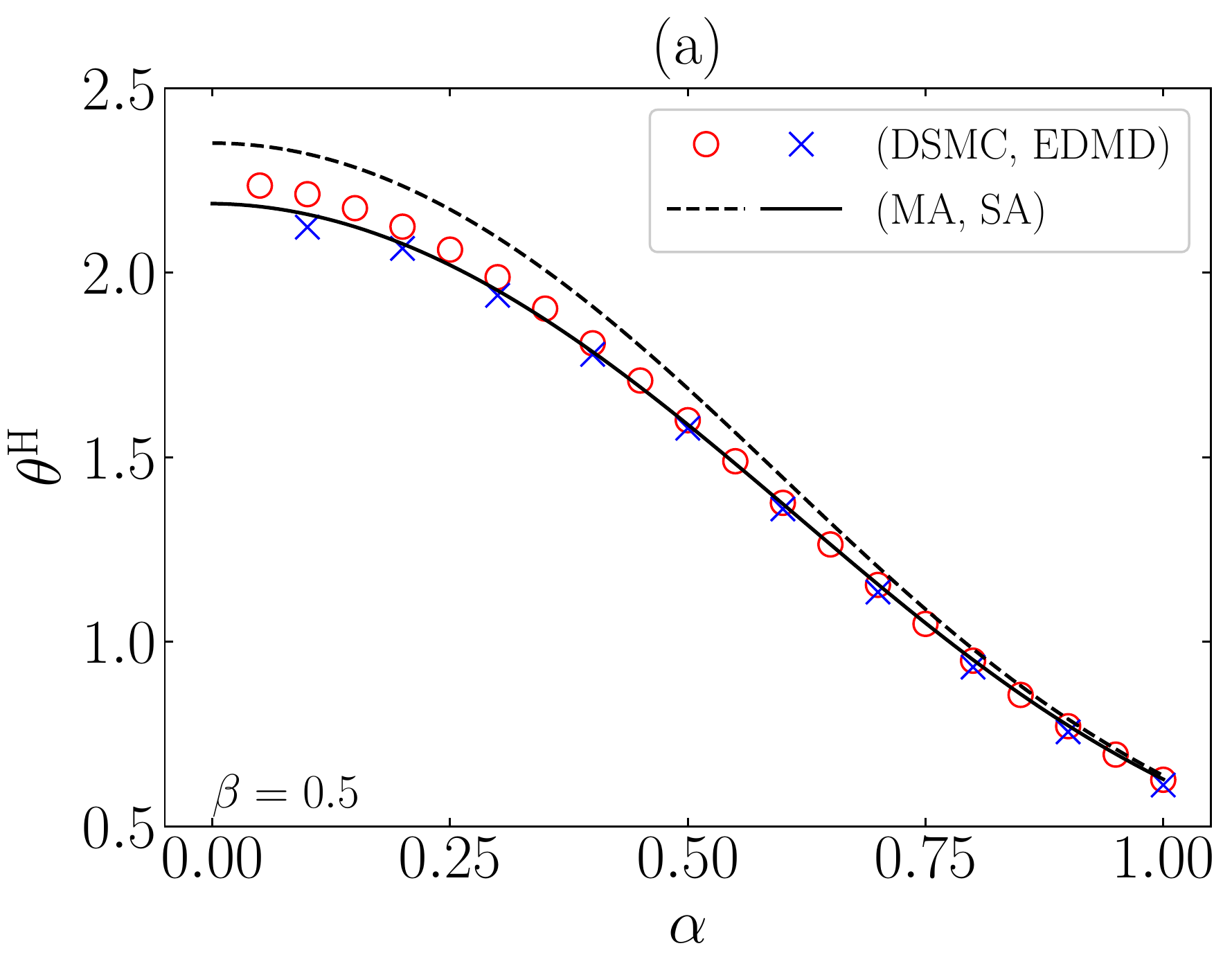}
    \includegraphics[width=0.3\textwidth]{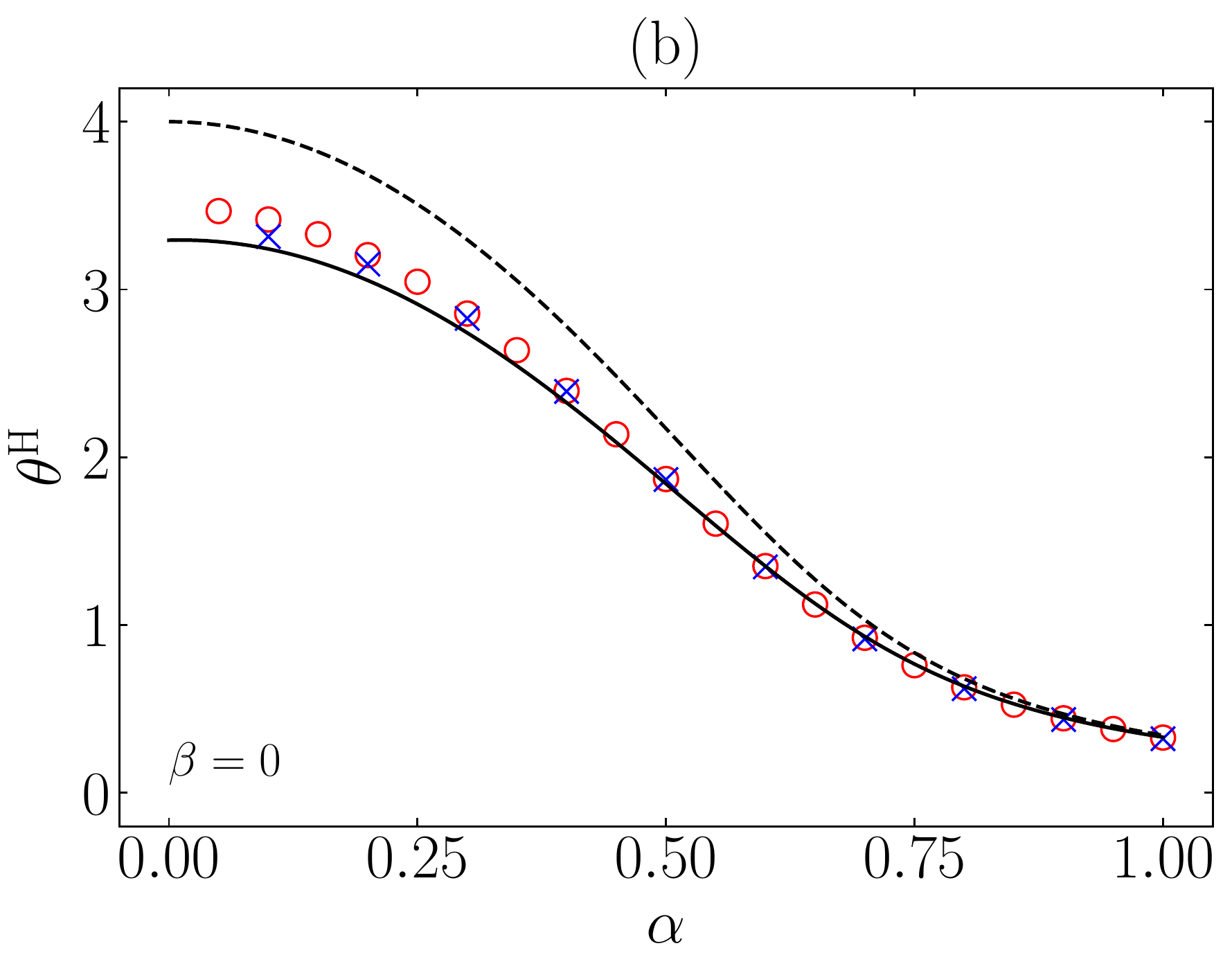}
    \includegraphics[width=0.3\textwidth]{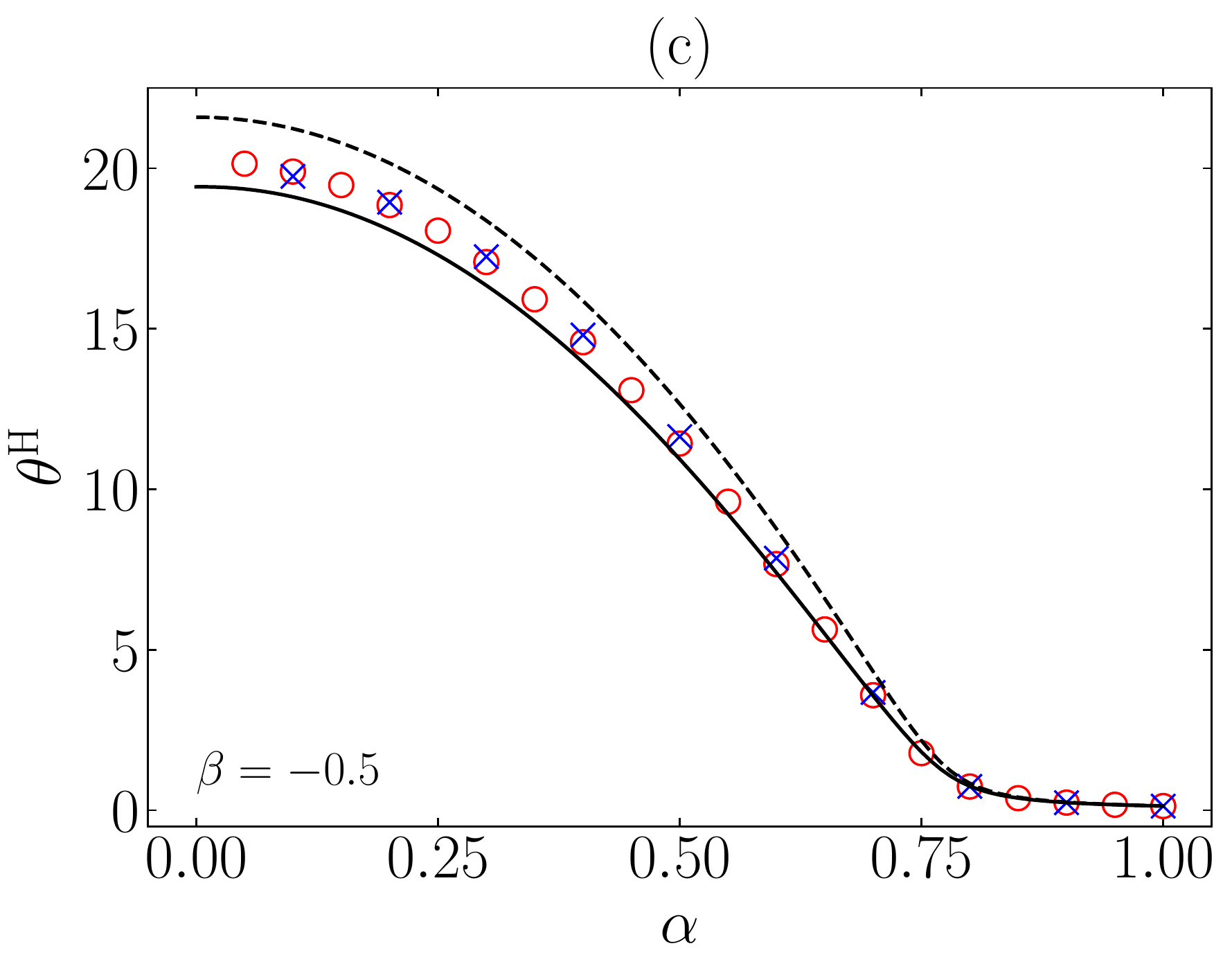}\\
    \includegraphics[width=0.3\textwidth]{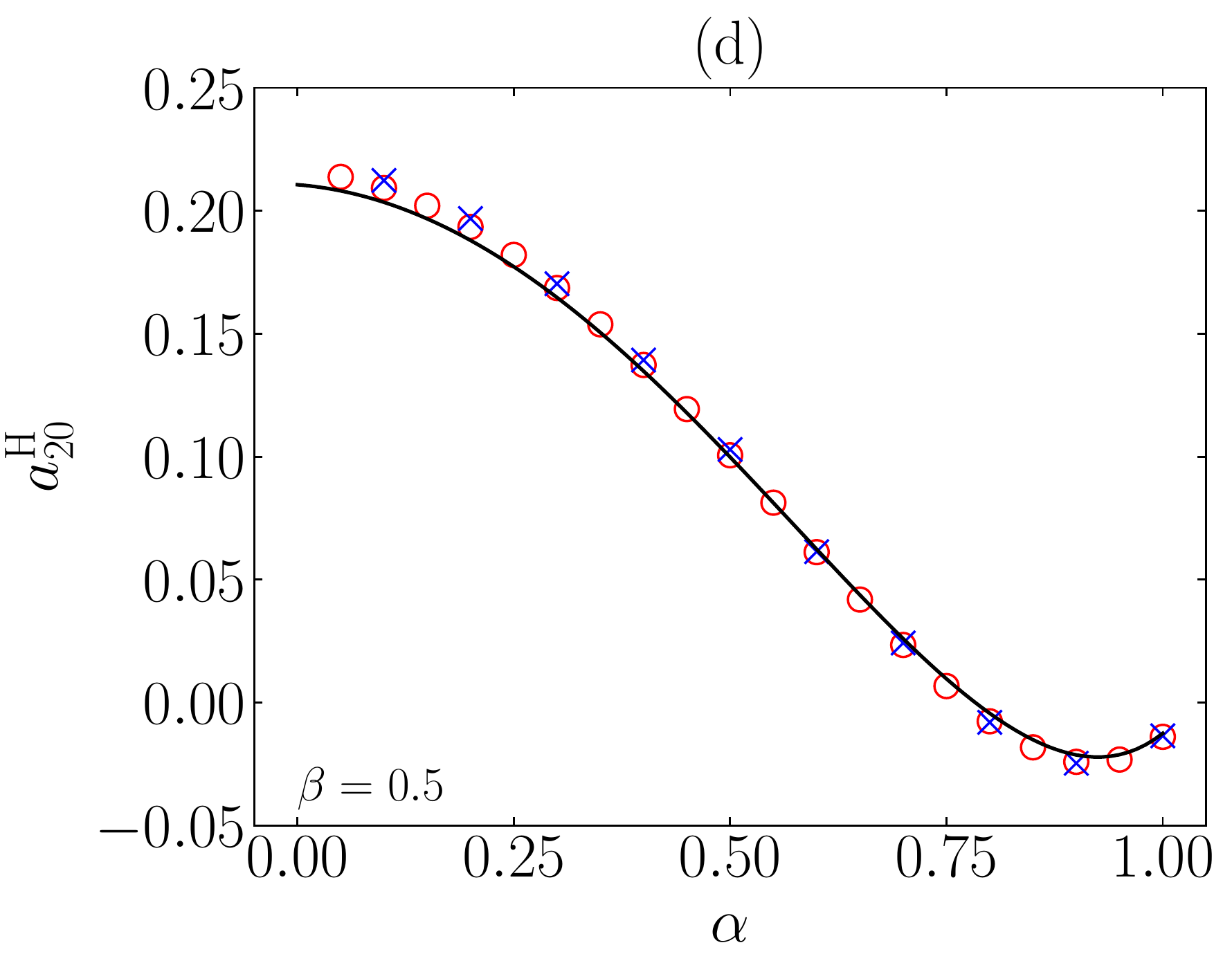}
    \includegraphics[width=0.3\textwidth]{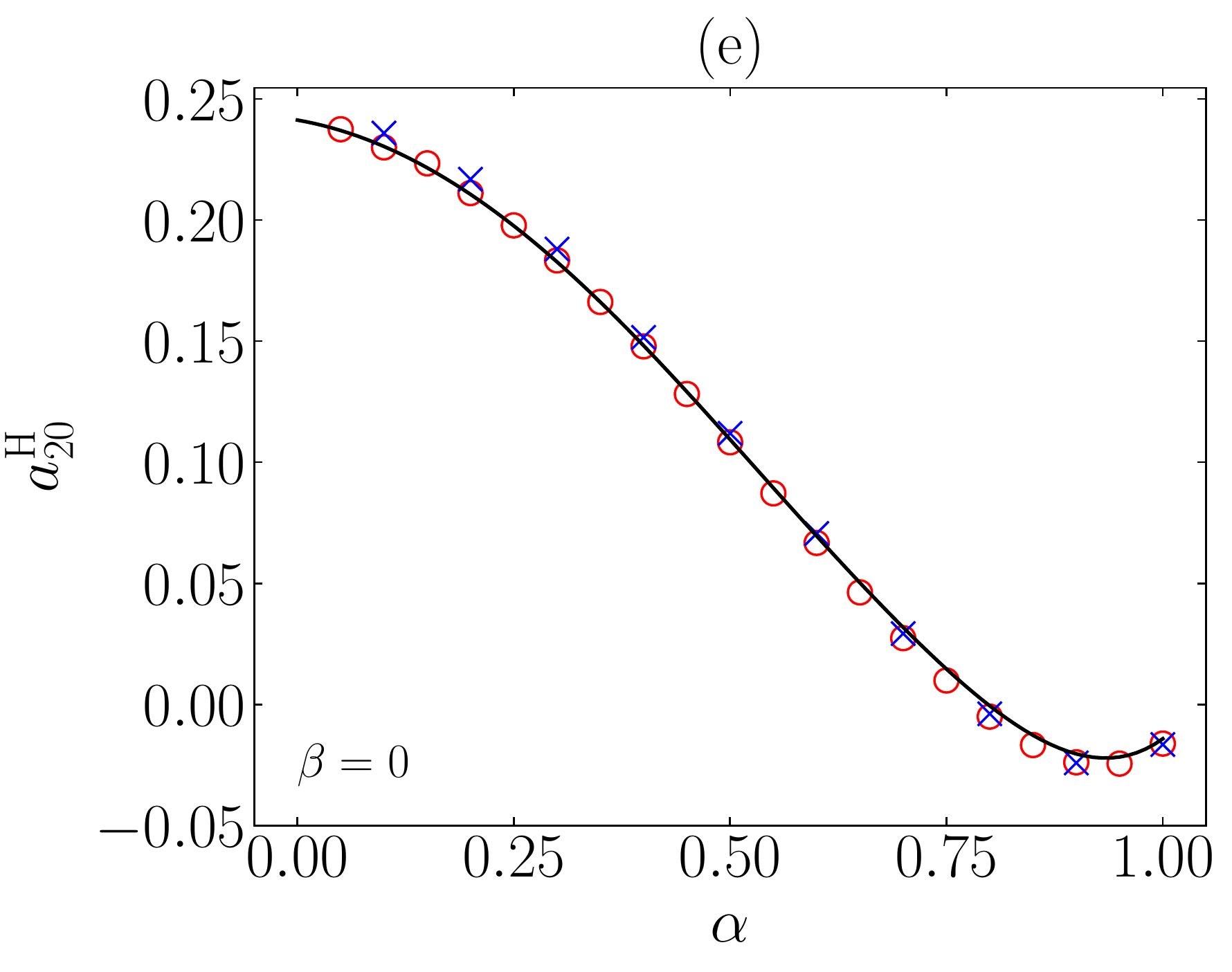}
    \includegraphics[width=0.3\textwidth]{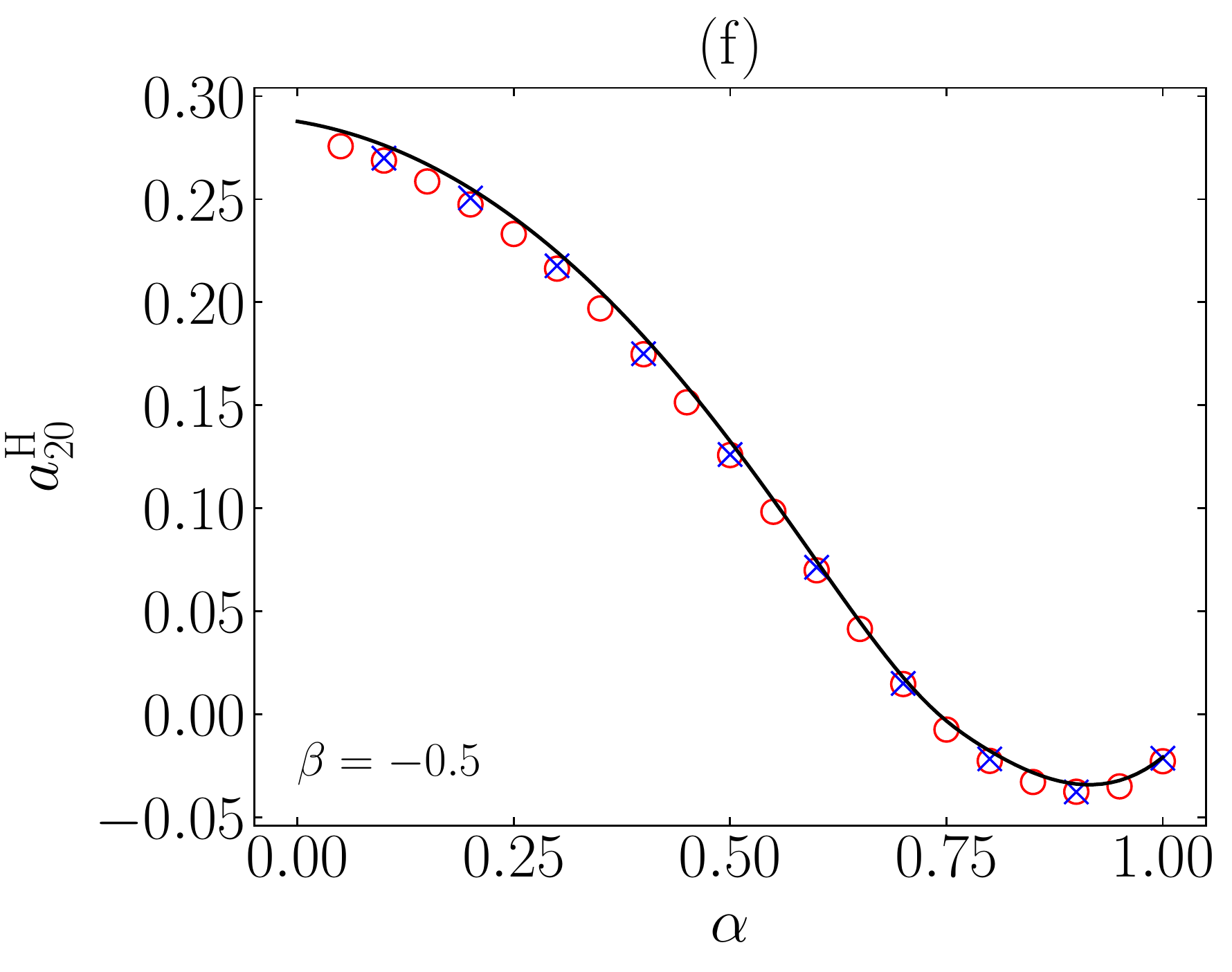}\\
    \includegraphics[width=0.3\textwidth]{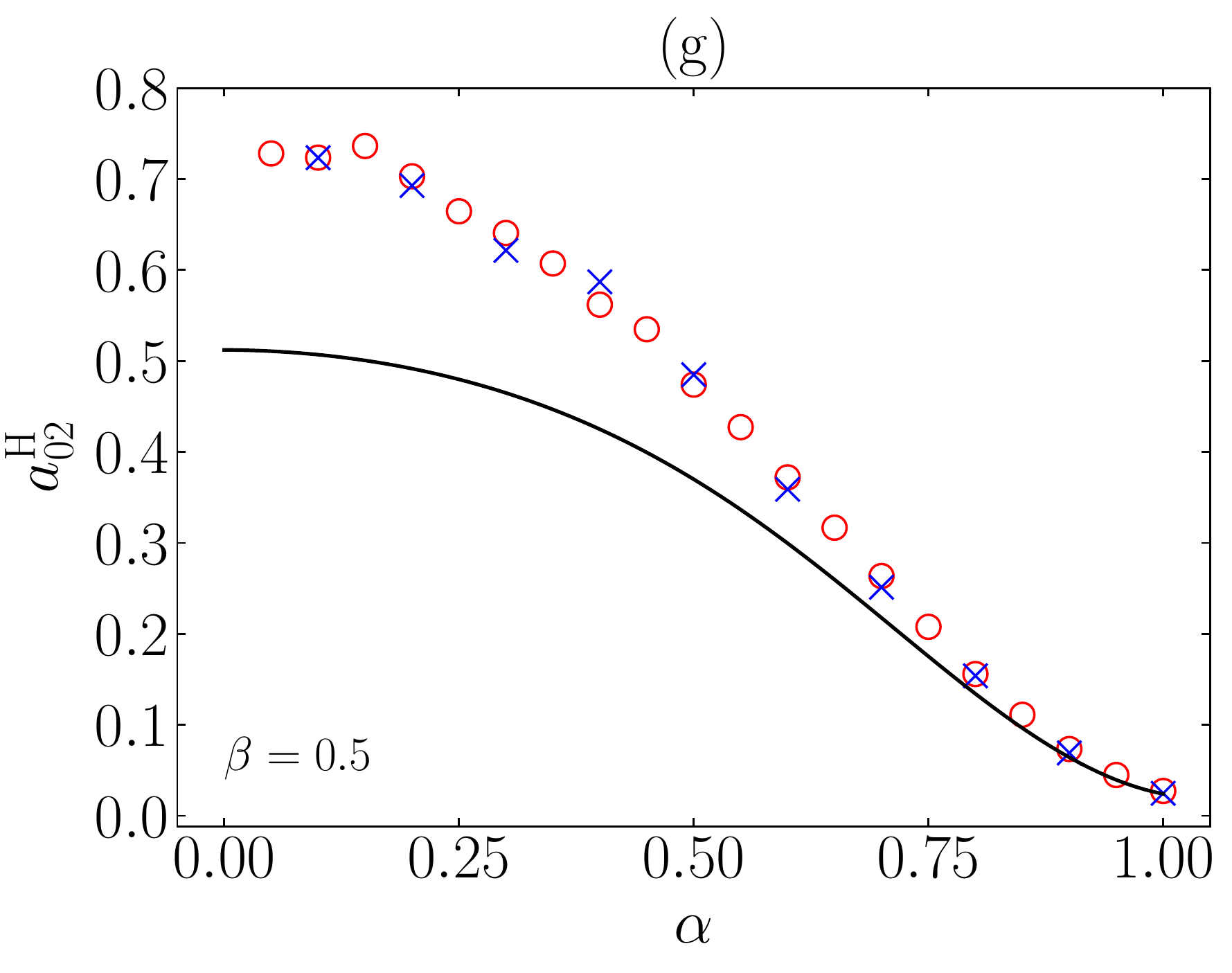}
    \includegraphics[width=0.3\textwidth]{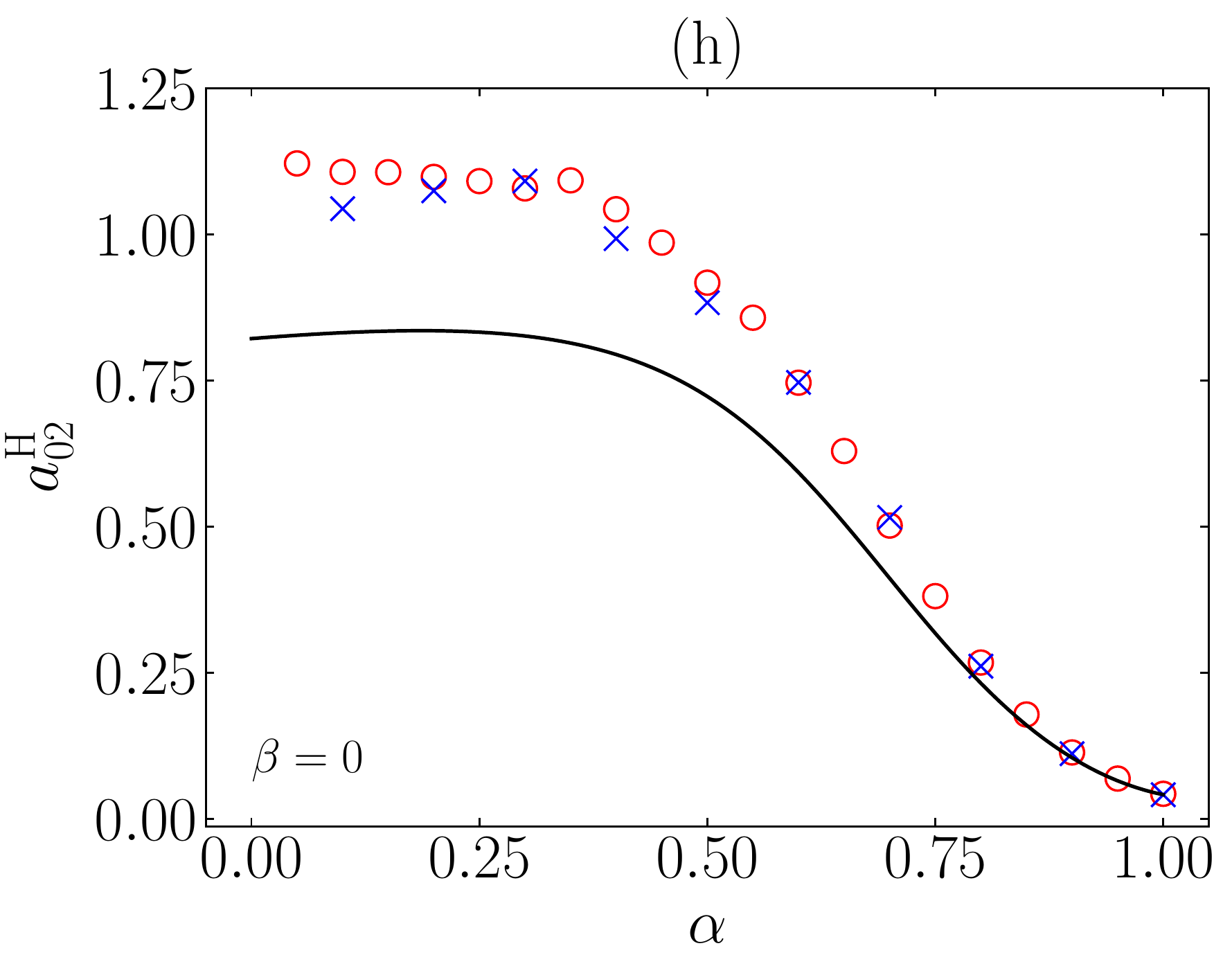}
    \includegraphics[width=0.3\textwidth]{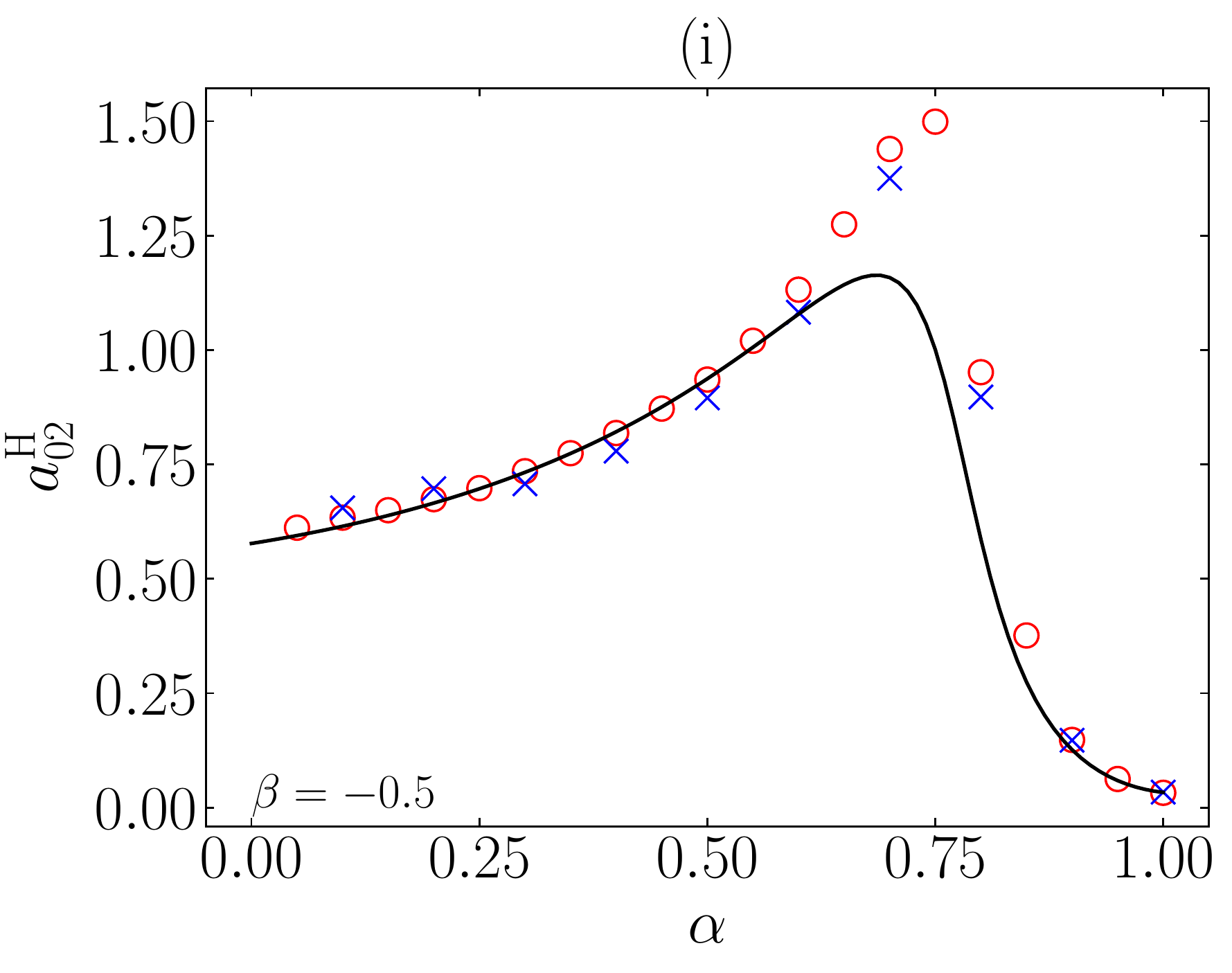}\\
    \includegraphics[width=0.3\textwidth]{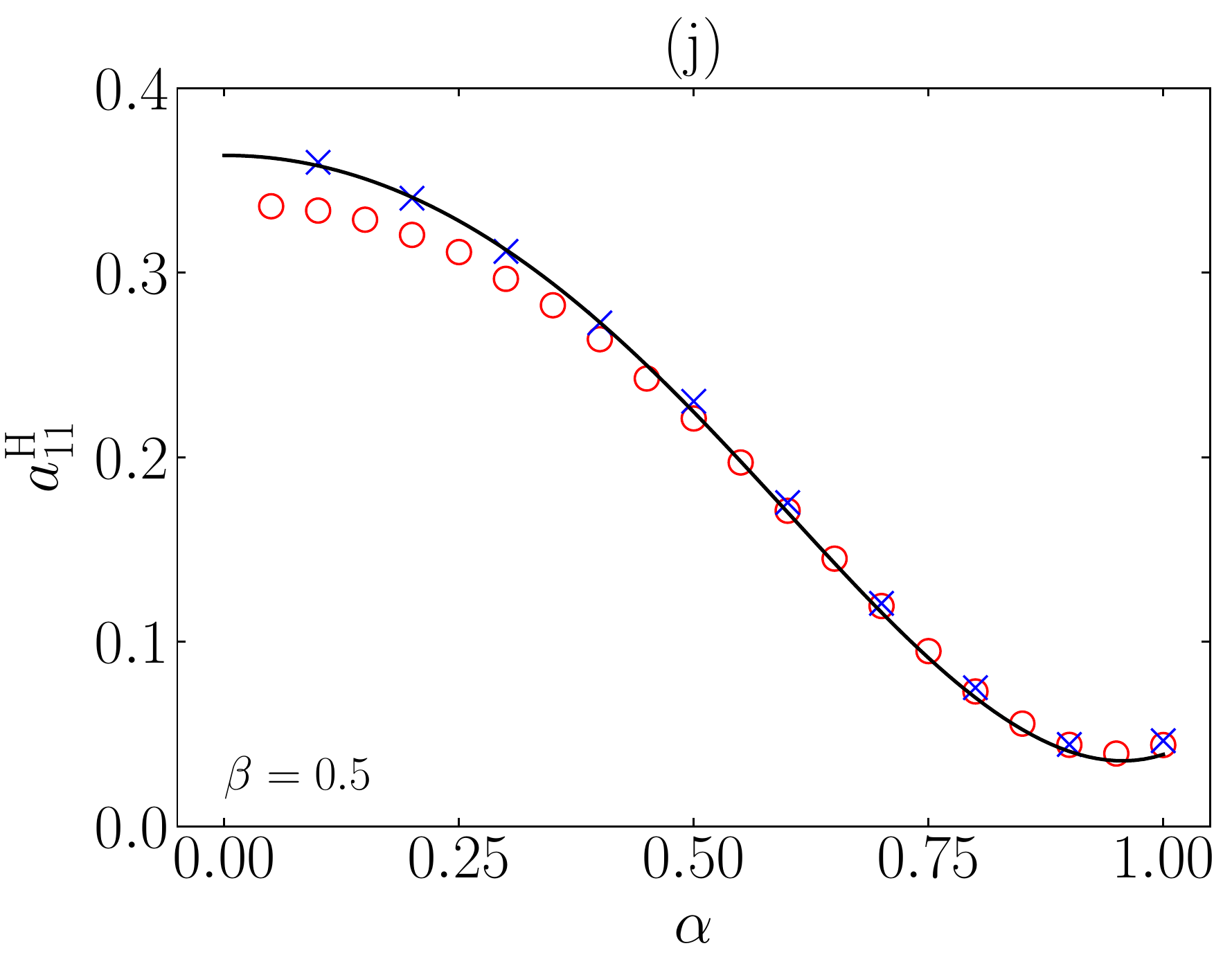}
    \includegraphics[width=0.3\textwidth]{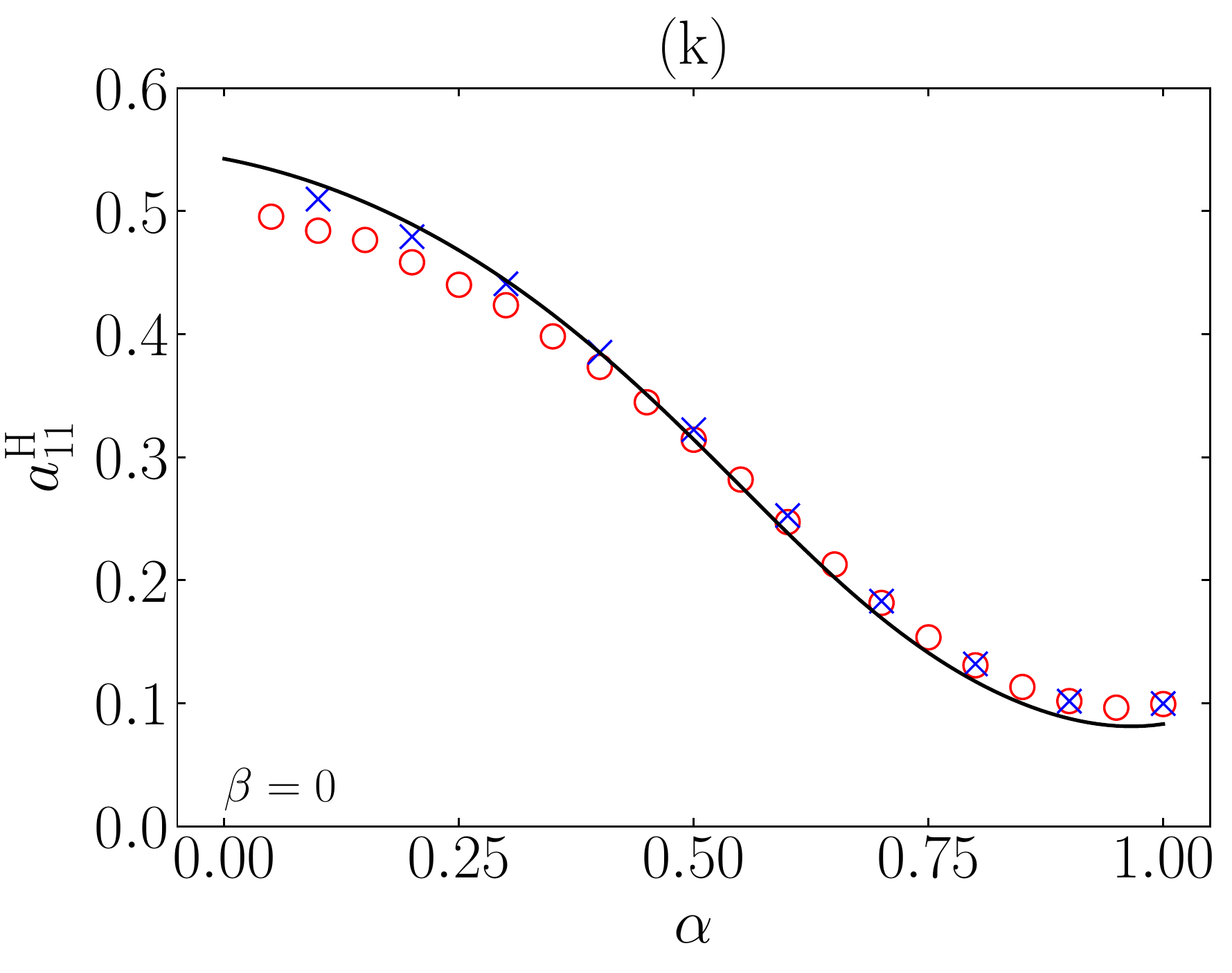}
    \includegraphics[width=0.3\textwidth]{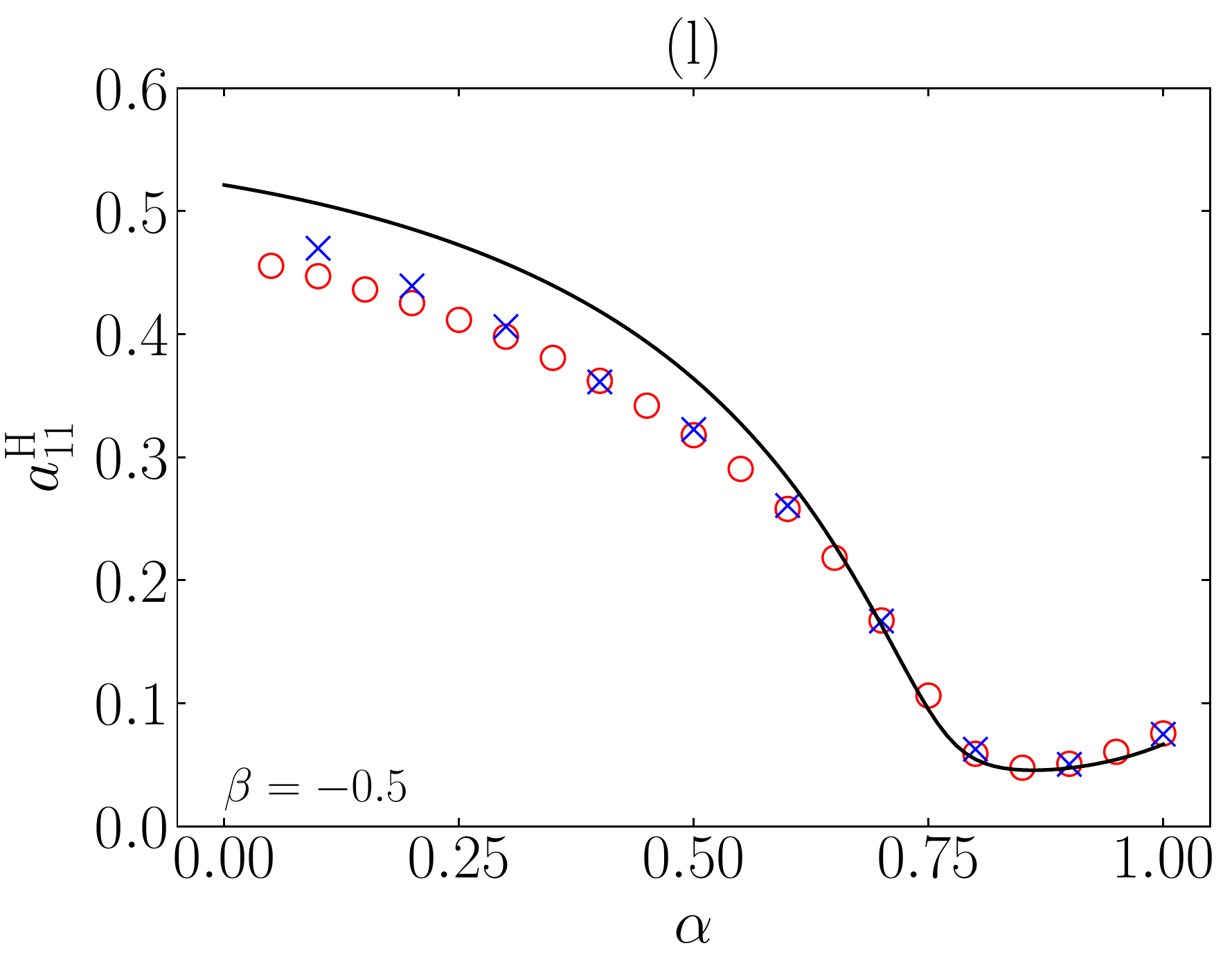}
    \caption{Same as in Fig.~\ref{fig:sim_res_1}, except that the quantities are plotted versus the coefficient of normal restitution $\een$ and now the left [(a), (d), (g), (j)], middle [(b), (e), (h), (k)], and right [(c), (f), (i), (l)] panels correspond to $\eet=0.5$, $0$, and $-0.5$, respectively. }
    \label{fig:sim_res_2}
\end{figure*}

\section{\label{sec:Sim_Res} Simulation Results}

\begin{figure*}[ht]
	\centering
	\includegraphics[width = 0.3\textwidth]{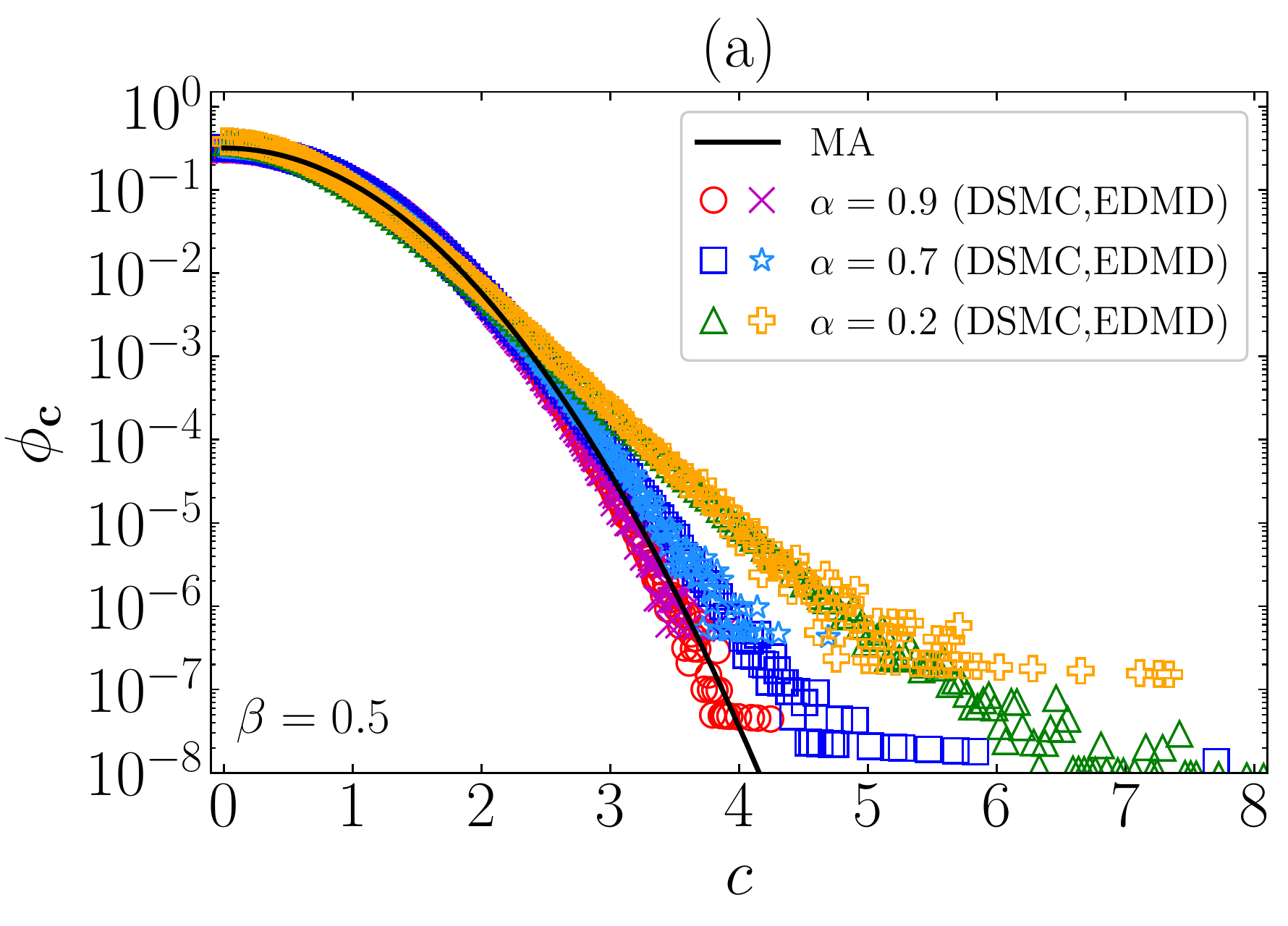}
	\includegraphics[width = 0.3\textwidth]{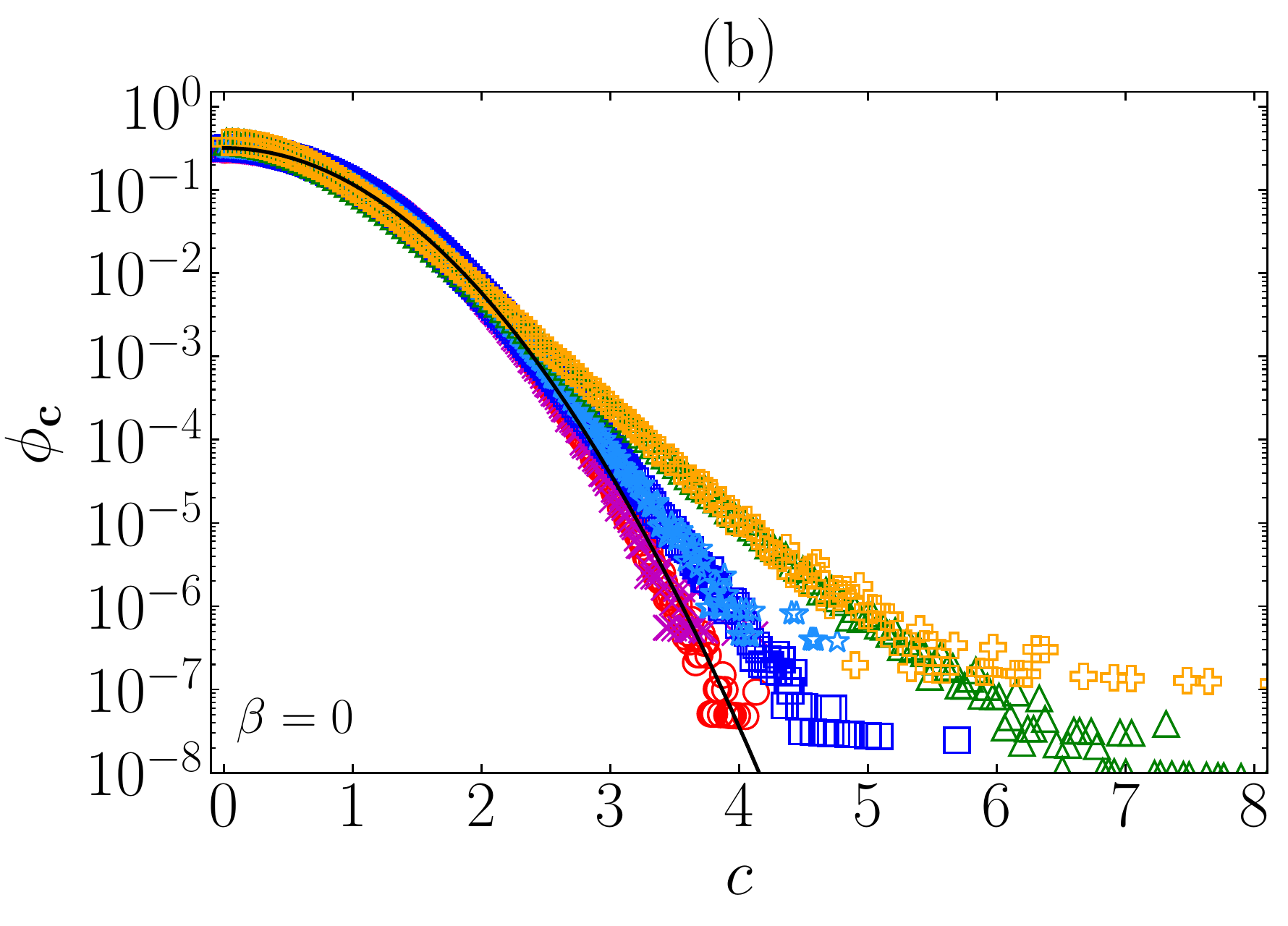}
	\includegraphics[width = 0.3\textwidth]{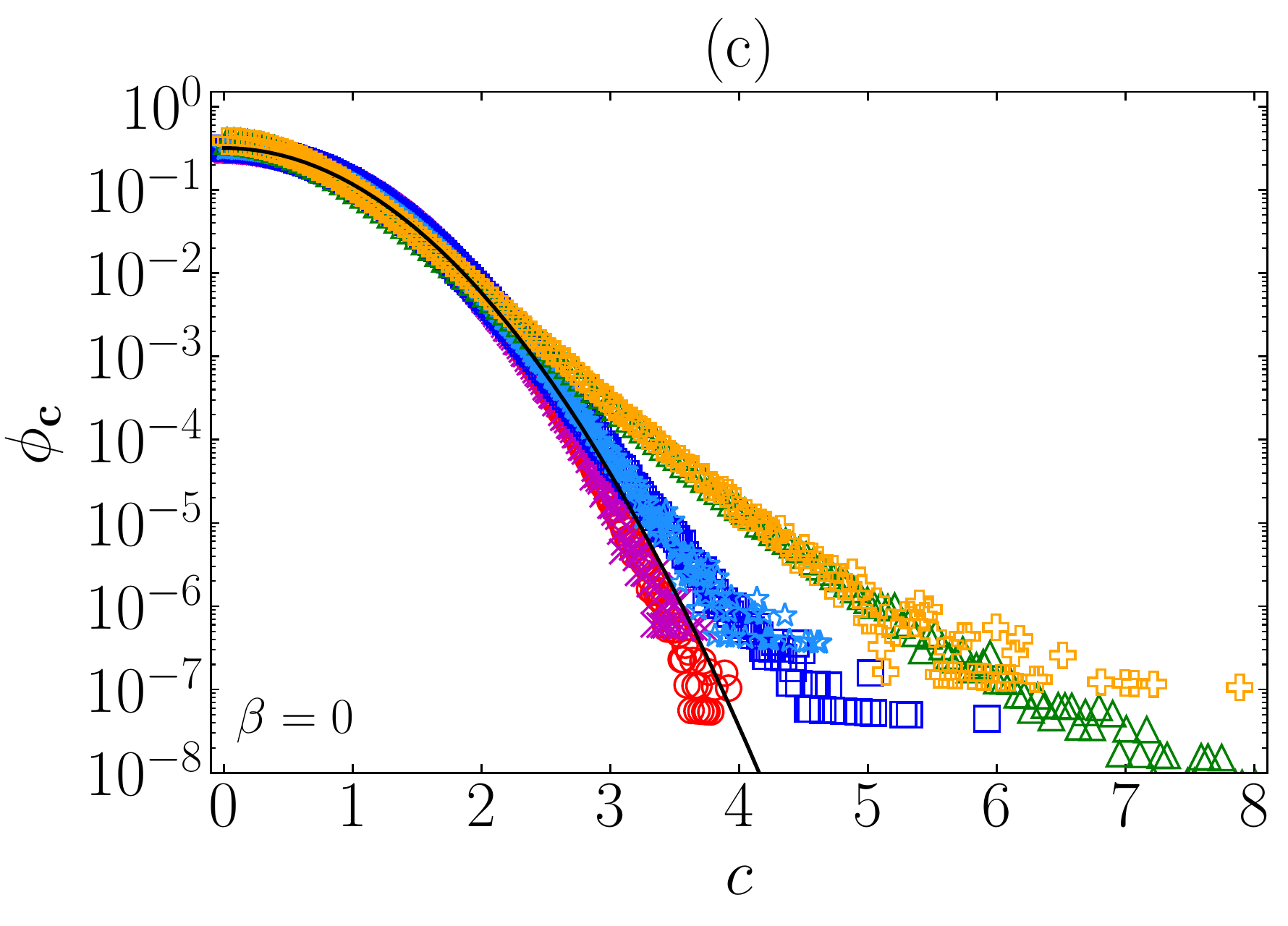}\\
	\includegraphics[width = 0.3\textwidth]{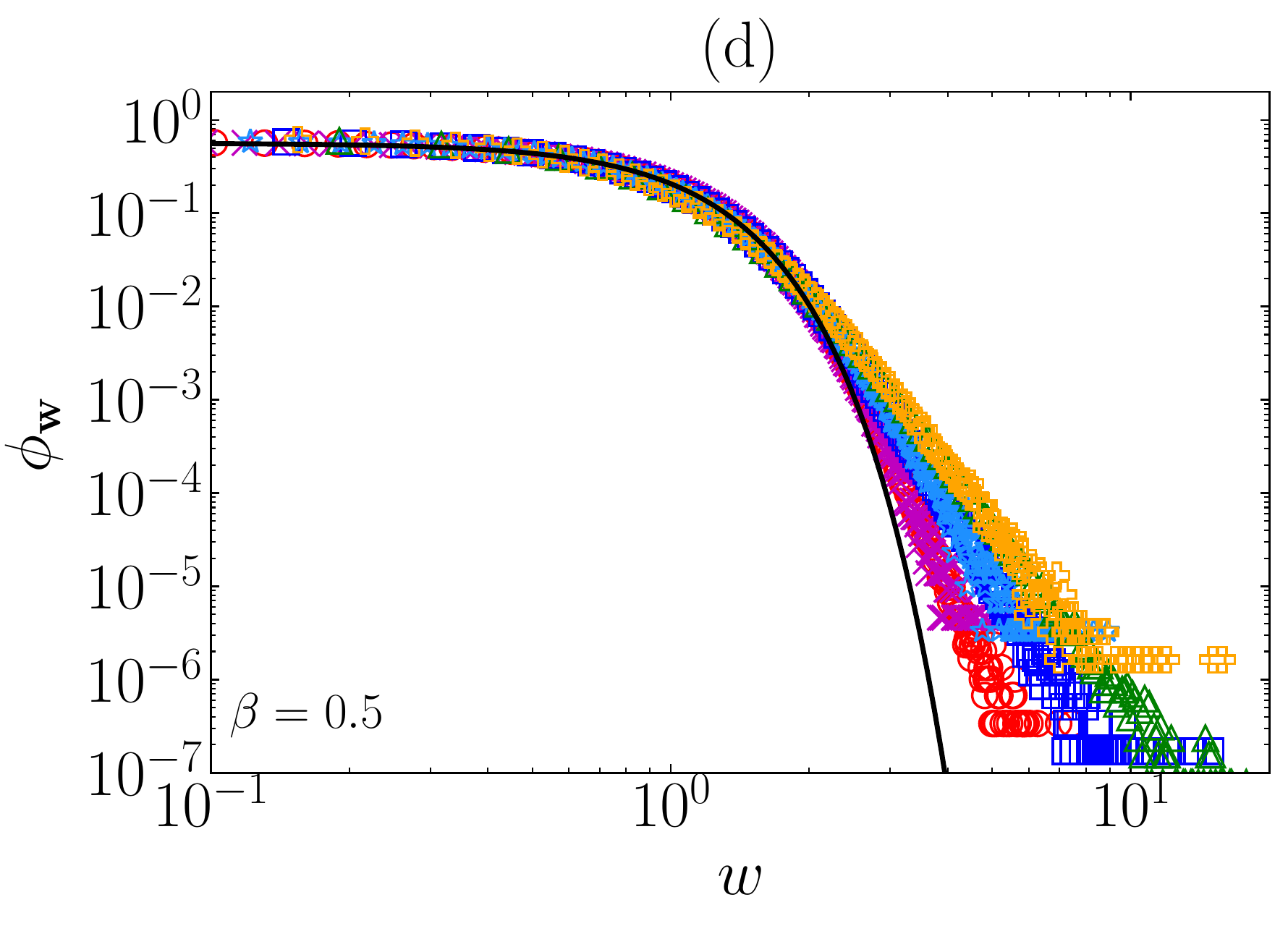}
	\includegraphics[width = 0.3\textwidth]{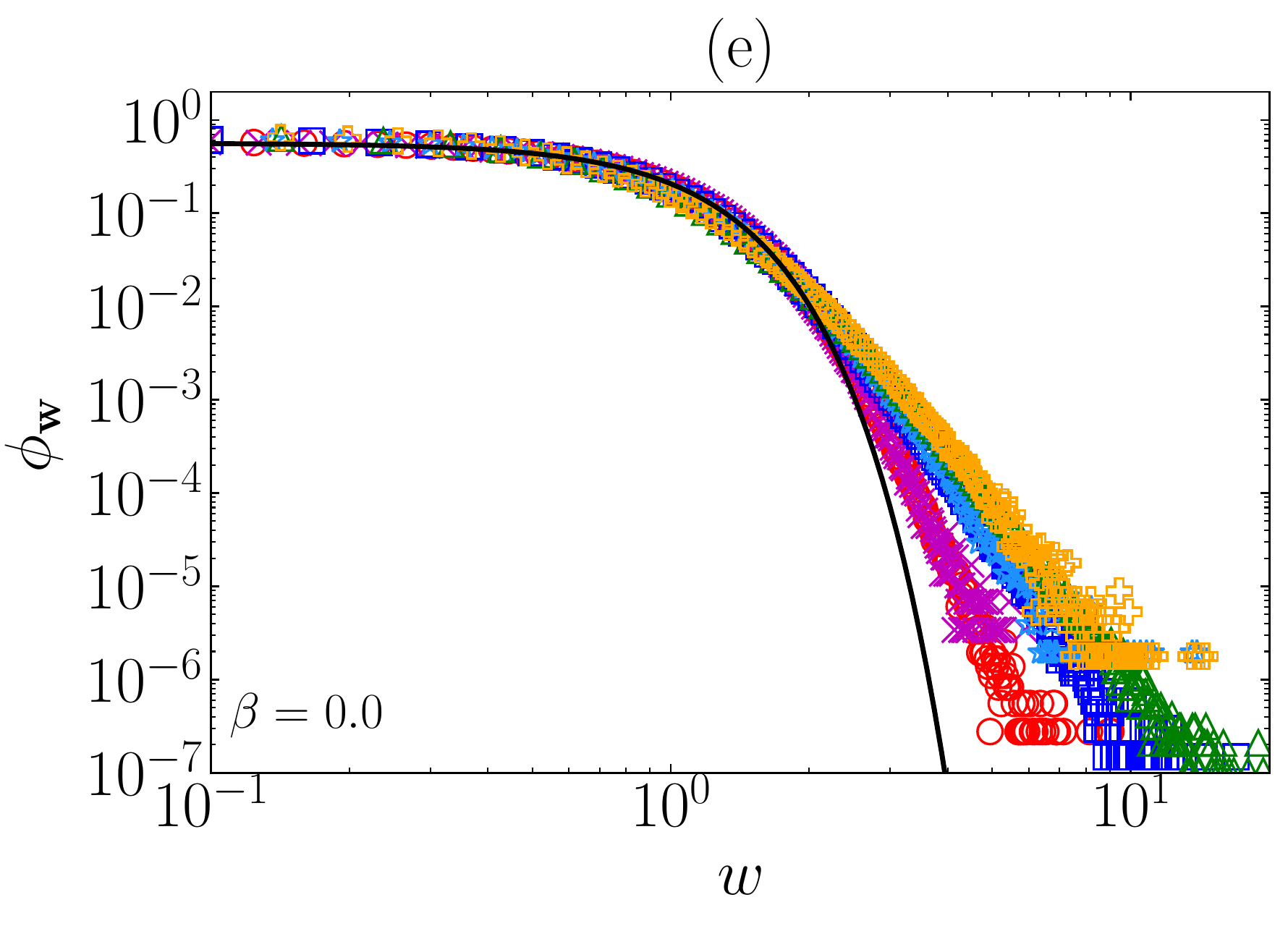}
	\includegraphics[width = 0.3\textwidth]{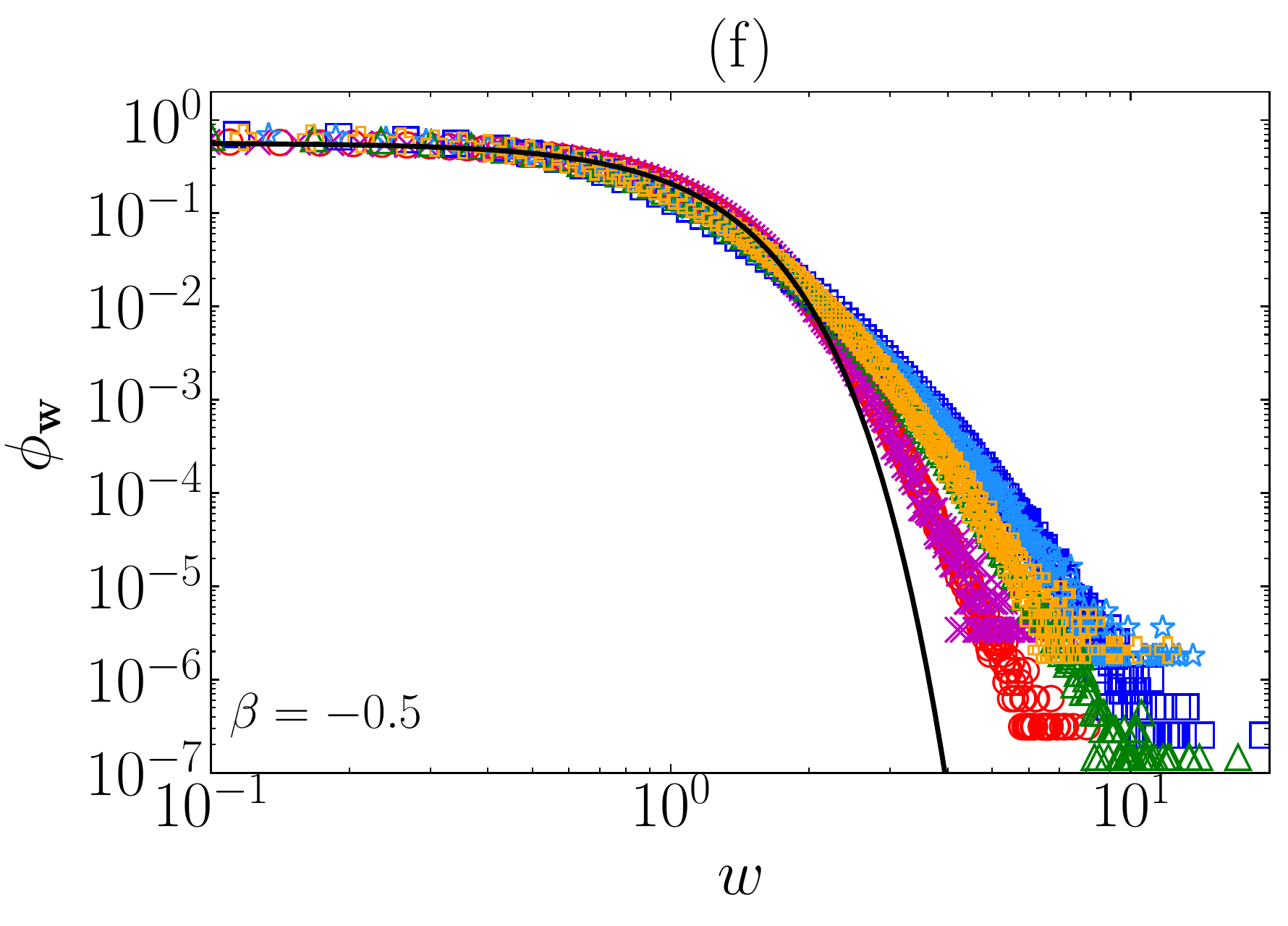}\\
	\includegraphics[width = 0.3\textwidth]{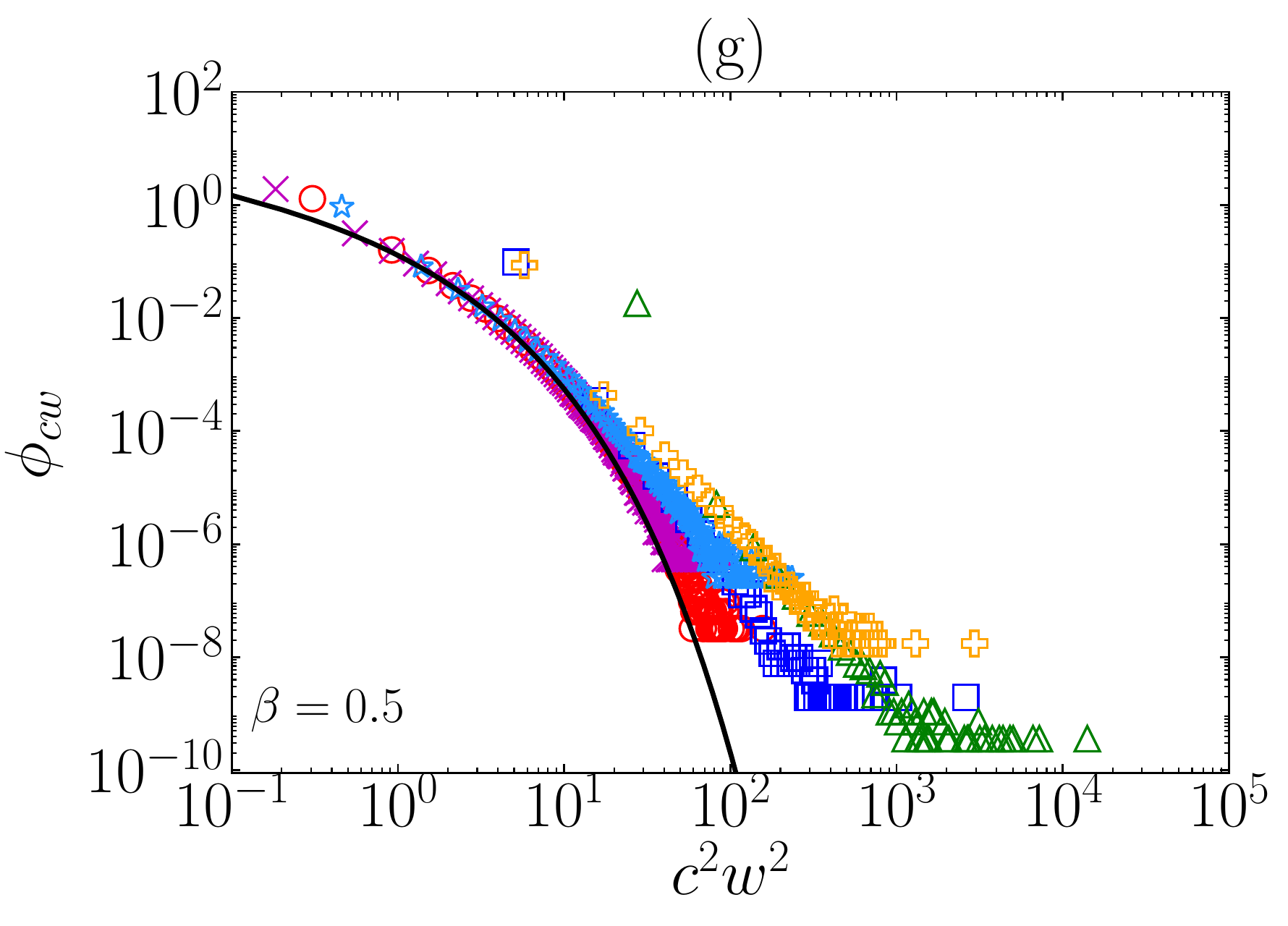}
	\includegraphics[width = 0.3\textwidth]{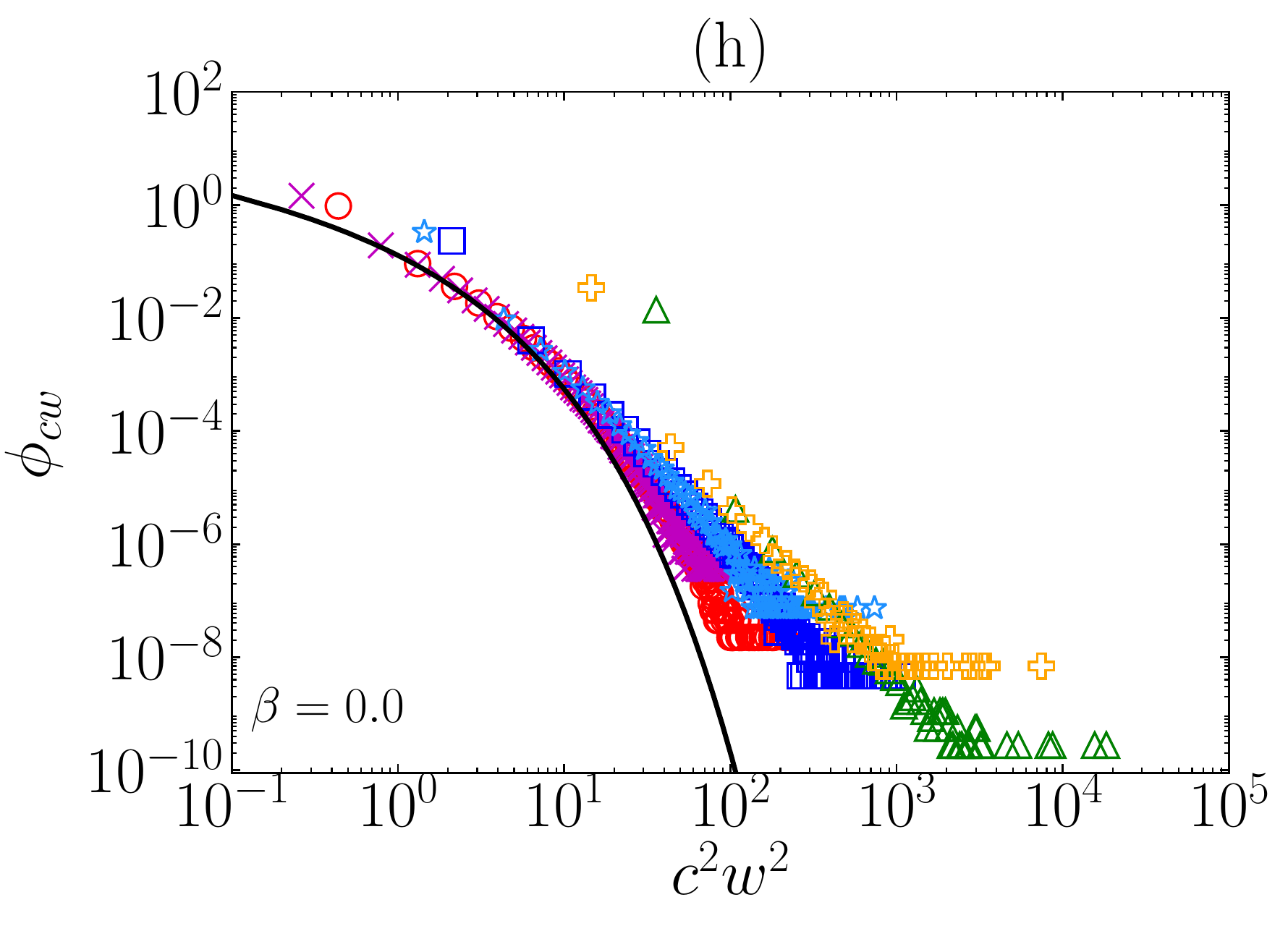}
	\includegraphics[width = 0.3\textwidth]{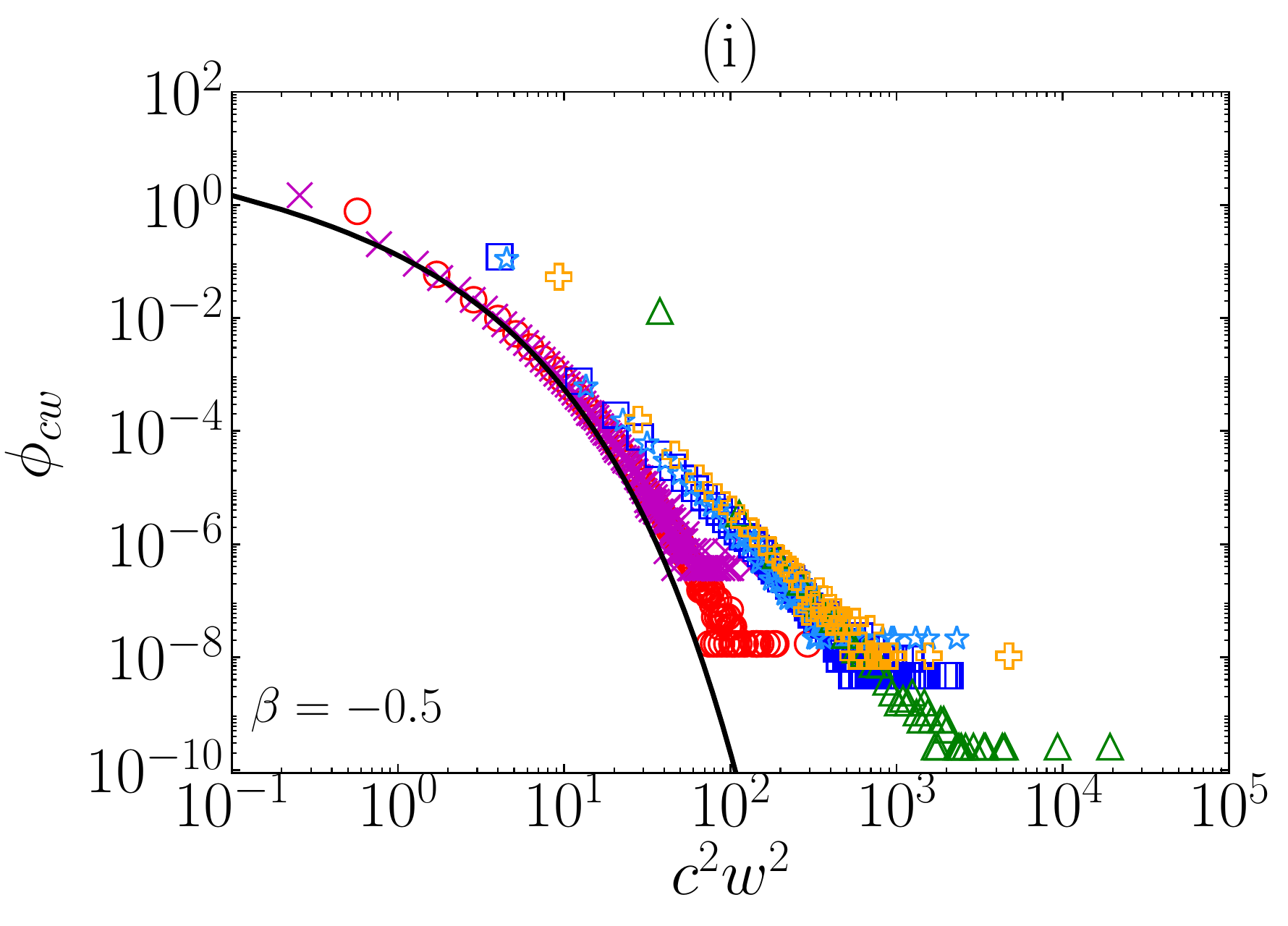}
	\caption{Simulation histograms for the marginal distributions (a)--(c) $\phi_\cc$, (d)--(f) $\phi_\ww$, and (g)--(i) $\phi_{cw}$, for uniform disks ($\kappa=\frac{1}{2}$). The left [(a), (d), (g)], middle [(b), (e), (h)], and right [(c), (f), (i)] panels correspond to $\eet=0.5$, $0$, and $-0.5$, respectively. In each panel, three values of $\een$ are considered: $0.9$ (DSMC: $\circ$; EDMD:  $\times$), $0.7$ (DSMC: $\square$; EDMD:  $\star$), and $0.2$ (DSMC: $\triangle$; EDMD:  $+$). The solid lines represent the marginal distributions in the Maxwellian approximation [see Eqs.~\eqref{eq:marginal_MA}]. Note that a log-linear scale is used in (a)--(c) and a log-log scale in (d)--(i).}
	\label{fig:histo_sim}
\end{figure*}

\begin{figure*}[ht]
    \centering
    \includegraphics[width=0.3\textwidth]{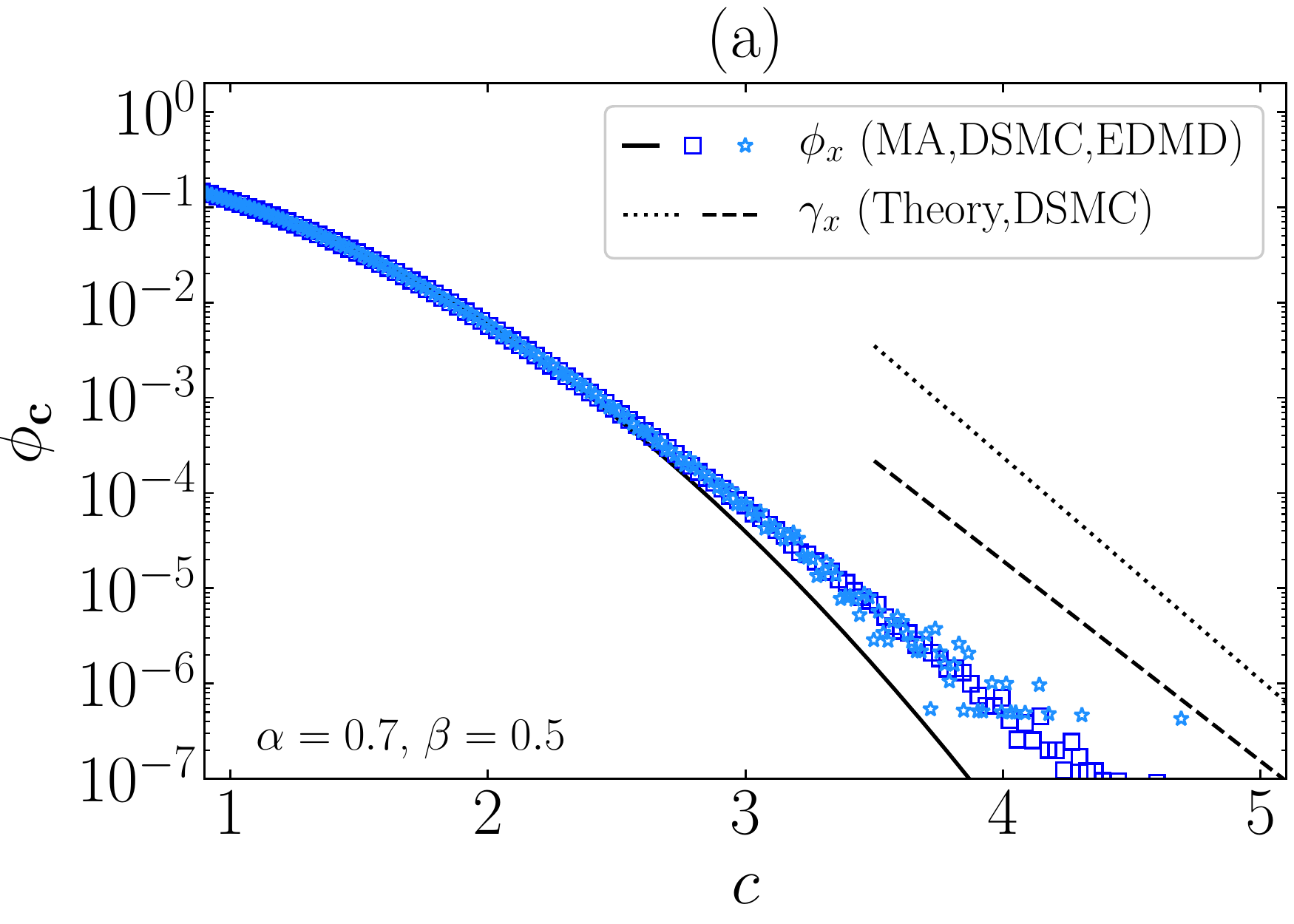}
    \includegraphics[width=0.3\textwidth]{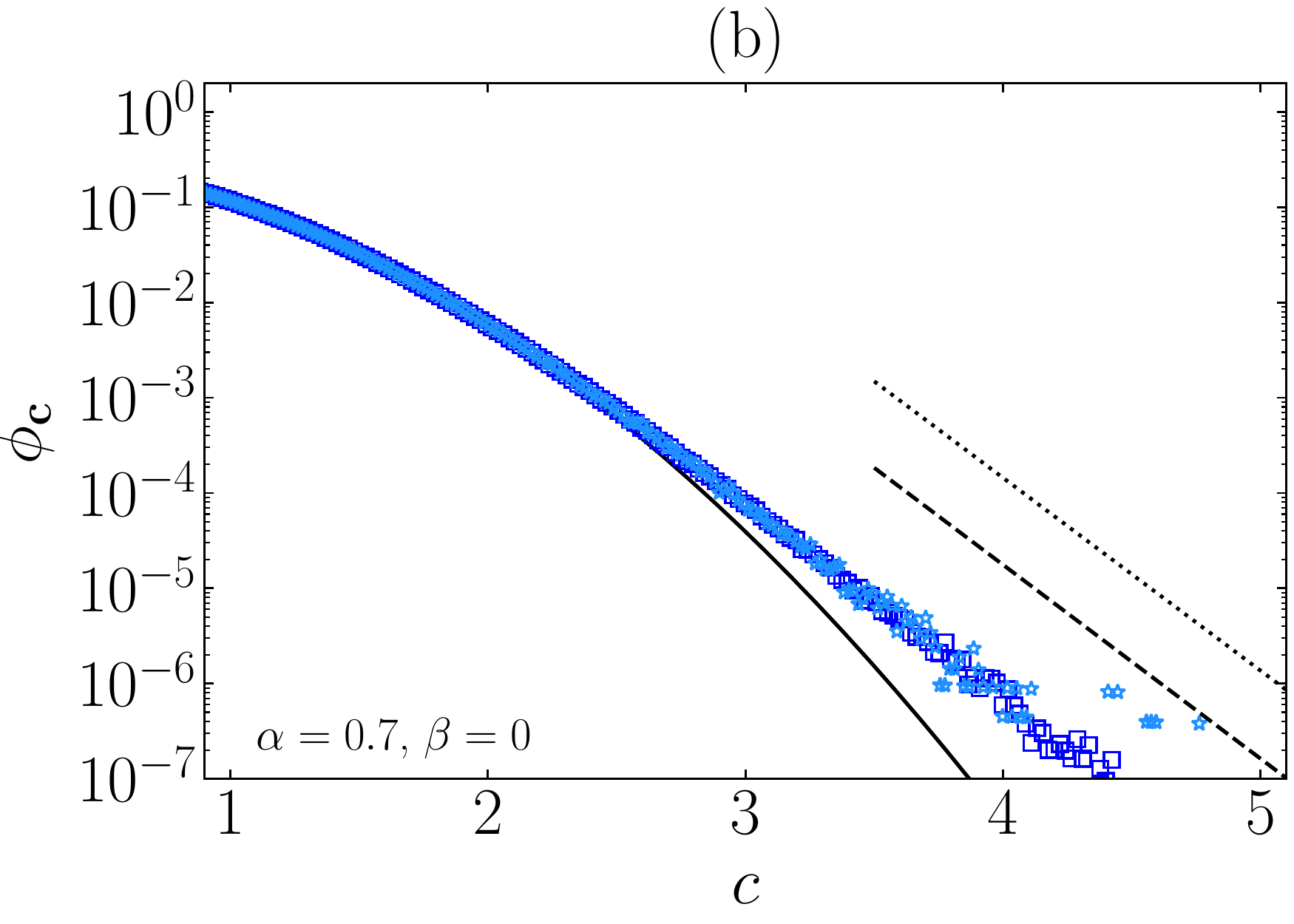}
    \includegraphics[width=0.3\textwidth]{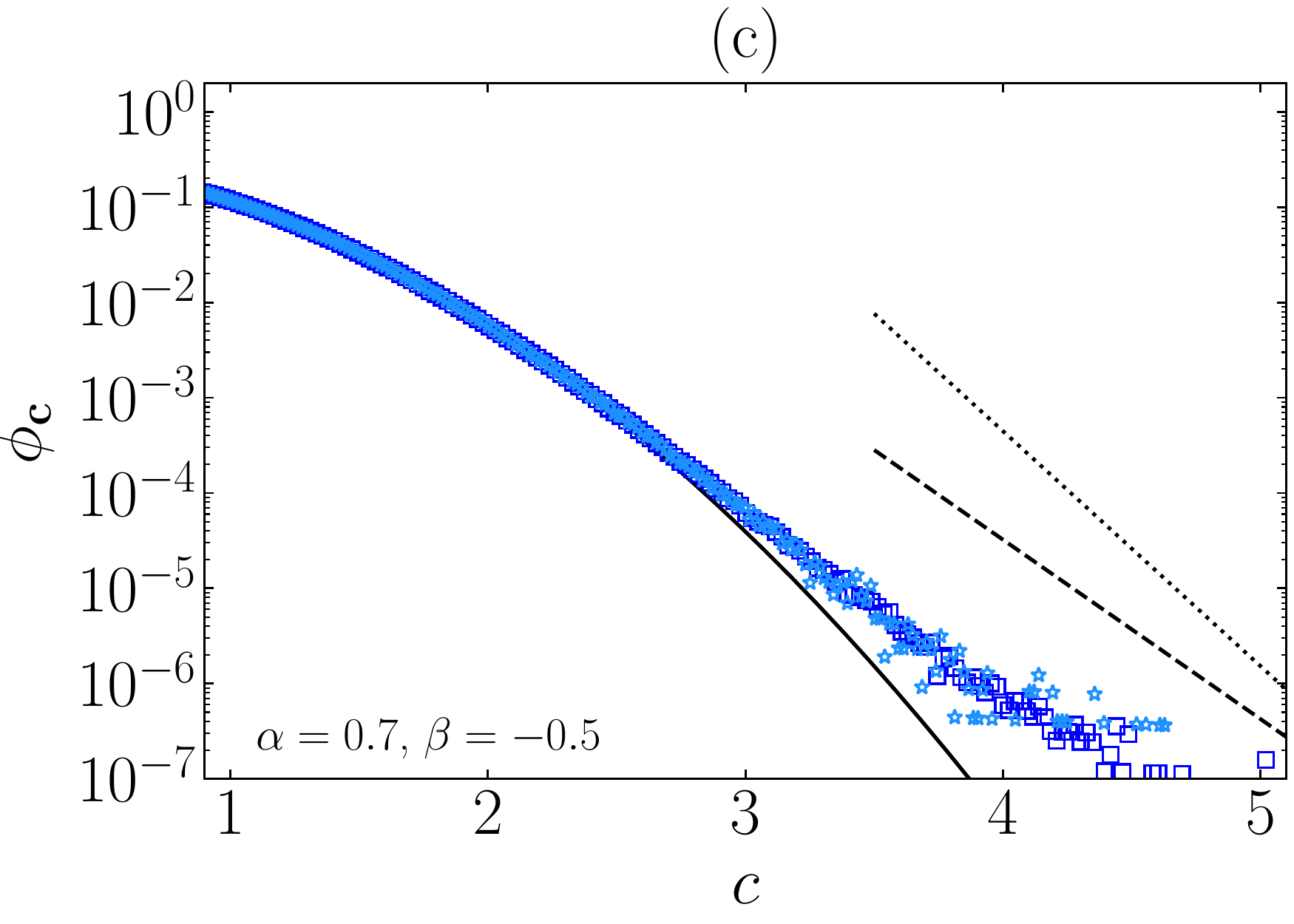}\\
    \includegraphics[width=0.3\textwidth]{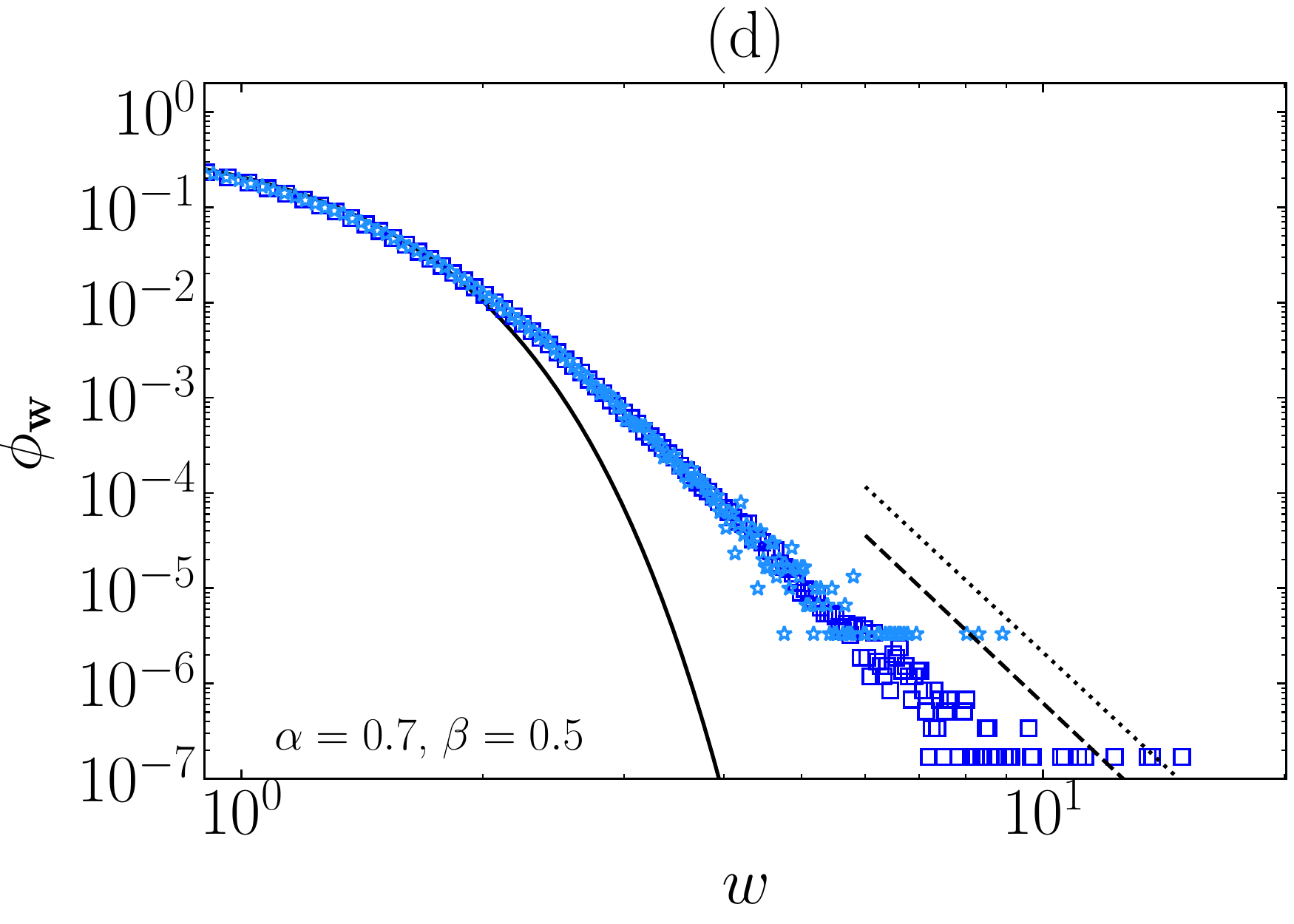}
    \includegraphics[width=0.3\textwidth]{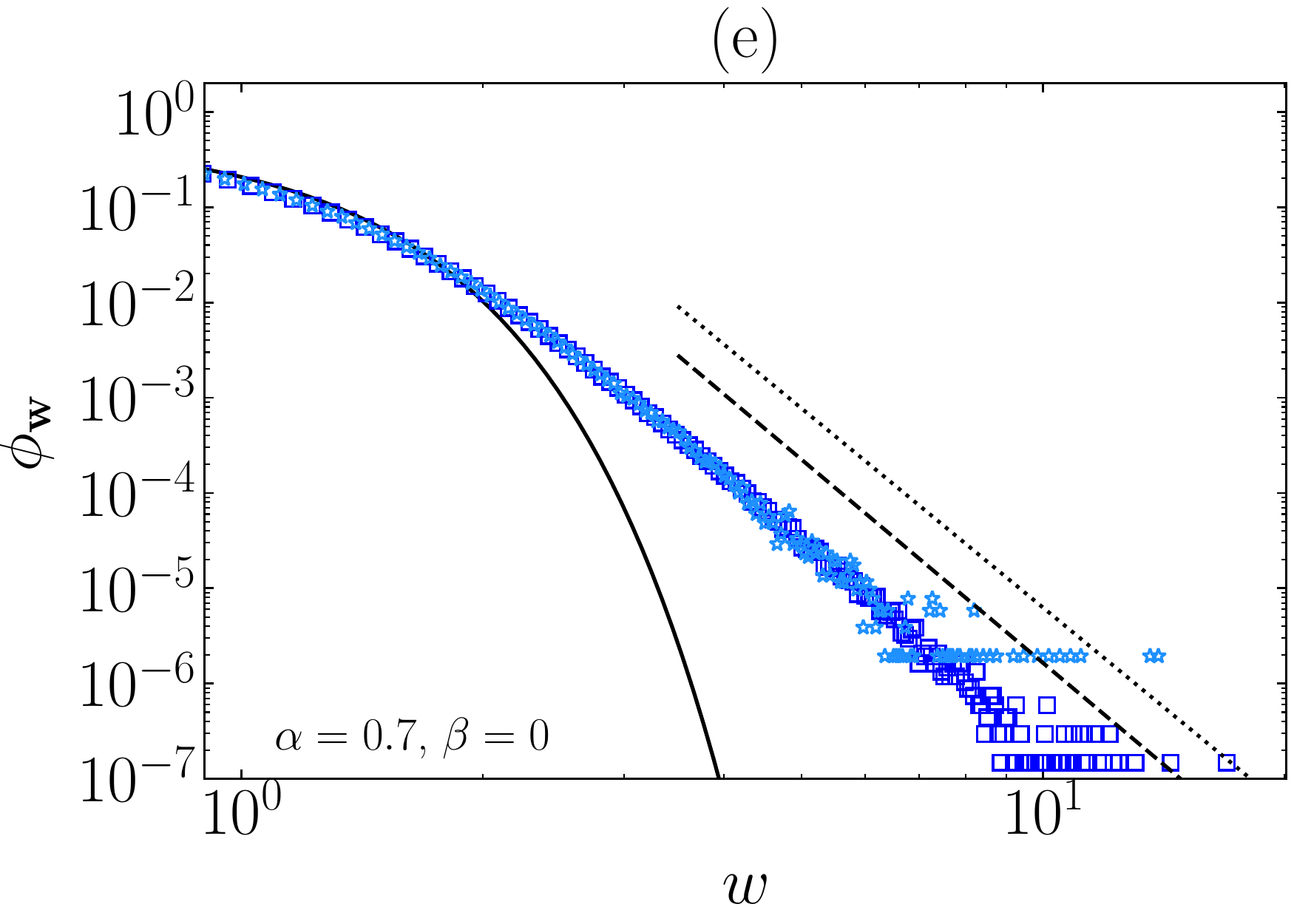}
    \includegraphics[width=0.3\textwidth]{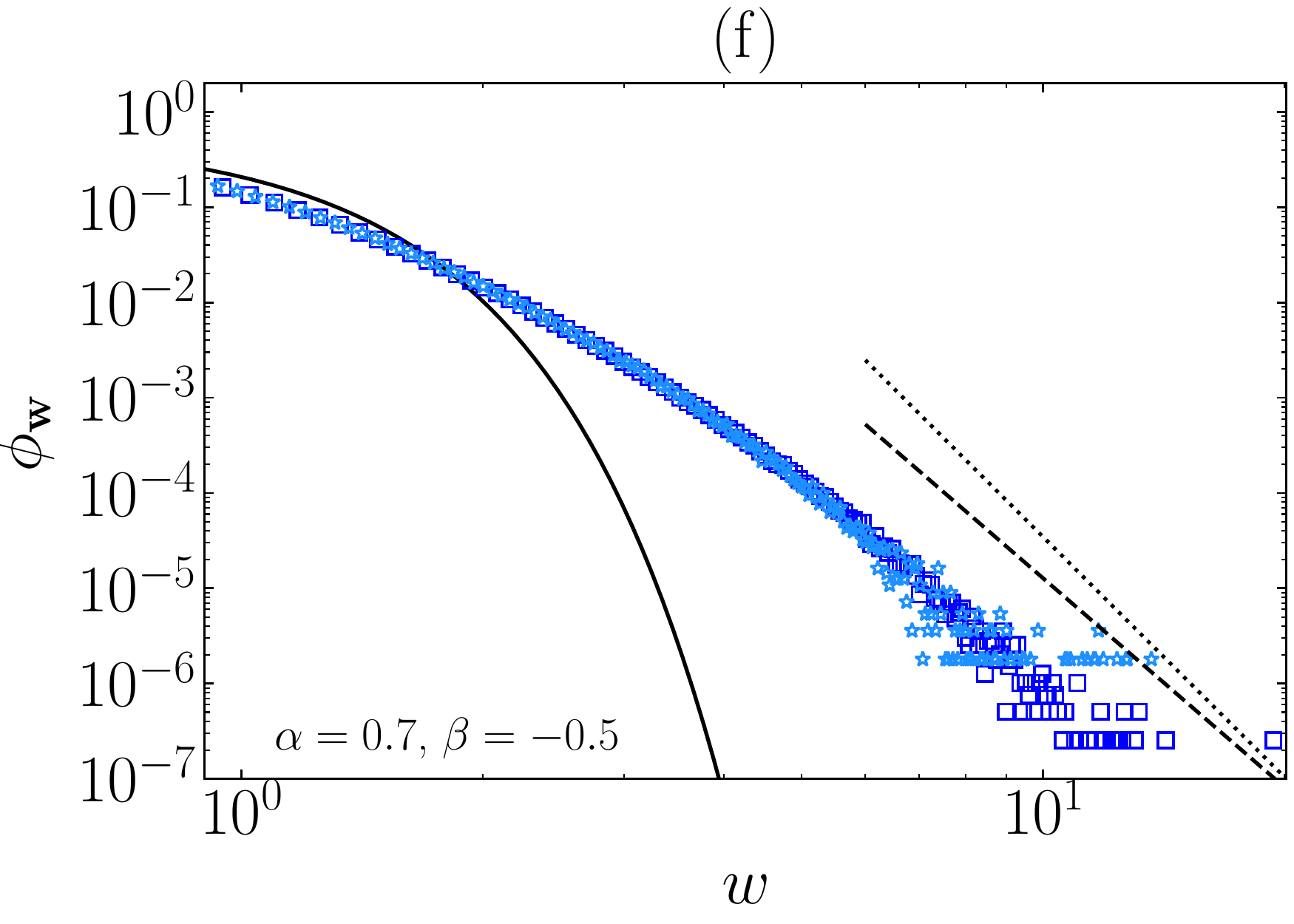}\\
    \includegraphics[width=0.3\textwidth]{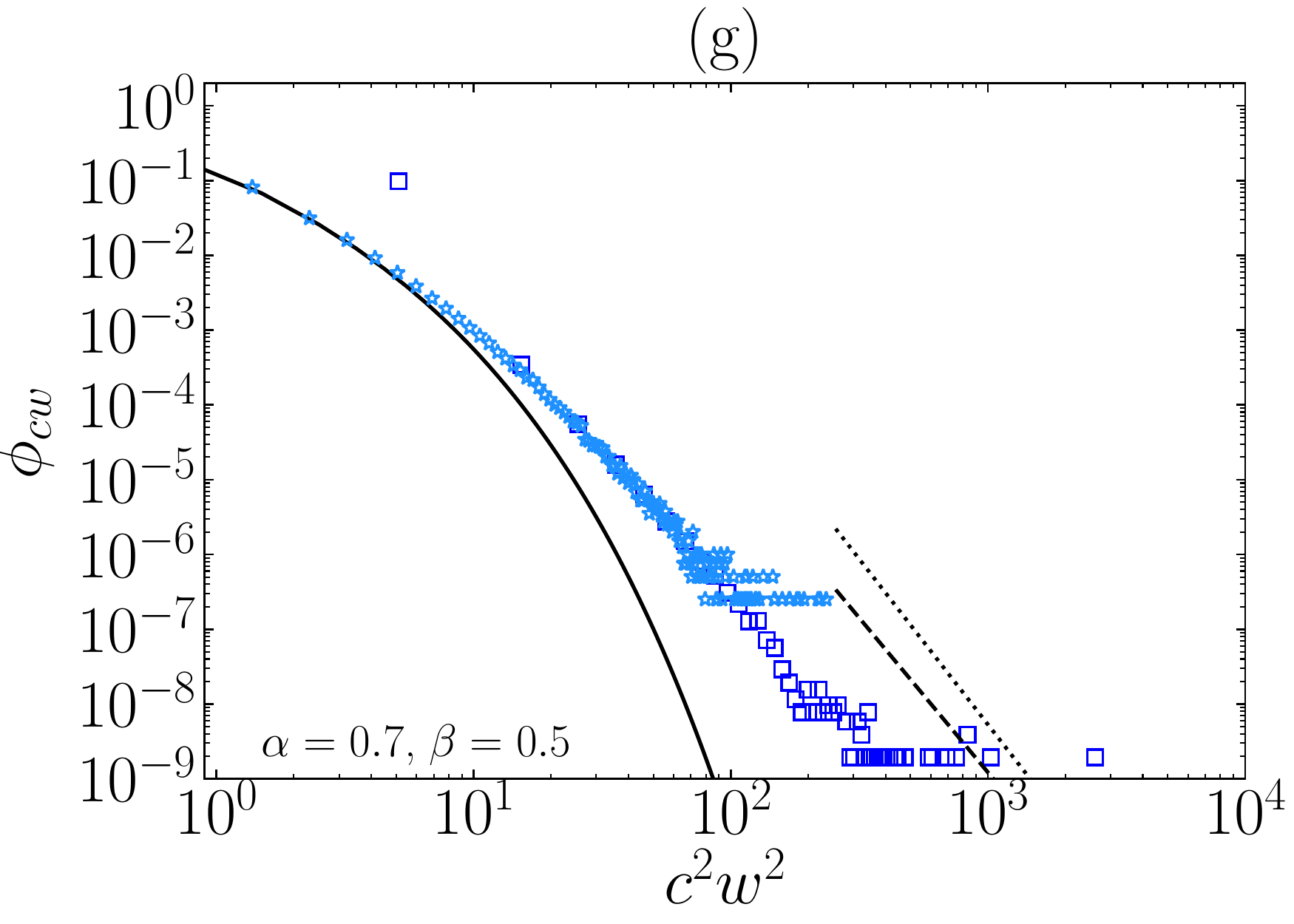}
    \includegraphics[width=0.3\textwidth]{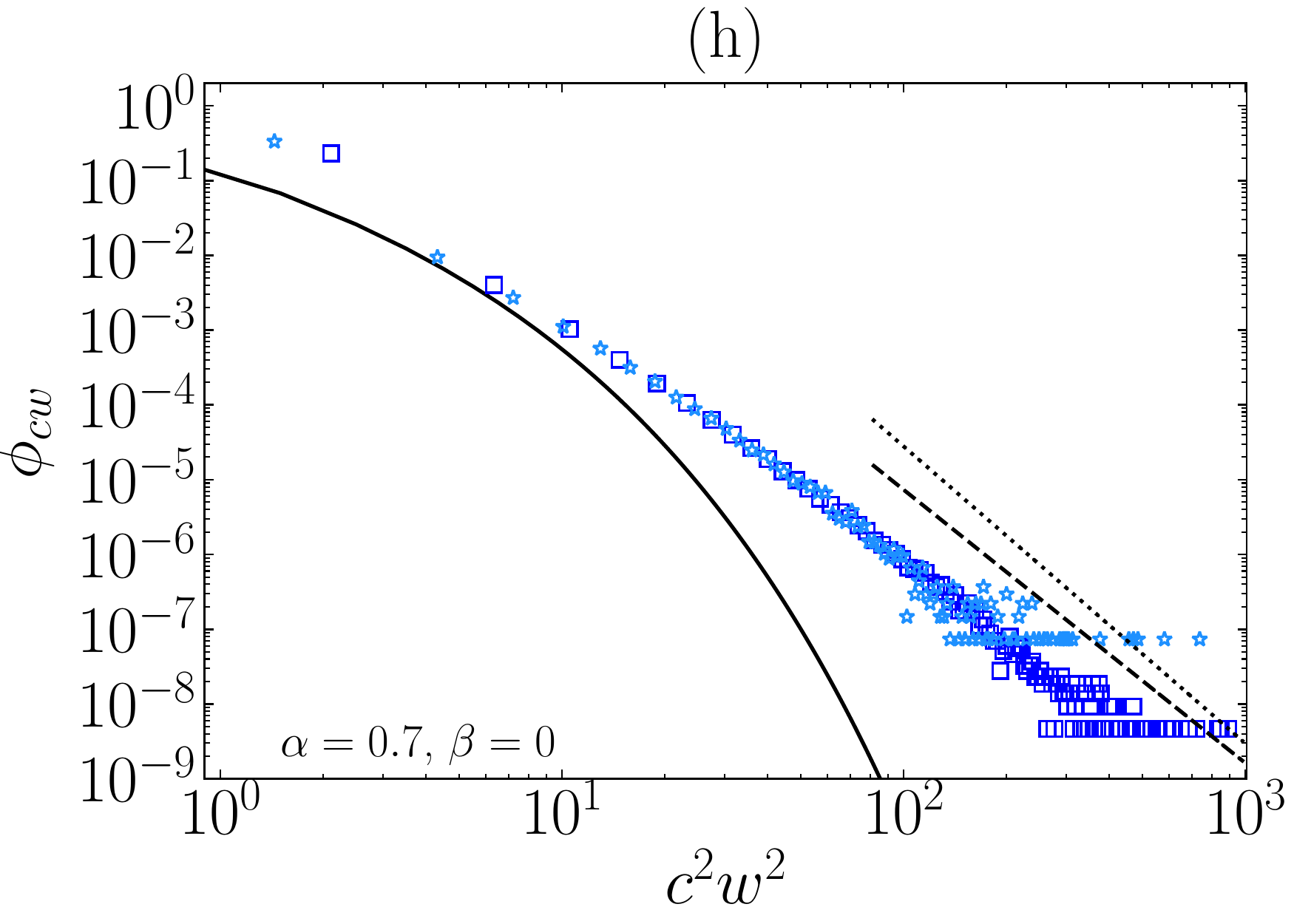}
    \includegraphics[width=0.3\textwidth]{{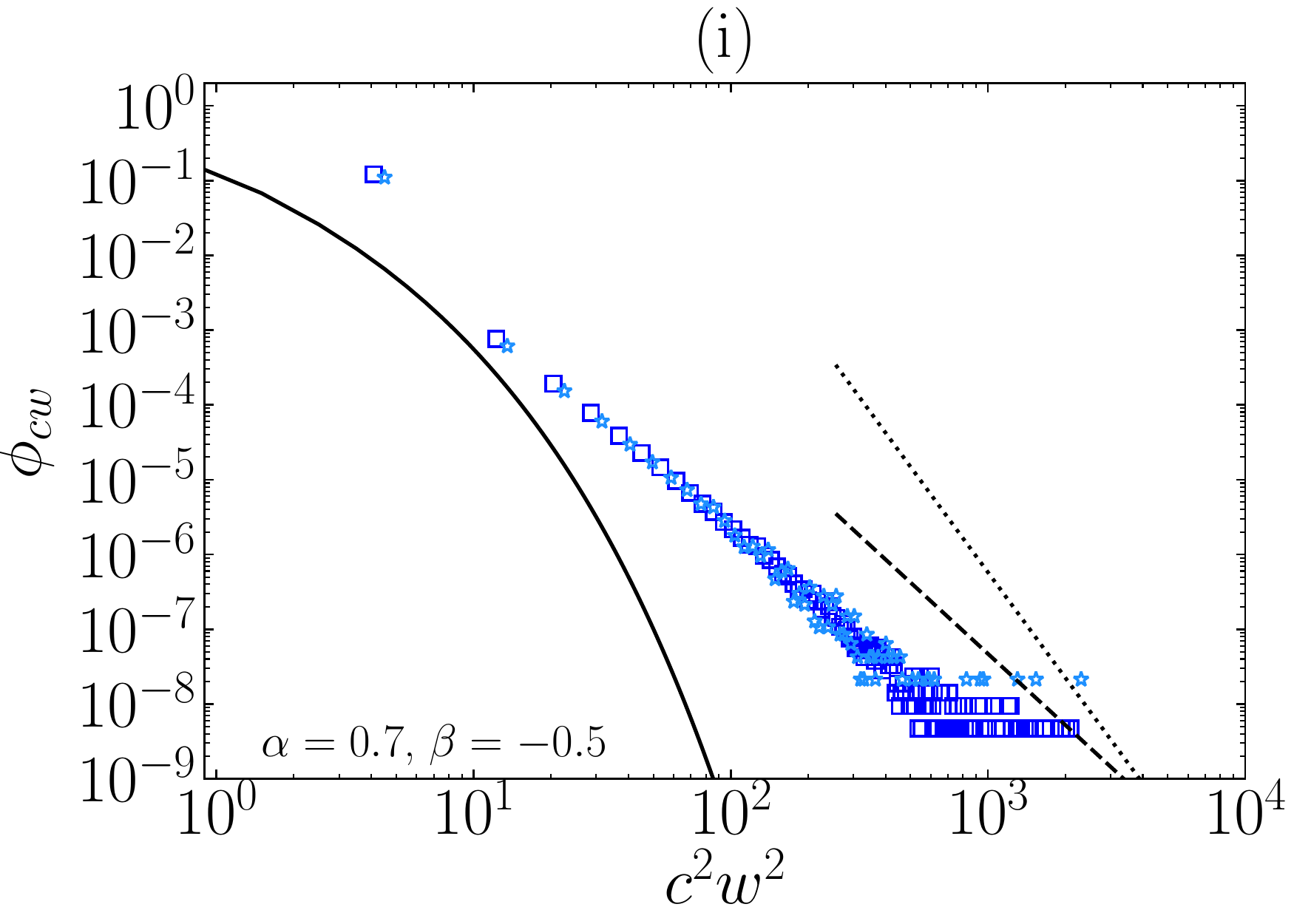}}
    \caption{Same as in Fig.~\ref{fig:histo_sim}, except that only the cases with $\een=0.7$ are shown. The dashed lines represent the exponents $\gamma_c$, $\gamma_w$, and $\gamma_{cw}$ obtained by a fit of the DSMC data. The dotted lines represent the theoretical exponents, as given by Eqs.~\eqref{eq:HVT_c}, \eqref{eq:phi_w(w)}, and \eqref{eq:sol_cw_1}, with the approximations $\mu_{20}^\HCS\approx\mu^{\HCS}_{20,\text{M}}$ [see Eq.~\eqref{eq:mu20M}] and $\llangle c_{12}\rrangle^\HCS\approx\llangle c_{12}\rrangle_{\text{M}}=\sqrt{\pi/2}$. }
    \label{fig:tails_sim}
\end{figure*}

\begin{figure}[ht]
    \centering
    \includegraphics[width=0.23\textwidth]{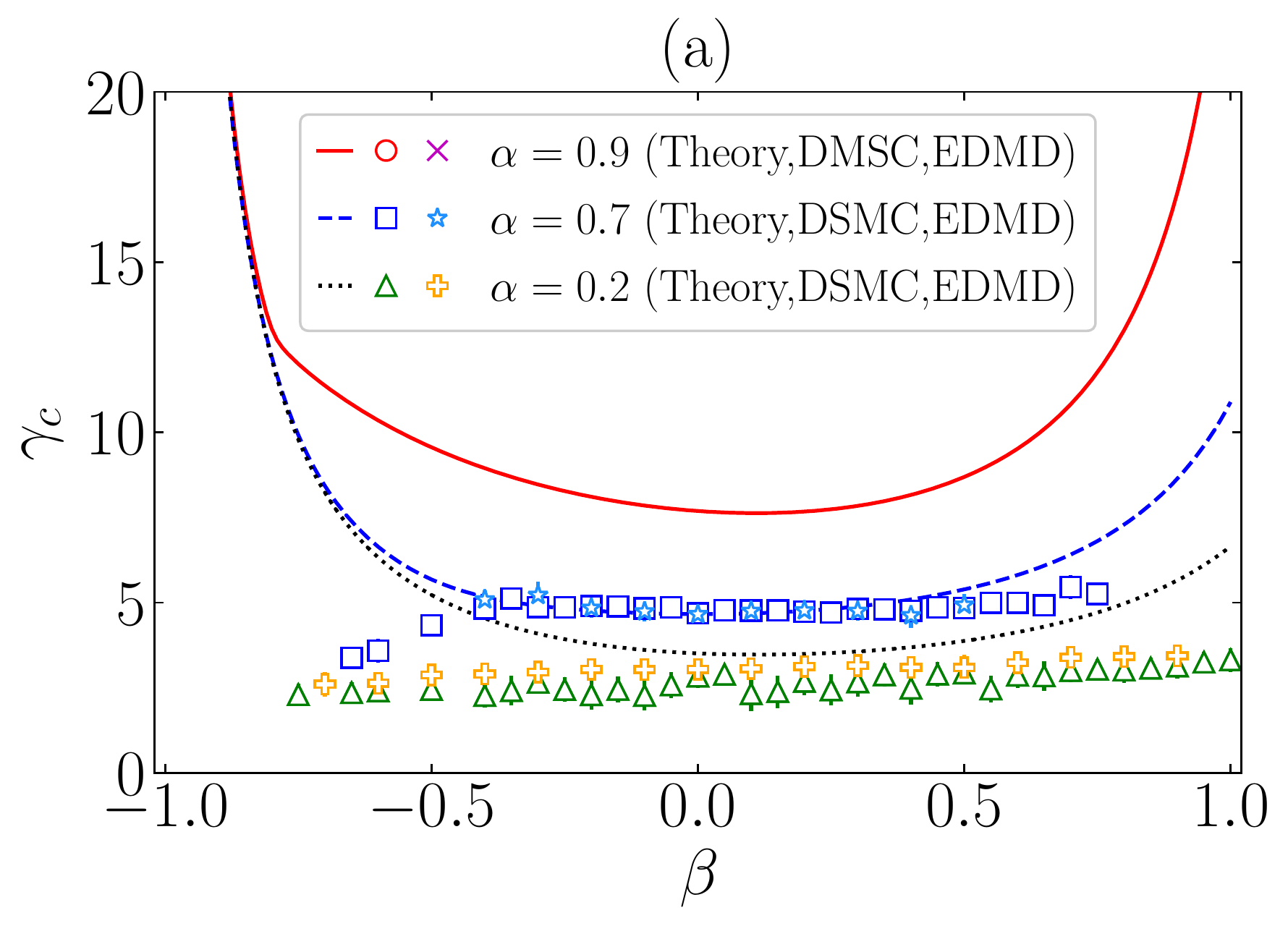}
    \includegraphics[width=0.23\textwidth]{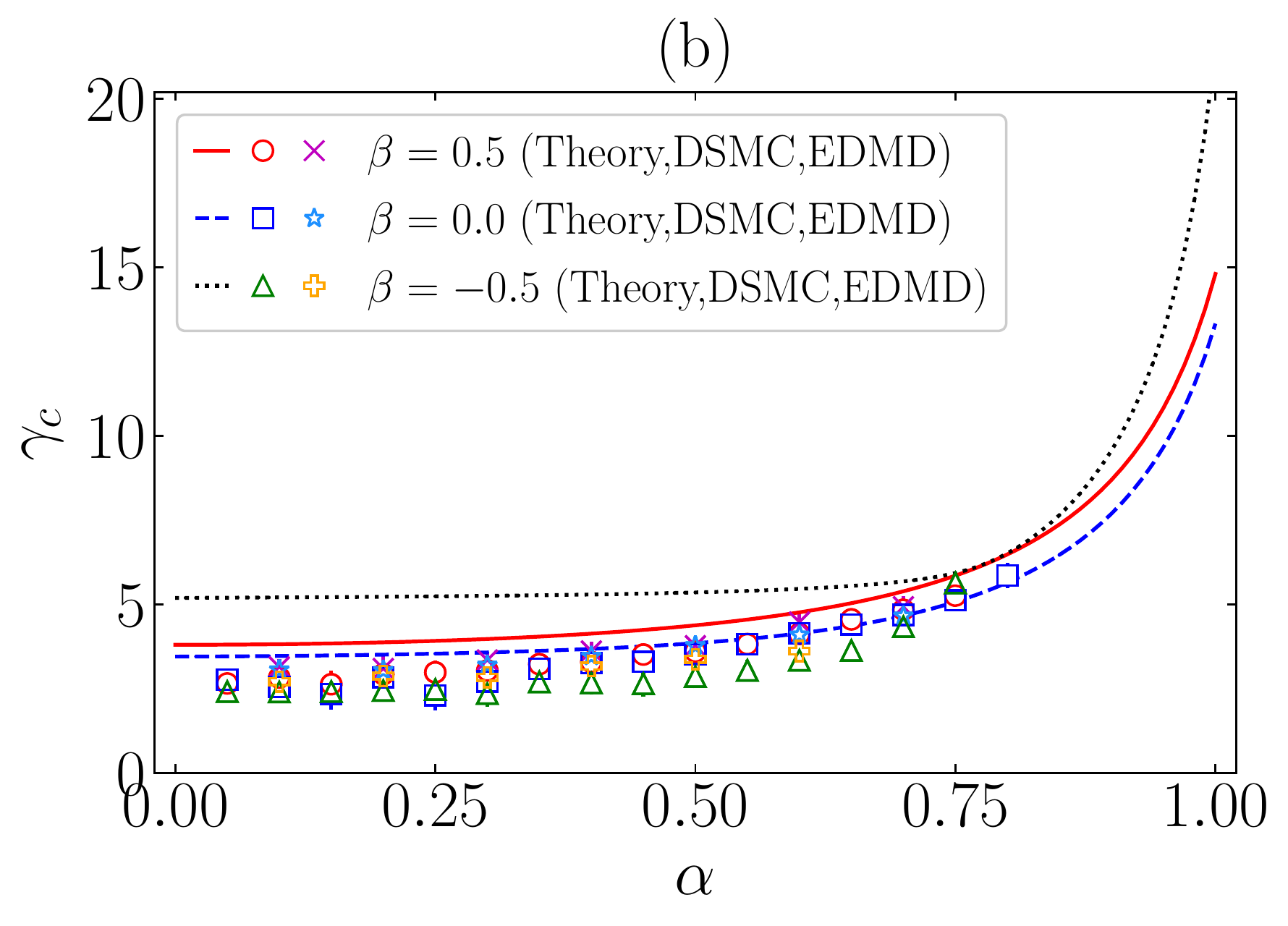}\\
    \includegraphics[width=0.23\textwidth]{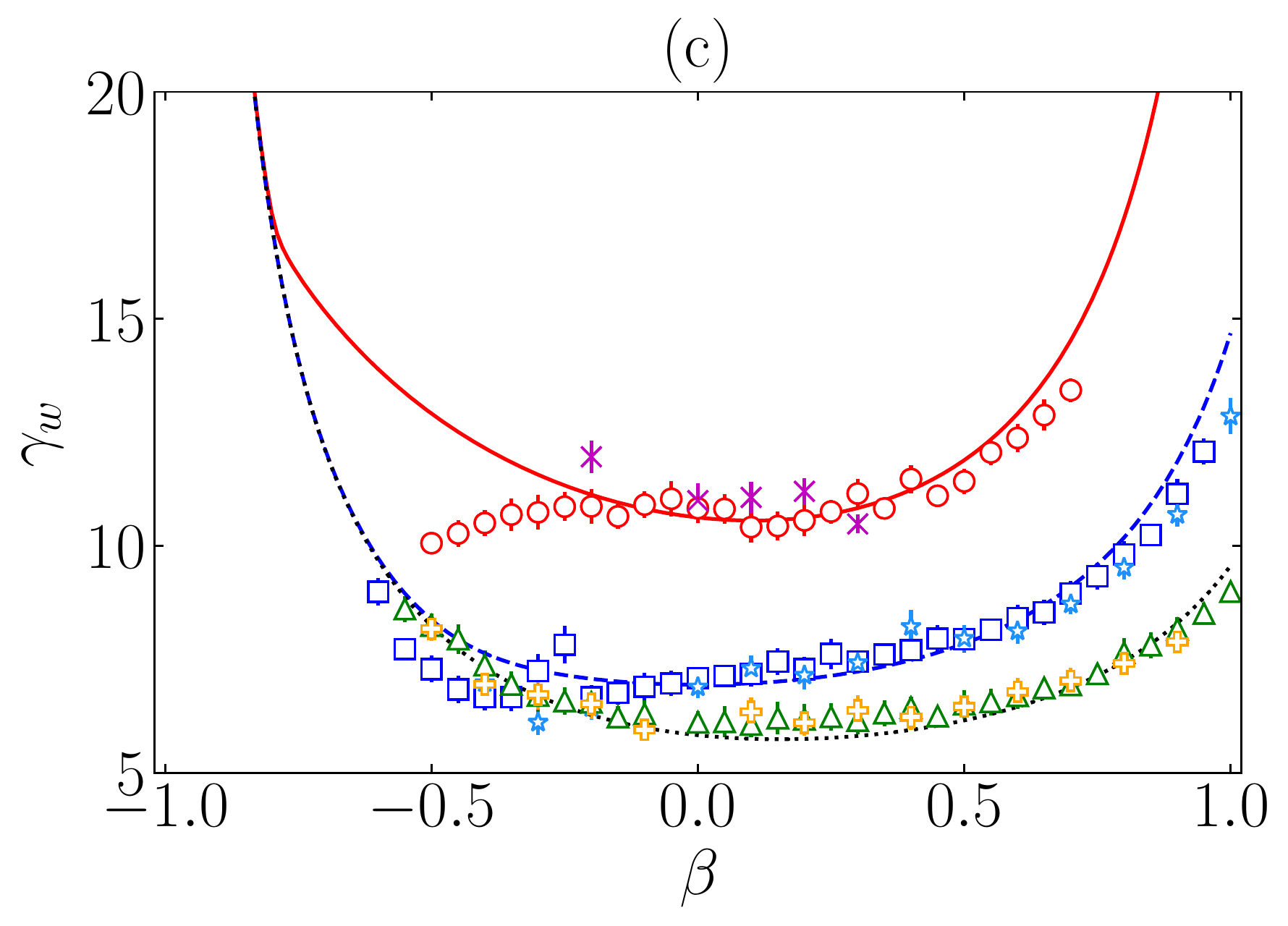}
    \includegraphics[width=0.23\textwidth]{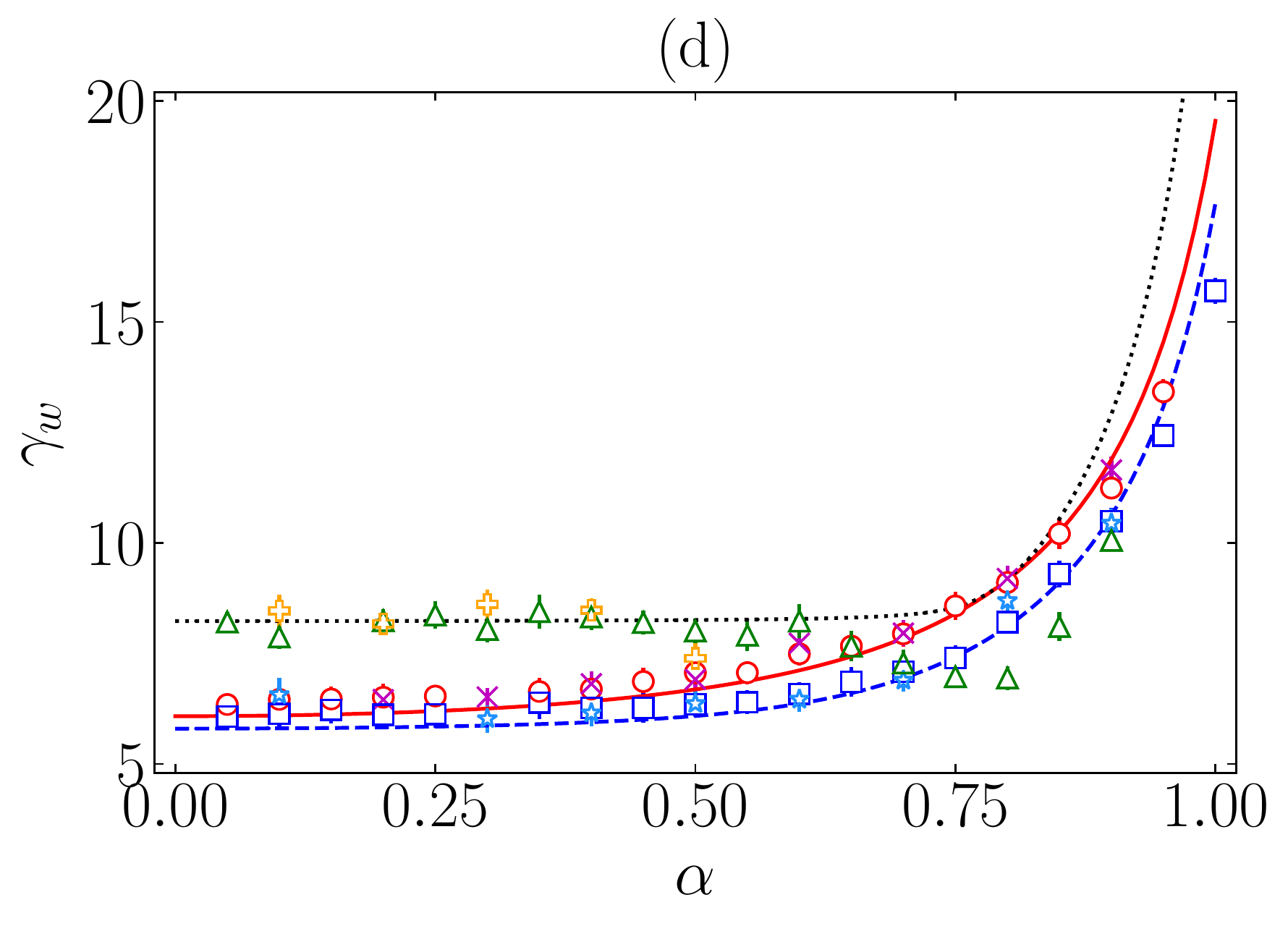}\\
    \includegraphics[width=0.23\textwidth]{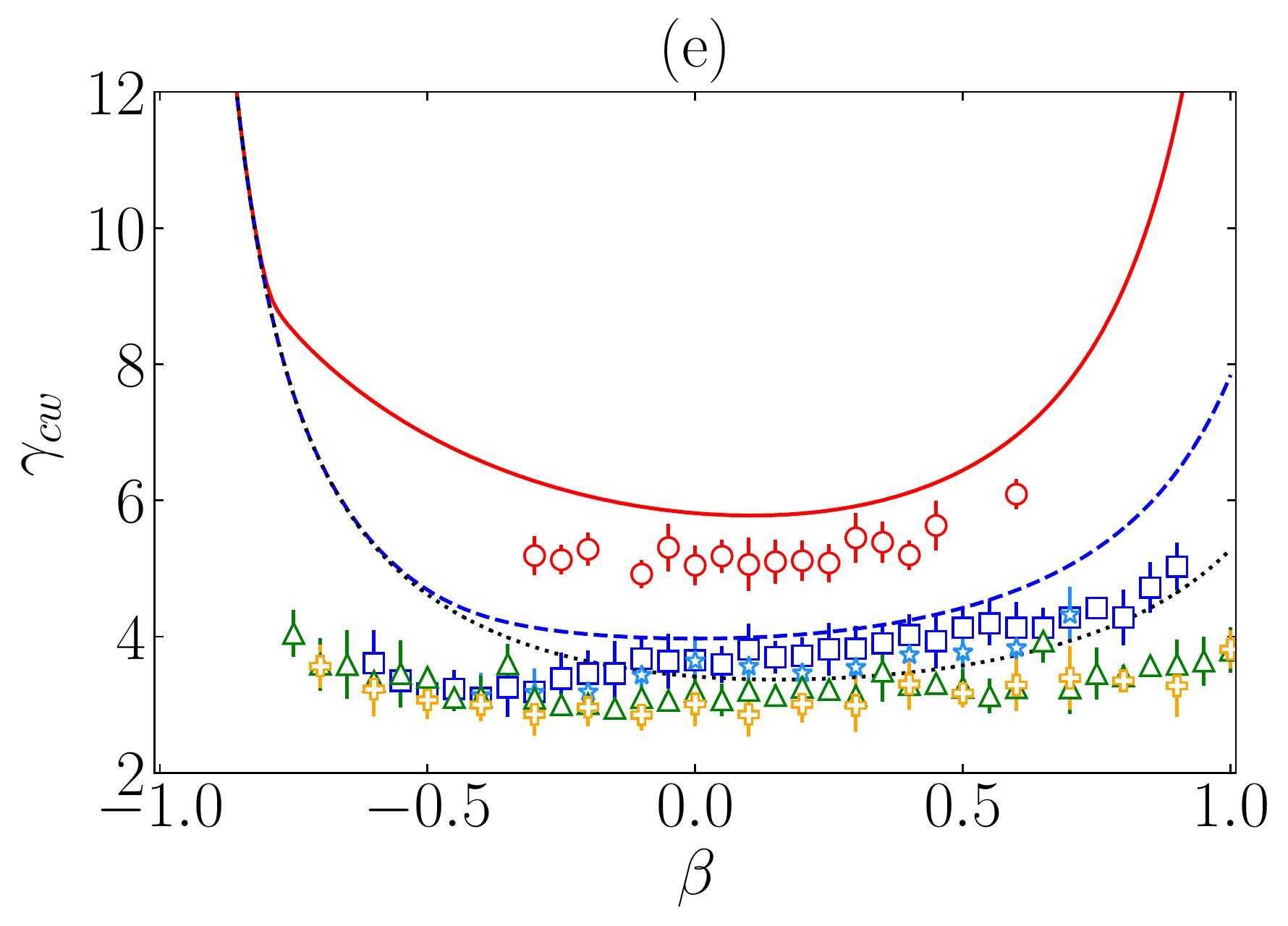}
    \includegraphics[width=0.23\textwidth]{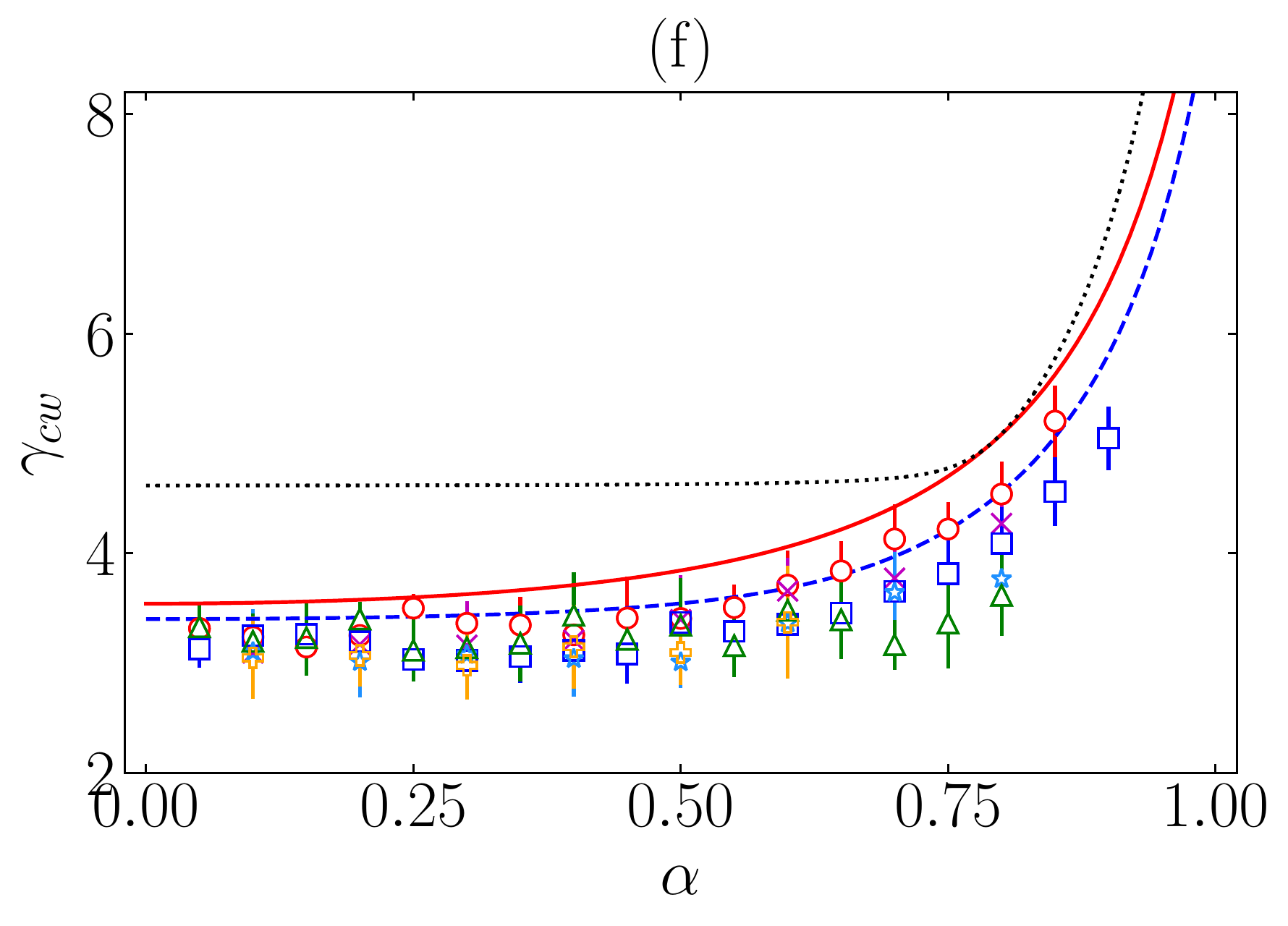}
    \caption{
    Plots of (a), (b)  $\gamma_c$; (c), (d)  $\gamma_w$;  and (e), (f)  $\gamma_{cw}$, for uniform disks ($\kappa=\frac{1}{2}$). The exponents are plotted versus $\eet$ in the left [(a), (c), (e)] panels and versus $\een$ in the right [(b), (d), (f)] panels. Symbols correspond to DSMC ($\circ$, $\square$, $\triangle$) and EDMD ($\times$, $\star$, $+$) fitting values, while lines  represent the theoretical exponents, as given by Eqs.~\eqref{eq:HVT_c}, \eqref{eq:phi_w(w)}, and \eqref{eq:sol_cw_1}, with the approximations $\mu_{20}^\HCS\approx\mu_{20,\text{M}}^\HCS$ [see Eq.~\eqref{eq:mu20M}] and $\llangle c_{12}\rrangle^\HCS\approx\llangle c_{12}\rrangle_{\text{M}}=\sqrt{\pi/2}$.}
    \label{fig:Bx_sim}
\end{figure}

To test the theoretical results, we have run two types of computer simulation algorithms for a dilute and homogeneous granular gas of inelastic and rough hard disks ($\dt=2$, $\dr=1$) with different values of the coefficients of restitution $\alpha$ and $\beta$. In all cases, the disks are assumed to have a uniform mass distribution, so the reduced moment of inertia is $\kappa=\frac{1}{2}$.

First, we used DSMC, as proposed by Bird~\cite{B94,B13} and conveniently adapted to the  granular case~\cite{MS00,VSK14}, to simulate a homogenenous and dilute granular gas of inelastic and rough hard disks, using $N=10^4$ representative particles. Additionally, we carried out EDMD computer simulations with $N=1600$ disks in a square box of side length $L/\sigma = 565.7$, which correspond to a number density  $n\sigma^2=0.005$, thus avoiding spatial instabilities~\cite{MS21b}. Whereas the EDMD system has nonzero density, the solid fraction $\varphi=\frac{\pi}{4}n\sigma^2\simeq 3.9\times10^{-3}$ is small enough to expect good agreement with the diluteness assumption. We ran $100$ and $50$ replicas for DSMC and EDMD, respectively, for each pair  $(\alpha,\beta)$,  not observing instabilities in the EDMD simulations. In addition to averaging over replicas, the stationary HCS values were measured by averaging over instantaneous values at $s=s_{\text{ini}}, s_{\text{ini}}+\delta s,s_{\text{ini}}+2\delta s,\ldots, s_{\text{fin}}$ with $(s_{\text{ini}},s_{\text{fin}},\delta s)=(500,1500,5)$ and $(150,200,1)$ for DSMC and EDMD, except in the case $\alpha=0.9$, $\beta =-0.8$, where we took $(450,500,1)$ in the EDMD simulations.
In the construction of histograms for the marginal distributions, we considered $2^8$ bins in the associated velocity variable.

In the Supplemental Material~\cite{note_23_01}, we present a comparison between the Sonine-approximation results [see Eqs.~\eqref{eq:ev_theta} and \eqref{eq:ev_eqs}] and simulation data for the temporal evolution toward the HCS of the temperature ratio and the cumulants, starting from an equipartioned Maxwellian state.  A generally good agreement is observed, except for  $a_{02}$ near the HCS if $a_{02}^\HCS$ reaches relatively high values.
Now we present results for the relevant quantities in the HCS.

\subsection{Temperature ratio and cumulants}

Figure~\ref{fig:sim_res_1} shows the HCS values of $\theta^\HCS$, $a_{20}^\HCS$, $a_{02}^\HCS$, and $a_{11}^\HCS$ versus $\eet$ for some representative values of $\een$. Figure~\ref{fig:sim_res_2} presents the same quantities versus $\een$ for some illustrative values of $\eet$.
We observe that the Maxwellian approximation provides a good description of $\theta^\HCS$, although it tends to overestimate it if $\een\lesssim 0.7$ [see Figs.~\ref{fig:sim_res_2}(a)--\ref{fig:sim_res_2}(c)]. Those deviations are satisfactorily corrected by the Sonine approximation.

In the case of the cumulants,  their qualitative shape as functions of both $\een$ and $\eet$ are well accounted for by the Sonine approximation. The quantitative agreement is good as long as the magnitude of the cumulants is small, thus validating the Sonine approximation in those cases. On the other hand, whenever the Sonine approximation predicts values $a_{ij}^\HCS=O(1)$, the approximation is itself signaling its breakdown. This situation, which is similar to that already reported in the case of HS~\cite{VSK14}. is especially noteworthy in the cases of $a_{02}^\HCS$ and, to a lesser extent, $a_{11}^\HCS$, and is clearly indicative of the high-velocity tails discussed in Sec.~\ref{subsec:HET} and confirmed below.

\subsection{High-velocity tails}

To further observe the non-Gaussianities of the HCS state, Fig.~\ref{fig:histo_sim} displays the histograms from simulation data  of $\phi^\HCS_\cc$, $\phi^\HCS_\ww$, and $\phi^\HCS_{cw}$, for nine combinations of coefficients of restitution ($\alpha=0.9$,  $0.7$, $0.2$, and $\beta=0.5$, $0$, $-0.5$). Except in the case of $\phi^\HCS_\cc$ for $\alpha=0.9$ (where $a_{20}^\HCS$ is small), the deviations  from the Maxwellian tail are quite apparent. In fact, the high-velocity tails observed in Fig.~\ref{fig:histo_sim} are consistent with an exponential tail for  $\phi^\HCS_\cc$ and power-law tails for $\phi^\HCS_\ww$ and $\phi^\HCS_{cw}$, in agreement with the analysis in Sec.~\ref{subsec:HET}.

A more quantitative test is presented in Fig.~\ref{fig:tails_sim}, where the three cases with $\alpha=0.7$ have been selected and the straight lines representing the asymptotic tails are included. The theoretical predictions for the exponents derived in Sec.~\ref{subsec:HET} (with additional Maxwellian estimates for $\mu_{20}^\HCS$ and  $\llangle c_{12}\rrangle^\HCS$) agree reasonably well with the fitted values, except for $\eet=-0.5$, in which case the actual decays are slower than predicted.

Figure~\ref{fig:Bx_sim} shows the exponents $\gamma_c$, $\gamma_w$, and $\gamma_{cw}$ as functions of $\eet$ (for $\een=0.9$, $0.7$, $0.2$) and $\een$ (for $\eet=0.5$, $0$, $-0.5$). There exists very good agreement between the theoretical estimates and the fitting simulation values in the case of the exponent  $\gamma_w$, which seems to worsen as $\beta$ decreases. However, in the case of $\gamma_c$ and $\gamma_{cw}$, the agreement is mainly qualitative. This might be due to the fact that the tails of $\phi_\cc^\HCS$ and $\phi_{cw}^\HCS$ are much less populated than that of  $\phi_\ww^\HCS$ (see Figs.~\ref{fig:histo_sim} and \ref{fig:tails_sim}) and, therefore, it is much more difficult to reach values of $c$ and $c^2w^2$  high enough to accurately measure the exponents $\gamma_c$ and $\gamma_{cw}$ in the simulations. If that were the case, then the values of $\gamma_c$ and $\gamma_{cw}$ empirically determined would characterize an intermediate velocity regime previous to the true asymptotic behavior. Of course, one cannot discard that our analysis becomes more limited as $\eet$ decreases.

Before closing this section, it is worth remarking the excellent mutual agreement between DSMC and EDMD results. There are, however, some small deviations for low values of $\een$, which  might be a consequence of the smaller number of disks in the EDMD simulations and also a reflection of possible violations of the molecular chaos ansatz in those highly dissipative systems~\cite{SM01}.

\section{\label{sec:conclusions} Conclusions}

In this paper, we have studied low-density, monodisperse, and homogeneous granular gases of hard disks and hard spheres from a kinetic-theory point of view, using a general framework to express the results in terms of the number of translational and rotational degrees of freedom, $\dt$ and $\dr$, respectively. Special attention has been paid to the non-Gaussian features of the HCS, as measured by the fourth-order cumulants and the high-velocity tails of the marginal distributions. The theory has been complemented  by DSMC and EDMD computer simulations.

The theoretical approach is based on the Boltzmann equation. First, we have expressed the collisional moments as formally exact functions of the parameters of the system ($\een$, $\eet$, and $\kappa$) and two-body averages. Next, we have employed a Grad--Sonine expansion of the complete one-body VDF,  Eq.~\eqref{eq:phi_Sonine_expansion}. Then, in analogy to Ref.~\cite{VSK14}, we have defined the Sonine approximation from the truncation of the Sonine expansion beyond the first nontrivial cumulants defined in Eq.~\eqref{eq:cum_dim}. This contrasts with the Maxwellian approximation, which is based on approximating the VDF by a two-temperature Maxwellian distribution, i.e., $\phi\approx \phi_{\mathrm{M}}$.

Within the Sonine approximation, and neglecting quadratic terms, the relevant collisional moments have been evaluated, thus recovering previous results for hard spheres ($\dt=\dt=3$)~\cite{VSK14}, and obtaining results for hard disks ($\dt=2$, $\dr=1$), as presented in Table~\ref{table:2}. Cumulant-linearization in Eqs.~\eqref{eq:ev_theta} and \eqref{eq:ev_eqs} allows us to deal with a closed set of differential equation for the evolution of the rotational-to-translational temperature ratio ($\theta$) and the fourth-order cumulants. Analogously, the stationary HCS values in the Sonine approximation have been obtained by linearization  in Eqs.~\eqref{eq:HCS}. As a consistency test, we have checked in Appendix \ref{sec:LSA} that the HCS is linearly stable with respect to homogeneous and isotropic perturbations.
The HCS quantities have been shown in Figs.~\ref{fig:HCS_HD} and~\ref{fig:HCS_HS} for uniform disks ($\kappa=\frac{1}{2}$) and uniform spheres ($\kappa=\frac{2}{5}$), respectively.
At a qualitative level, their dependence on $\alpha$ and $\beta$  is very similar for disks and spheres, but the values are generally more extreme in the former system than in the latter. In both cases, the kurtosis for the angular velocity, $a_{02}^\HCS$, reaches values of $O(1)$ in a lobular region of the parameter space with a vertex at $(\een,\eet)=(1,-1)$, thus announcing a breakdown of the Sonine approximation in that region.

Moreover, the non-Gaussianities of the HCS have been studied not only in the context of the first nontrivial cumulants, but also analyzing the tails of the marginal VDF $\phi_\cc^\HCS(\cc)$, $\phi_\ww^\HCS(\ww)$, and $\phi_{cw}^\HCS(c^2w^2)$ defined by Eqs.~\eqref{eq:marginal_defs}. Using previous methods developed for the smooth case~\cite{EP97,vNE98,BRC99}, which are based on the prevalence of the collisional loss term  with respect to the gain term, we have obtained  the expected exponential tail $\phi_\cc^\HCS(\cc)\sim e^{-\gamma_c c}$, with formally the same expression  for the exponent coefficient $\gamma_c$ as in the smooth case [see Eq.~\eqref{eq:HVT_c}].
On the other hand, we have found  much slower scale-free decays $\phi_\ww^\HCS(\ww)\sim  w^{-\gamma_w}$ and $\phi_{cw}^\HCS(c^2w^2)\sim  (c^2w^2)^{-\gamma_{cw}}$, with exponents given by Eqs.~\eqref{eq:phi_w(w)} and \eqref{eq:sol_cw_1}, respectively. These algebraic tails of  $\phi_\ww^\HCS(\ww)$ and $\phi_{cw}^\HCS(c^2w^2)$ explain the relatively large values attained by the cumulants $a_{02}^\HCS$ and $a_{11}^\HCS$, especially in the hard-disk case, and predict divergences in higher-order cumulants which recall the ones already observed in the case of the three-dimensional inelastic and rough Maxwell model~\cite{KS22}.

To test the theoretical predictions, we have run DSMC and EDMD computer simulations for hard disks (with $\kappa=\frac{1}{2}$), as described in Sec.~\ref{sec:Sim_Res}. First, the quantities $\theta^\HCS$, $a_{20}^\HCS$, $a_{02}^\HCS$, and $a_{11}^\HCS$ have been studied for different values of $\een$ and $\eet$, as depicted in Figs.~\ref{fig:sim_res_1} and~\ref{fig:sim_res_2}. The agreement between the Sonine approximation and simulation is rather good, except when the values of the cumulants are not small. Even in those cases, it is remarkable that the Sonine approximation reproduces qualitatively well the shape of the curves. Second, we have extended the comparison to the three marginal distributions in Figs.~\ref{fig:histo_sim} and \ref{fig:tails_sim}, finding that  the predicted exponential tail of $\phi_\cc^\HCS(\cc)$ and algebraic tails of $\phi_\ww^\HCS(\ww)$ and $\phi_{cw}^\HCS(c^2w^2)$ are supported by simulation data. The theoretical and fitting exponents have been compared in Fig.~\ref{fig:Bx_sim}, where a good agreement for $\gamma_{w}$ has been observed, while the agreement is more qualitative for $\gamma_{c}$ and $\gamma_{cw}$. This might be due to a lack of statistically reliable simulation data in the high-velocity regime.

To sum up, the HCS VDF of a monodisperse granular gas of inelastic and rough hard particles is, in general, strongly non-Maxwellian. Moreover, the non-Gaussianities exposed in this paper might solve some inconsistencies reported in the stability analysis of Navier--Stokes hydrodynamics from a Maxwellian approximation in hard-disk systems~\cite{MS21b} and improve the predictions of the inelastic hard-sphere model for real experimental systems, such as the one of Ref.~\cite{YSS20}. As a follow-up of the study presented in this paper, we plan to extend it to driven hard-disk systems (in analogy to a previous study for hard spheres~\cite{VS15}), whose dynamics has very interesting implications~\cite{TLLVPS19,MS22b}. Finally, we hope this paper could stimulate further research in all these issues, not only from theoretical and simulation points of view, but also from experimental setups.

The data that support the findings of this study are openly available in Ref.~\cite{M23a}.

\acknowledgments

The authors acknowledge financial support from Grant No.~PID2020-112936GB-I00 funded by MCIN/AEI/10.13039/501100011033, and from Grant No.~IB20079  funded by Junta de Extremadura (Spain) and by ERDF, ``A way of making Europe.'' A.M. is grateful to the Spanish Ministerio de Ciencia, Innovaci\'on y Universidades for a predoctoral fellowship No.~FPU2018-3503. The authors are grateful to the computing facilities of the Instituto de Computaci\'on Cient\'ifica Avanzada of the University of Extremadura (ICCAEx), where the simulations were run.


\appendix



\section{\label{sec:LSA} Linear stability analysis of the homogeneous cooling state}
In this Appendix we show that, within the Sonine approximation, the HCS for hard disks is linearly stable under uniform and isotropic perturbations.
Let us define the time-dependent set of perturbed quantities:
\begin{equation}
  \delta\mathbf{Y}(s)=\begin{pmatrix}
    \theta(s)-\theta^\HCS\\
    a_{20}(s)-a_{20}^\HCS\\
    a_{02}(s)-a_{02}^\HCS\\
    a_{11}(s)-a_{11}^\HCS
  \end{pmatrix}.
\end{equation}
Insertion into the Sonine approximation versions of Eqs.~\eqref{eq:ev_theta} and \eqref{eq:ev_cum_theta2}--\eqref{eq:ev_cum_theta4}, and linearization around the HCS values, yield
\begin{equation}\label{eq:LSS}
    \partial_s \delta \mathbf{Y}(s) = -\mathsf{L}\cdot \delta \mathbf{Y}(s),
\end{equation}
where $\mathsf{L}$ is a constant matrix, its four eigenvalues, $\{\ell_i;i=1,2,3,4\}$, determining the evolution of $\delta\mathbf{Y}(s)$ from an arbitrary initial perturbation $\delta\mathbf{Y}(0)$.

\begin{figure}[h!]
	\centering
	\includegraphics[width=0.23\textwidth]{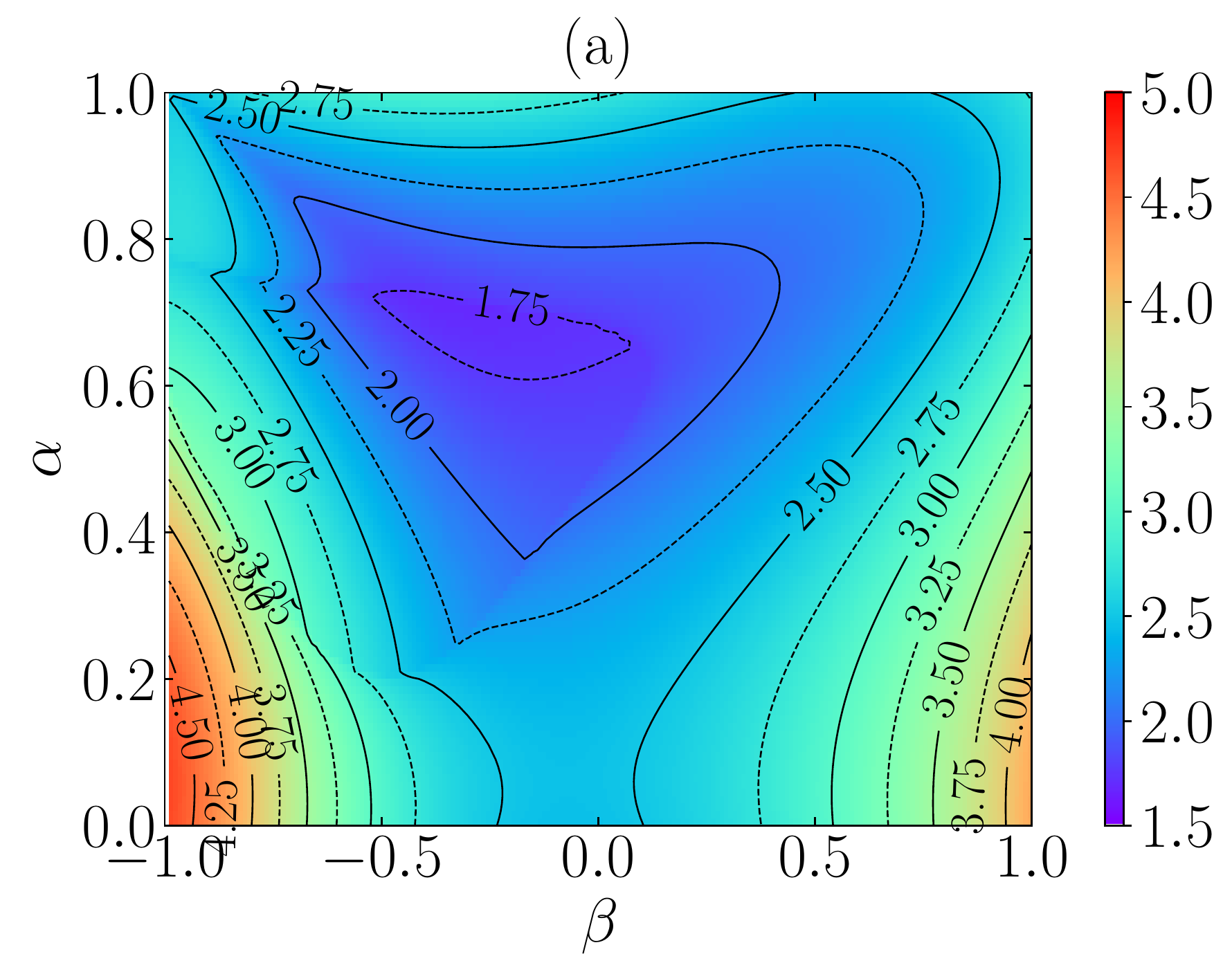}
	\includegraphics[width=0.23\textwidth]{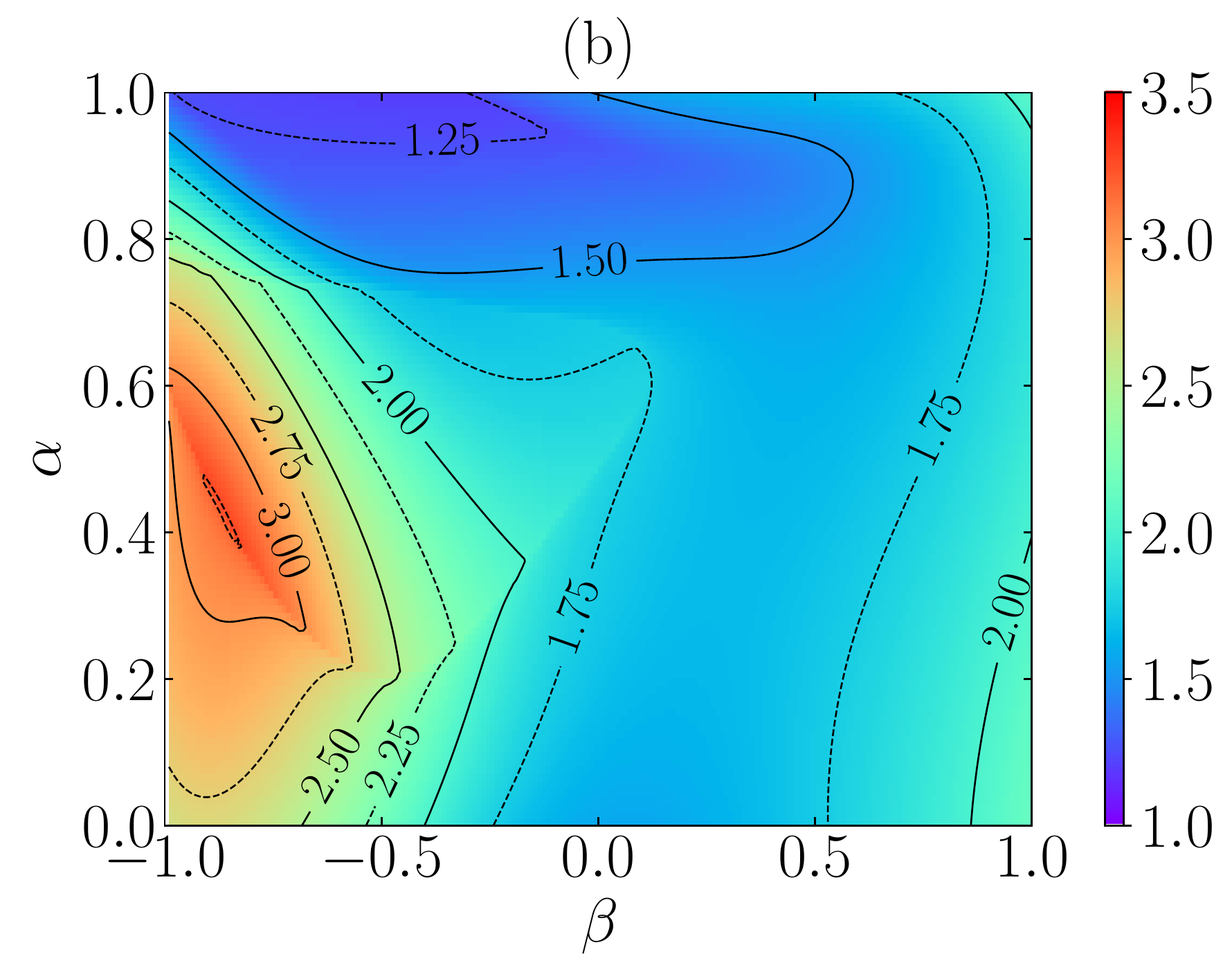}\\
	\includegraphics[width=0.23\textwidth]{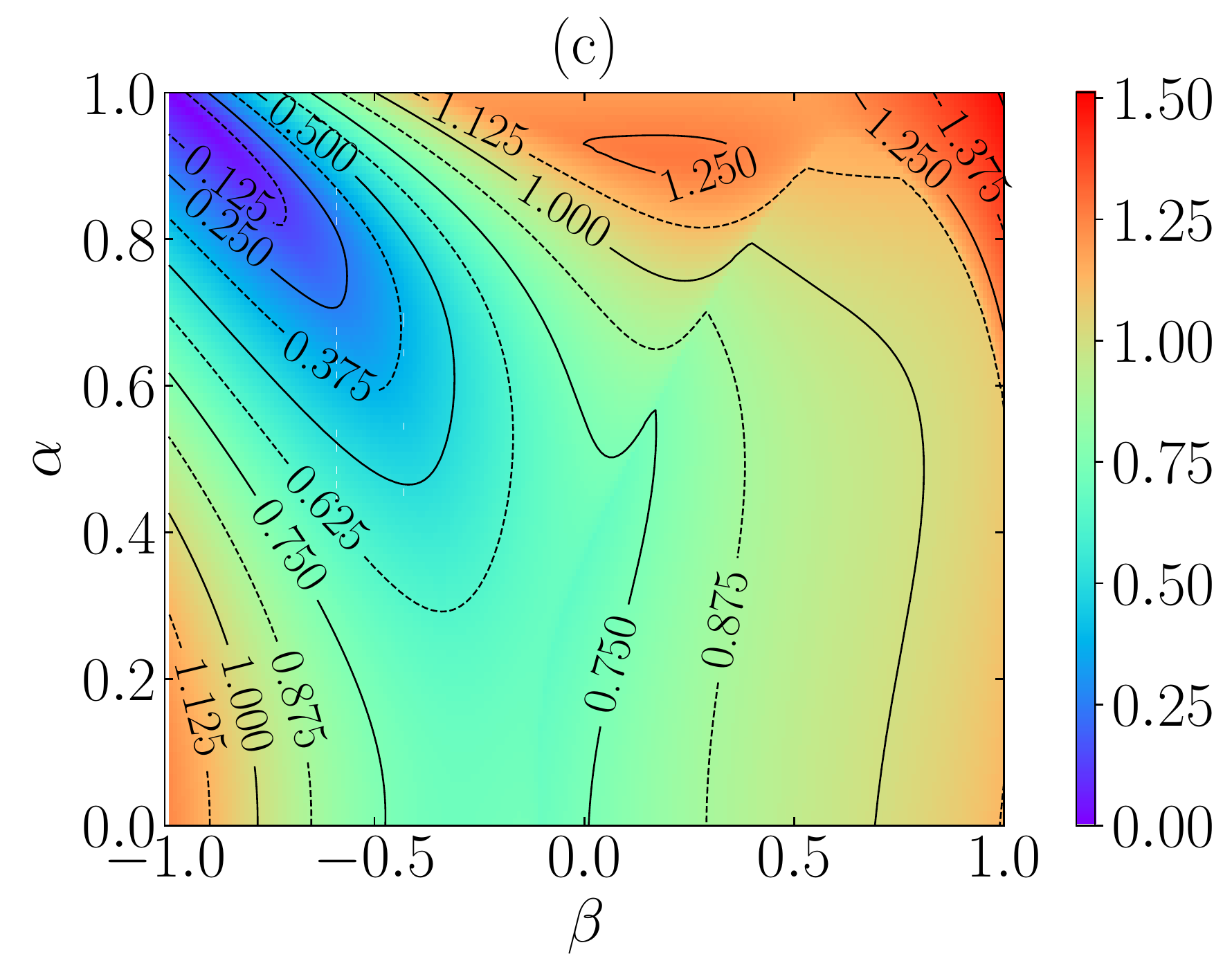}
	\includegraphics[width=0.23\textwidth]{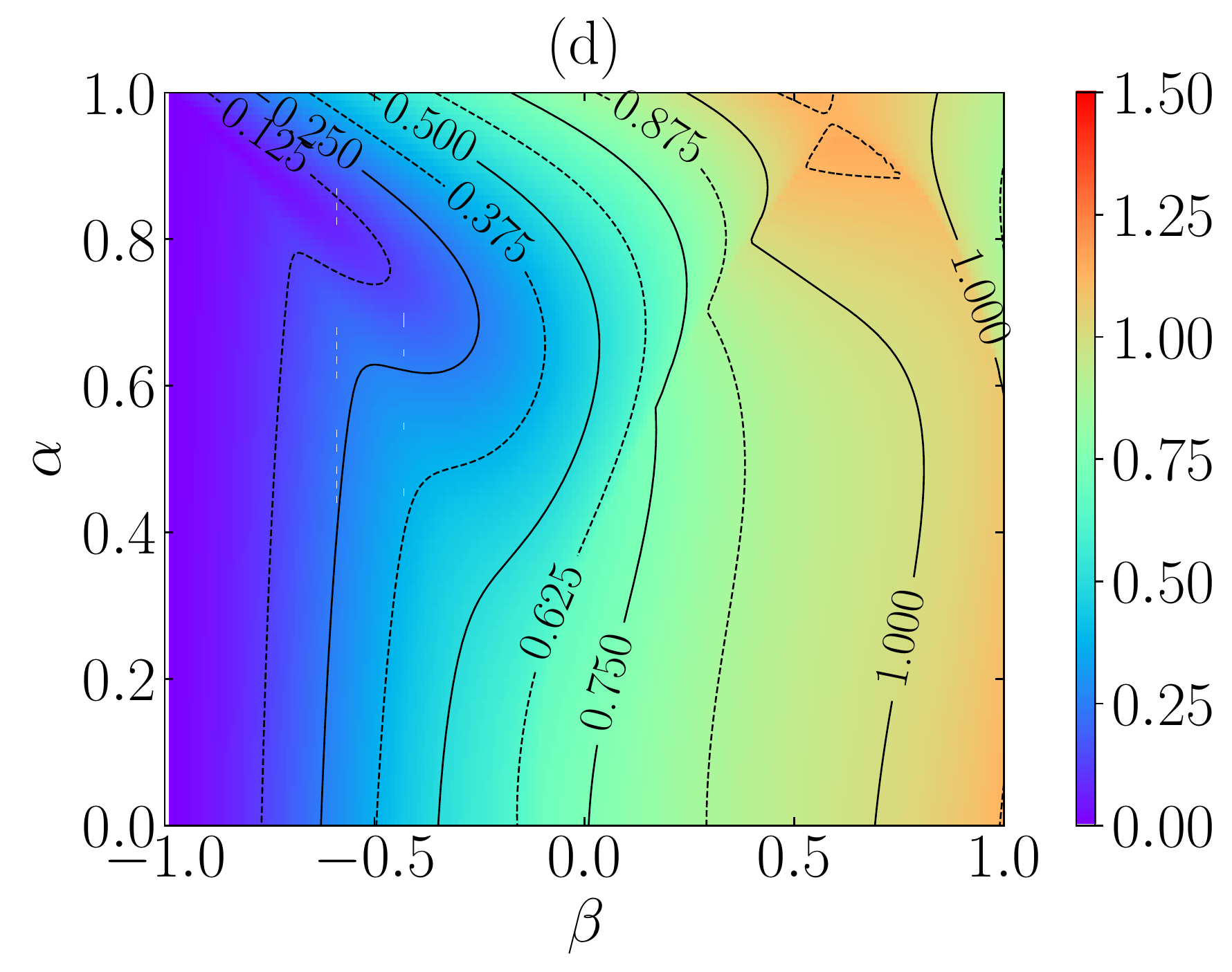}\\
\includegraphics[width=0.3\textwidth]{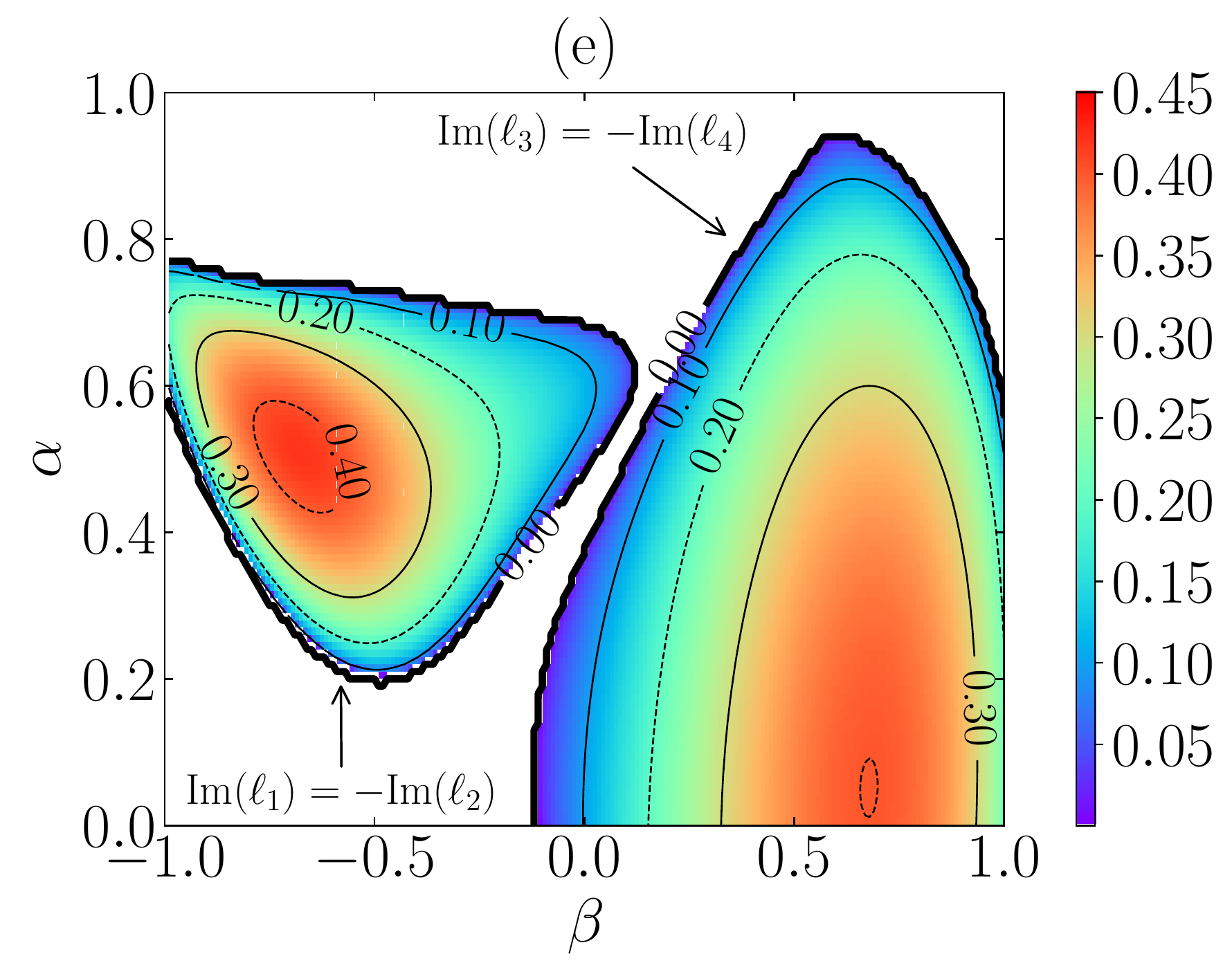}
	\caption{Plot of the four eigenvalues of the matrix $\mathsf{L}$ as functions of $\alpha$ and $\beta$ for uniform disks ($\kappa=\frac{1}{2}$). (a)--(d) show $\mathrm{Re}(\ell_i)$ for $\ell_1$--$\ell_4$, respectively. The imaginary parts, $\mathrm{Im}(\ell_i)$, are plotted in (e), where all the eigenvalues are real-valued inside the blank region.}
	\label{fig:Re_li}
\end{figure}

The dependence of the four eigenvalues on the coefficients of restitution is displayed in Fig.~\ref{fig:Re_li} for uniform disks ($\kappa=\frac{1}{2}$). As can be seen, the real parts are always positive, thus signaling the linear stability and attractor character of the HCS under uniform perturbations, as expected on physical grounds.
In turn, since the cumulant-linearization scheme within the Sonine approximation is not univocally defined~\cite{MS00,SM09}, the fact that we get $\mathrm{Re}(\ell_i)>0$ reinforces the reliability of the  linearization criterion applied to the right-hand sides of Eqs.~\eqref{eq:ev_eqs} and \eqref{eq:HCS}.

The imaginary parts plotted in Fig.~\ref{fig:Re_li}(e) show the regions of the parameter space where the decay toward the HCS is oscillatory. In this respect, the plane $(\alpha,\beta)$ turns out to be split into three disjoint regions: a region where $(\ell_1,\ell_2)$ make a pair of complex conjugates but $(\ell_3,\ell_4)$ are real, a region where $(\ell_3,\ell_4)$ make a pair of complex conjugates but $(\ell_1,\ell_2)$ are real, and, finally, the blank region in Fig.~\ref{fig:Re_li}(e), where the four eigenvalues are real.

\section{\label{ap:ansatz_het} Consistency of the high-velocity tails}

In Sec.~\ref{subsec:HET}, the high-velocity tails of the HCS marginal distributions $\phi_\cc^\HCS(\cc)$, $\phi_\ww^\HCS(\ww)$, and $\phi_{cw}^\HCS(c^2w^2)$ were obtained by assuming Eq.~\eqref{eq:HVT_test}. Here, we test the self-consistency of that assumption.

\subsection{$\phi_\cc^\HCS(\cc)$}

Let us insert Eqs.~\eqref{eq:HVT_c} into the ratio resulting from the replacement $\phi^\HCS(\cw)\to\phi_\cc^\HCS(\cc)$ in Eq.~\eqref{eq:HVT_test}:
\begin{equation}
\label{eq:B1}
    \frac{\phi_\cc^\HCS(\cc_1^{\prime\prime})\phi_\cc^\HCS(\cc_2^{\prime\prime})}{\phi_\cc^\HCS(\cc_1)\phi_\cc^\HCS(\cc_2)}\approx \exp\left[-\gamma_c\left(c_1^{\prime\prime}+c_2^{\prime\prime}-c_1-c_2\right)\right].
\end{equation}
Assuming  $c_1\gg \{1,c_2,w_1,w_2\}$ in the inverse binary collisional rules, Eq.~\eqref{eq:col_rules_inv}, and after some algebra, one gets
\begin{subequations}
\begin{align}
    c_1^{\prime\prime} \approx& c_1\sqrt{1+\frac{\en}{\een}\left(\frac{\en}{\een}-2\right)\cos^2\vartheta_c+\frac{\et^2}{\eet^2}\sin^2\vartheta_c},\\
    c_2^{\prime\prime} \approx& c_1\sqrt{\frac{\en^2}{\een^2}\cos^2\vartheta_c+\frac{\et^2}{\eet^2}\sin^2\vartheta_c},\label{eq:c2primeprime_Ap}
\end{align}
\end{subequations}
where $\vartheta_c = \cos^{-1}\left|\widehat{\cc}_1\cdot\widehat{\boldsymbol{\sigma}} \right|$.
Therefore, the exponent in Eq.~\eqref{eq:B1} is strictly negative, except for smooth particles ($\eet=-1$) and grazing collisions ($\cos\vartheta_c=0$). Thus, apart from those cases with zero Lebesgue measure, $\lim_{c_1\to\infty}{\phi_\cc^\HCS(\cc_1^{\prime\prime})\phi_\cc^\HCS(\cc_2^{\prime\prime})}/{\phi_\cc^\HCS(\cc_1)\phi_\cc^\HCS(\cc_2)}=0$.

\subsection{$\phi_\ww^\HCS(\ww)$}
In the rotational case, from Eqs.~\eqref{eq:phi_w(w)} we have
\begin{equation}
    \frac{\phi_\ww^\HCS(\ww_1^{\prime\prime})\phi_\ww^\HCS(\ww_2^{\prime\prime})}{\phi_\ww^\HCS(\ww_1)\phi_\ww^\HCS(\ww_2)}\approx \left(\frac{w_1^{\prime\prime} w_2^{\prime\prime}}{w_1 w_2}\right)^{-\gamma_w}.
\end{equation}
Let us take $w_1\gg \{1,w_2,c_1,c_2\}$. Then,
\begin{subequations}
\begin{align}
    w_1^{\prime\prime} \approx& w_1\sqrt{1+\frac{\et}{\kappa\eet}\left(\frac{\et}{\kappa\eet}-2\right)\sin^2\vartheta_w},\\
    w_2^{\prime\prime} \approx& w_1\frac{\et}{\kappa|\eet|}|\sin\vartheta_w|,\label{eq:w2primeprime_Ap}
\end{align}
\end{subequations}
where $\vartheta_w = \cos^{-1}\left|\widehat{\ww}_1\cdot\widehat{\boldsymbol{\sigma}} \right|$.
Therefore, $\lim_{w_1\to\infty}\phi^\HCS_\ww(\ww_1^{\prime\prime})\phi^\HCS_\ww(\ww_2^{\prime\prime})/\phi^\HCS_\ww(\ww_1)\phi^\HCS_\ww(\ww_2)=0$, except if $\sin\vartheta_w=0$, which  has zero Lebesgue measure in its continuous domain.

\subsection{$\phi_{cw}^\HCS(x)$}
From Eqs.~\eqref{eq:sol_cw_1}, one has
\begin{equation}
    \frac{\phi_{cw}^\HCS(x_1^{\prime\prime})\phi_{cw}^\HCS(x_2^{\prime\prime})}{\phi_{cw}^\HCS(x_1)\phi_{cw}^\HCS(x_2)}\approx \left(\frac{x_1^{\prime\prime} x_2^{\prime\prime}}{x_1 x_2}\right)^{-\gamma_{cw}}.
\end{equation}
If both $c_1$ and $w_1$ are much larger than $\{1,c_2, w_2\}$, it is possible to obtain
\begin{subequations}
\begin{align}
	x_1^{\prime\prime} \approx & x_1\Bigg[1+\frac{\en}{\een}\left(\frac{\en}{\een}-2\right)\cos^2\vartheta_c+\frac{\et}{\eet}\left(\frac{\et}{\eet}-2\right)\sin^2\vartheta_c\nonumber \\
	 &+\frac{\et^2}{\eet^2}\frac{\theta}{\kappa}\sin^2\vartheta_w-2\frac{\et}{\eet}\left(\frac{\et}{\eet}-1\right)\sqrt{\frac{\theta}{\kappa}}\cos\vartheta_{cw}\Bigg]\nonumber\\
	&\times\Bigg[1+\frac{\et^2}{\eet^2\kappa\theta}\sin^2\vartheta_c-\frac{2\et}{\eet\sqrt{\kappa\theta}}\left(\frac{\et}{\eet\kappa}-1\right) \cos\theta_{cw}\nonumber\\
	&+\frac{\et}{\eet\kappa}\left(\frac{\et}{\eet\kappa}-2\right)\sin^2\vartheta_w\Bigg],
\end{align}
\begin{align}
	x_2^{\prime\prime} \approx & \frac{x_1\et^2}{\kappa\theta\eet^2}\left(\sin^2 \vartheta_c+\frac{\theta}{\kappa}\sin^2\vartheta_{w}
-2\sqrt{\frac{\theta}{\kappa}}\cos\vartheta_{cw} \right)\Bigg[\frac{\en^2}{\een^2}\cos^2\vartheta_c\nonumber \\
	& +\frac{\et^2}{\eet^2}\left(\sin^2 \vartheta_c+\frac{\theta}{\kappa}\sin^2\vartheta_{w}-2\sqrt{\frac{\theta}{\kappa}}\cos\vartheta_{cw} \right)\Bigg],
\end{align}
\end{subequations}
where $\vartheta_{cw}=\cos^{-1}\left[\widehat{\cc}_1\cdot\left(\widehat{\boldsymbol{\sigma}}\times\widehat{\ww}_1 \right)\right]$.
Thus, $\lim_{x_1\to\infty}{\phi_{cw}^\HCS(x_1^{\prime\prime})\phi_{cw}^\HCS(x_2^{\prime\prime})}/{\phi_{cw}^\HCS(x_1)\phi_{cw}^\HCS(x_2)}=0$, except if  $\sin\vartheta_c=\sin\vartheta_w=\cos\vartheta_{cw}=0$, which again have zero Lebesgue measure.



%

\newpage
\clearpage

\foreach \x in {1,...,17}
{%
\clearpage
\includepdf[pages={\x}]{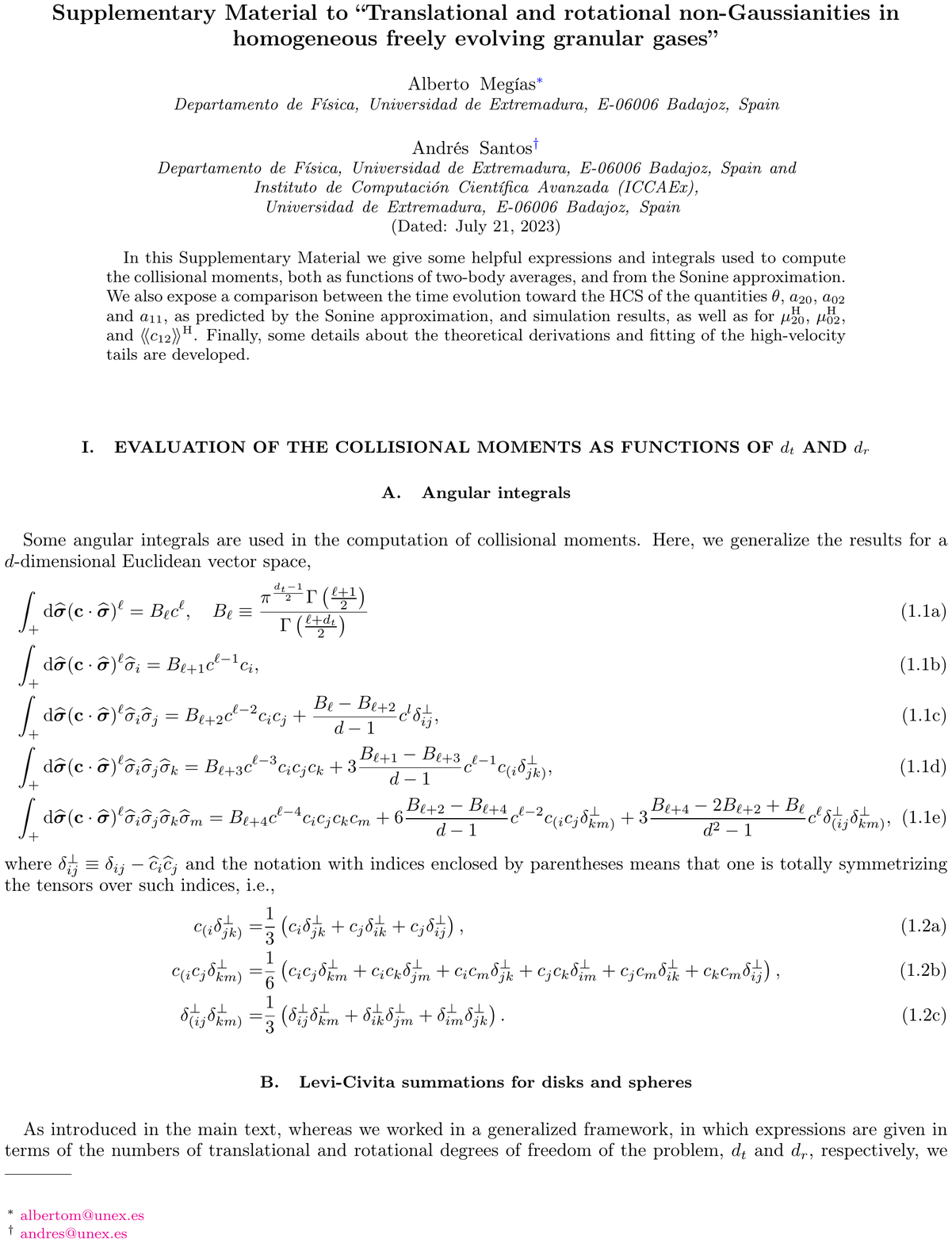}
}

\end{document}